\documentclass[10pt,preprint]{aastex}
\usepackage{amssymb,amsmath,graphicx,lscape}
\usepackage[T1]{fontenc}

\begin{document}

\title{A High Resolution Study of the H{\textsc i}--H$\boldsymbol{_{2}}$ Transition \\ 
across the Perseus Molecular Cloud}

\author{Min-Young Lee\altaffilmark{1}, 
Sne\v{z}ana Stanimirovi\'c\altaffilmark{1},
Kevin A. Douglas\altaffilmark{2}, 
Lewis B. G. Knee\altaffilmark{3,4,5}, \\  
James Di Francesco\altaffilmark{4}, 
Steven J. Gibson\altaffilmark{6},  
Ayesha Begum\altaffilmark{1},
Jana Grcevich\altaffilmark{7}, 
Carl Heiles\altaffilmark{8}, 
Eric J. Korpela\altaffilmark{9}, \\
Adam K. Leroy\altaffilmark{5}, 
J. E. G. Peek\altaffilmark{7,10},
Nickolas M. Pingel\altaffilmark{1}, 
Mary E. Putman\altaffilmark{7}, 
and Destry Saul\altaffilmark{7}}

\altaffiltext{1}{Department of Astronomy, University of Wisconsin-Madison, 475 North Charter Street, Madison, WI 53706, USA}
\altaffiltext{2}{Arecibo Observatory, NAIC, HC3 Box 53995, Arecibo, Puerto Rico, PR 00612, USA}
\altaffiltext{3}{Joint ALMA Observatory, Alonso de Cordova 3107, Vitacura, Santiago, Chile}
\altaffiltext{4}{National Research Council of Canada, Herzberg Institute of Astrophysics, 
5071 West Saanich Road, Victoria, BC, V9E 2E7 Canada} 
\altaffiltext{5}{National Radio Astronomy Observatory, Charlottesville, VA 22903, USA}
\altaffiltext{6}{Department of Physics and Astronomy, Western Kentucky University, Bowling Green, KY 42101, USA}
\altaffiltext{7}{Department of Astronomy, Columbia University, New York, NY 10027, USA}
\altaffiltext{8}{Radio Astronomy Lab, University of California-Berkeley, 601 Campbell Hall, Berkeley, CA 94720, USA}
\altaffiltext{9}{Space Sciences Laboratory, University of California-Berkeley, Berkeley, CA 94720, USA}
\altaffiltext{10}{Hubble Fellow}

\begin{abstract}
To investigate the fundamental principles of H$_{2}$ formation in a giant molecular cloud,
we derive the H\textsc{i} and H$_{2}$ surface density ($\Sigma_{\rm H\textsc{i}}$ and $\Sigma_{\rm H2}$) images
of the Perseus molecular cloud on sub-pc scales ($\sim$0.4 pc).
We use the far-infrared data from the Improved Reprocessing of the \textit{IRAS} Survey
and the $V$-band extinction image provided by the COMPLETE Survey to estimate the dust column density image of Perseus.
In combination with the H\textsc{i} data from the Galactic Arecibo L-band Feed Array H\textsc{i} Survey
and an estimate of the local dust-to-gas ratio,
we then derive the $\Sigma_{\rm H2}$ distribution across Perseus. 
We find a relatively uniform $\Sigma_{\rm H\textsc{i}} \sim$ 6--8 M$_{\odot}$ pc$^{-2}$ for both dark and star-forming regions, 
suggesting a minimum H\textsc{i} surface density required to shield H$_{2}$ against photodissociation.
As a result, a remarkably tight and consistent relation is found between 
$\Sigma_{\rm H2}/\Sigma_{\rm H\textsc{i}}$ and $\Sigma_{\rm H\textsc{i}}+\Sigma_{\rm H2}$.
The transition between the H\textsc{i}- and H$_{2}$-dominated regions occurs 
at $N{\rm (H\textsc{i})}+2N(\rm{H_{2}})$ $\sim$ (8--14) $\times$ 10$^{20}$ cm$^{-2}$. 
Our findings are consistent with predictions for H$_{2}$ formation in equilibrium, 
suggesting that turbulence may not be of primary importance for H$_{2}$ formation.
However, the importance of warm neutral medium for H$_{2}$ shielding,
internal radiation field, and the timescale of H$_{2}$ formation still remain as open questions.
We also compare H$_{2}$ and CO distributions and estimate the fraction of ``CO-dark'' gas, $f_{\rm DG}$ $\sim$ 0.3.
While significant spatial variations of $f_{\rm DG}$ are found,
we do not find a clear correlation with the mean $V$-band extinction.
\end{abstract}

\section{Introduction}

Most of the dense molecular gas in galaxies is assembled into giant molecular clouds (GMCs)
with masses from 10$^{4}$ M$_{\odot}$ to 10$^{7}$ M$_{\odot}$ 
and radii from 50 pc to a few hundred pc (e.g., Fukui \& Kawamura 2010).
As stars, the ``atoms'' of galaxies, are exclusively formed in these GMCs, 
physical processes that govern the conversion of H\textsc{i} into H$_{2}$ 
clearly play an important role in determining the properties of GMCs 
and consequently the initial conditions for star formation.
Therefore, understanding the properties of the interstellar regions
where gas transitions from H$\textsc{i}$--H$_{2}$-dominated (H\textsc{i}--H$_{2}$ transition hereafter) 
is an important step toward a complete theory of star formation.

From an observational point of view, the H\textsc{i}--H$_{2}$ transition has been investigated in the Galaxy 
mainly via ultraviolet (UV) absorption measurements toward background sources.
For example, Savage et al. (1977) and Gillmon et al. (2006) measured H$_{2}$ column densities, $N$(H$_{2}$),
along the lines of sight to early-type stars and AGNs using the \textit{Copernicus} and \textit{FUSE} satellites
and combined them with H\textsc{i} column densities, $N$(H\textsc{i}),
to estimate H$_{2}$ fraction, $f$ = 2$N$(H$_{2}$)/($N$(H\textsc{i}$)+2N$(H$_{2}$)).  
They found that $f$ sharply rises at $N$(H) = $N$(H\textsc{i}$)+2N$(H$_{2}$) $\sim$ (3--5) $\times$ 10$^{20}$ cm$^{-2}$.
Several far-infrared (FIR) studies of interstellar clouds have revealed 
the presence of H$_{2}$ based on the comparison between dust and atomic gas contents 
(e.g., Reach et al. 1994; Meyerdierks \& Heithausen 1996; Douglas \& Taylor 2007).
Reach et al. (1994) found a significant positive deviation from a linear relation 
between IR intensity at 100 $\mu$m and $N$(H\textsc{i}) for cirrus clouds,
indicating a substantial amount of H$_{2}$.
The threshold $N$(H\textsc{i}) at which $N$(H$_{2}$) implied by the FIR data becomes significant 
was found to be $\sim$4 $\times$ 10$^{20}$ cm$^{-2}$, 
consistent with the results from Savage et al. (1977) and Gillmon et al. (2006).
 
More recently, the H\textsc{i}--H$_{2}$ transition has been observationally 
inferred by an absence of high $N$(H\textsc{i}). 
For example, Barriault et al. (2010) observed two high Galactic latitude clouds
in the OH and H\textsc{i} transitions and found that $N$(H\textsc{i}) saturates
at $\sim$5 $\times$ 10$^{20}$ cm$^{-2}$ for high values of $N$(OH),
suggesting that there is molecular gas not traced by H\textsc{i}.
Extragalactic studies have found this saturation as well.
Wong \& Blitz (2002), Blitz \& Rosolowsky (2004, 2006), Bigiel et al. (2008), and Wong et al. (2009)
observed nearby galaxies in the CO and H\textsc{i} transitions 
and found that H\textsc{i} surface density, $\Sigma_{\rm H\textsc{i}}$, saturates at $\sim$10 M$_{\odot}$ pc$^{-2}$.
Most extragalactic studies have a spatial resolution of $\sim$700 pc at best, 
where most GMCs are unresolved and several are likely to be blended along a line of sight. 
In addition, at this resolution it is impossible to compare H$_{2}$ 
with the corresponding H\textsc{i} associated with the cloud envelopes.

From a theoretical point of view, the H\textsc{i}--H$_{2}$ transition has been studied
in the context of the structure of photodissociation regions (PDRs).
As UV photons filling up the bulk of the interstellar medium (ISM) photodissociate H$_{2}$,
predominantly molecular gas is only found in the dense regions 
where high enough shielding is achieved by both dust grains and H$_{2}$ self-shielding.
These molecular regions are bounded by PDRs where the gas predominantly exists 
in the atomic phase (e.g., Hollenbach \& Tielens 1997).
The structure of PDRs has been mainly investigated for 1D geometries that include
uni-directional or bi-directional beams of radiation propagating into semi-infinite slabs
or purely radial radiation fields hitting the surfaces of spherical clouds
(e.g., van Dishoeck \& Black 1986; Draine \& Bertoldi 1996; Browning et al. 2003).
However, this 1D approximation is not appropriate for GMCs
that are embedded in a radiation field produced by the combination of many single stars and star clusters.
In addition, when GMCs are surrounded by large H\textsc{i} envelopes, 
which have been frequently observed in the Galaxy (e.g., Wannier et al. 1983, 1991), 
the curvature of the PDRs and the finite size of the molecular regions are no longer negligible.
A higher dimensional approach is required in this case, 
but as it requires a treatment of the angular dependence of the radiation field,  
all previous studies have been purely numerical (e.g., Spaans \& Neufeld 1997; Liszt \& Lucas 2000; Liszt 2002).
These numerical studies do not provide a simple analytic estimate of the structure of PDRs
that allows us to extrapolate the models over a wide range of ISM environments 
and to easily compare with observations.

Recently, Krumholz et al. (2008, 2009; KMT08 and KMT09 hereafter) performed analytic studies 
on a PDR in a spherical cloud that is bathed in a uniform and isotropic interstellar radiation field (ISRF). 
They investigated the steady-state equilibrium formation and photodissociation of H$_{2}$ 
and approximated the H\textsc{i}--H$_{2}$ transition as being infinitely sharp. 
Their calculation of where exactly inside a molecular cloud this transition happens 
resulted in two important analytic predictions.
First, they found that a certain amount of $\Sigma_{\rm H\textsc{i}}$ is required 
to provide enough shielding for H$_{2}$ formation. 
For GMCs with solar metallicity, this is $\Sigma_{\rm H\textsc{i}}$ $\sim$ 10 M$_{\odot}$ pc$^{-2}$.
Once the minimum $\Sigma_{\rm H\textsc{i}}$ is achieved to shield H$_{2}$ against photodissociation,
all excess H\textsc{i} is converted into H$_{2}$ and the H$_{2}$ abundance increases.
Second, they derived an analytic function for H$_{2}$-to-H\textsc{i} ratio,  
$R_{\rm H2}$ = $\Sigma_{\rm H2}$/$\Sigma_{\rm H\textsc{i}}$,  
and showed that $R_{\rm H2}$ depends on the total gas surface density and metallicity 
but is interestingly almost independent of the strength of the ISRF. 
This is a very powerful prediction, suggesting that for a given ISM environment, 
the total gas surface density determines the H$_{2}$-to-H\textsc{i} ratio.  
If confirmed by observations, this ``easy'' prescription of H$_{2}$ formation
will be invaluable for numerical simulations of galaxy formation and evolution.
From a purely practical perspective, as the above predictions are functions of direct observables 
(total gas surface density and metallicity), they are much easier to test observationally.
Most of the previous models of H$_{2}$ formation 
(e.g., van Dishoeck \& Black 1986; Draine \& Bertoldi 1996; Browning et al. 2003) 
require the intensity of the UV radiation field and the number density of H nuclei as input parameters,
which cannot be easily determined by observations.

KMT09's predictions specifically focus on the properties of GMCs on galactic scales ($\sim$kpc). 
Several recent studies investigated the H\textsc{i} and H$_{2}$ properties of nearby spiral 
and dwarf galaxies on kpc scales and found reasonable agreement with KMT09's prediction 
of $R_{\rm H2}$ as a function of total gas surface density and metallicity (e.g., KMT09; Fumagalli et al. 2009, 2010).
On the other hand, KMT09's simplified treatment of the thermal and chemical processes in PDRs  
may be less suited for studies of individual PDRs.
For example, KMT09's model does not consider the temperature dependence of H$_{2}$ formation rate coefficients,
H$_{2}$ dissociation by cosmic rays, or absorption of Lyman-Werner (LW) photons by molecules other than H$_{2}$.
Nevertheless, KMT09's model provides strong physical insights into the fundamental principles of H$_{2}$ formation.

\begin{figure}
\begin{center}
\includegraphics[scale=0.25]{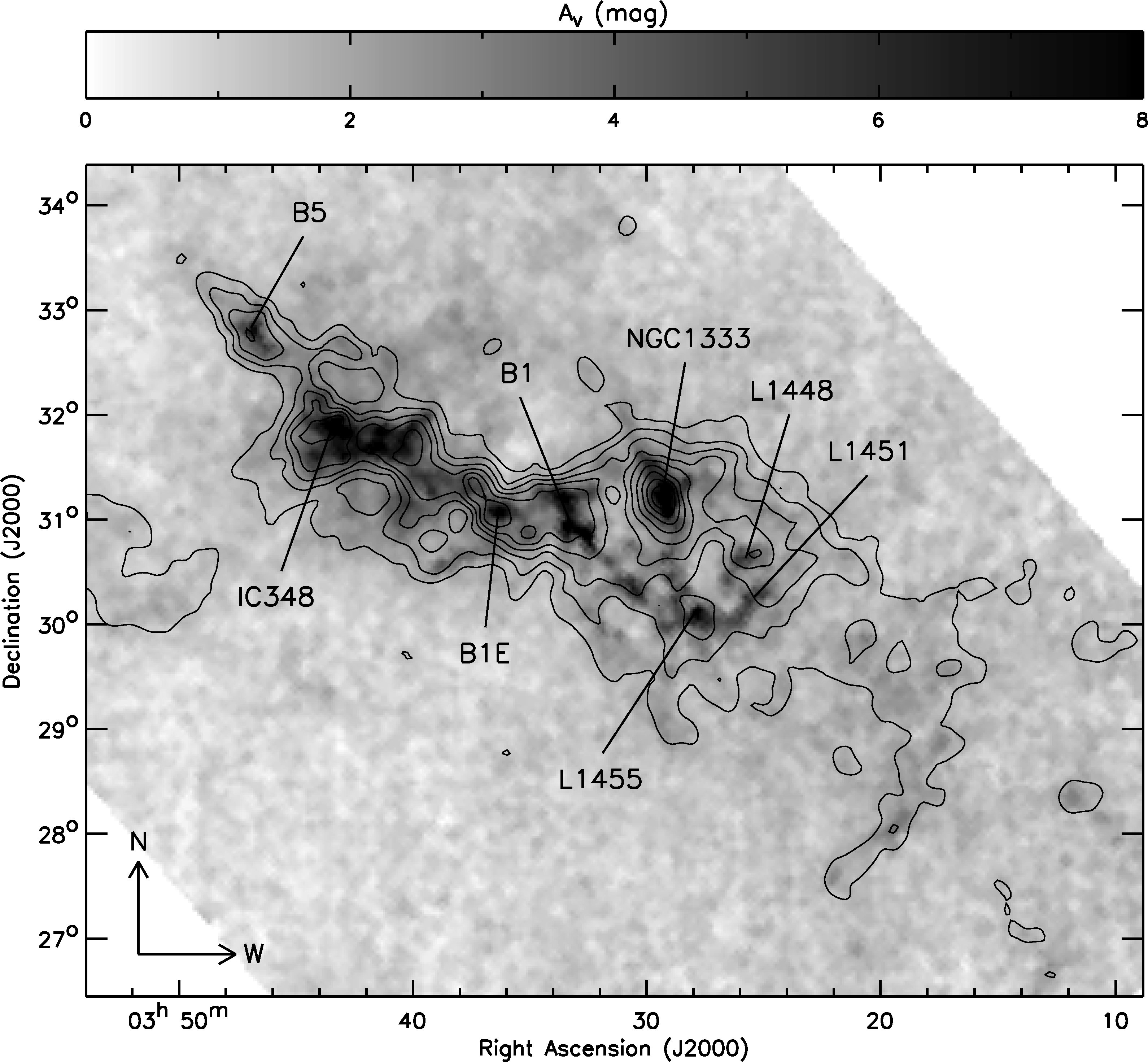}
\caption{\label{f:intro} 2MASS $A_{V}$ image of Perseus.  
The contours of CfA $I_{\rm CO}$ are overlaid in black. 
The contour levels range from 10\% to 90\% of the peak (69 K km s$^{-1}$) with 10\% steps. 
The angular resolution of the 2MASS $A_{V}$ and CfA $I_{\rm CO}$ images is 5$'$ and 8.4$'$, respectively.
Several prominent dark and star-forming regions are labeled.} 
\end{center}
\end{figure}

In this paper, we test KMT09's predictions on the Perseus molecular cloud (Perseus hereafter) 
on sub-pc scales ($\sim$0.4 pc) for the first time in an attempt to understand the fundamental principles of H$_{2}$ formation in GMCs. 
Perseus is a nearby molecular cloud located at a distance ranging from 200 pc to 350 pc 
(Herbig \& Jones 1983; \v Cernis 1990) and has roughly a projected angular size of 
6$^{\circ}$ by 3$^{\circ}$ on the sky (based on CO emission\footnote{In this paper, all $^{12}$CO(1--0) data are quoted as CO.}).  
We adopt the distance to Perseus of 300 pc throughout this paper.
Perseus is part of the large Taurus--Auriga--Perseus molecular cloud complex 
and lies below the Galactic plane ($b\sim-20^{\circ}$),
resulting in slightly less confused H\textsc{i} spectra compared to other Galactic GMCs. 
With $M \sim 10^4$ M$_{\odot}$ (Sancisi et al. 1974; Ungerechts \& Thaddeus 1987), 
this is a low mass cloud with an intermediate level of star formation (Bally et al. 2008). 
Figure \ref{f:intro} shows the $V$-band extinction image of Perseus
with CO contours to emphasize the locations of several prominent dark and star-forming regions.  
IC348 (age = 0.7--12 Myr; Herbig 1998) and NGC 1333 (age < 1 Myr; Lada et al. 1996) are active star-forming regions, 
while B5, B1E, B1, L1455, L1448, and L1451 are dark regions with low levels of star formation. 
A wealth of multi-wavelength observations exist for Perseus; we provide details in Section 2. 
One of the recent observational advances is the high resolution H\textsc{i} data set of Perseus 
and its outskirts provided by the GALFA--H\textsc{i} survey (Peek et al. 2011). 
This data set nicely matches the angular resolution of FIR images, $\sim$4$'$, 
allowing us to estimate $R_{\rm H2}$ across and around Perseus and to compare the results to KMT09's predictions.

This paper is organized in the following way. 
We start with a description of the data sets used in this study (Section 2).
We then describe details of our methodology to derive $N$(H\textsc{i}) and $N$(H$_{2}$) images (Sections 3 and 4). 
In Section 5, we examine the dust temperature image closely  
and discuss the implications for the strength of the radiation field across Perseus.  
In Section 6, we summarize KMT09's model and predictions.
In Section 7, we compare our $\Sigma_{\rm H\textsc{i}}$ and $R_{\rm H2}$ images with KMT09's predictions. 
We then compare the derived H$_{2}$ distribution with the existing CO observations
and investigate the existence of the ``CO-dark'' H$_{2}$ gas (Section 8). 
Finally, we discuss and summarize our conclusions (Sections 9 and 10). 

\section{Data}

\subsection{H\textsc{i}}

We use H\textsc{i} data from the Galactic Arecibo L-band Feed Array H\textsc{i} Survey (GALFA--H\textsc{i}).
GALFA--H\textsc{i} uses ALFA, a seven-beam array of receivers
mounted at the focal plane of the 305-m Arecibo telescope, to map H\textsc{i} emission in the Galaxy.
Each of seven dual polarization beams has an effective beamsize of 3.9$'$ $\times$ 4.1$'$ 
and a gain of 8.5--11 Jy K$^{-1}$ (Peek et al. 2011). 
The GALFA--H\textsc{i} spectrometer, GALSPECT, has a velocity resolution of 0.184 km s$^{-1}$ (872 Hz)
and covers $-700$ km s$^{-1}$  < $v$ < $+700$ km s$^{-1}$ (7 MHz) in the Local Standard of Rest (LSR) 
frame\footnote{In this paper, all velocities are quoted in the LSR frame.}.

GALFA--H\textsc{i} adopts two different scanning modes, the drift and basketweaving modes.  
In the drift mode, the telescope is fixed at azimuth of 0$^{\circ}$ or 180$^{\circ}$ 
and the sky is allowed to drift by.
In the basketweaving mode, on the other hand, the telescope is fixed at the meridian  
and is driven up and down in zenith angle over the chosen declination (Dec) range. 
For each day of observations, the starting point of the telescope is shifted in right ascension (RA) by $\sim$12$'$ 
such that the entire proposed region is covered.
ALFALFA (Giovanelli et al. 2005) and GALFACTS (Guram \& Taylor 2009), 
two large area surveys with which GALFA--H\textsc{i} commensally observes, 
use the drift and basketweaving mode, respectively.  

GALFA--H\textsc{i} combines H\textsc{i} data from many individual projects, both in solo and commensal modes.
Reduction of H\textsc{i} data is performed with the GALFA--H\textsc{i} Standard Reduction (GSR) pipeline. 
Full details of GSR are presented in Peek et al. (2011), but here we briefly describe its most important steps.  
First, H\textsc{i} spectra are corrected for the intermediate frequency (IF) gain 
by dividing them by a reference spectrum.
This reference spectrum is obtained via the least-squares frequency switching (LSFS) technique (Heiles 2005). 
The LSFS observation is performed at the beginning of each observing period 
and the obtained IF gain spectrum is used for the whole period's data. 
Second, the H\textsc{i} spectra are searched for baseline ripples 
produced by reflections in the signal chain, the primary reflector, and the Gregorian dome. 
This initial calibration is done by examining the H\textsc{i} data in the Fourier space 
and removing the Fourier components that correspond to the reflection patterns.   
Third, the gain calibration is applied to the H\textsc{i} spectra.  
The relative gains of each beam over each day of observations are determined 
by finding the points where beam tracks intersect and comparing the H\textsc{i} spectra of each beam.  
This gain calibration is most effective for the scans obtained in the basketweaving mode, 
where a large number of intersection points are produced.  
Lastly, the fully calibrated H\textsc{i} spectra are gridded onto the image plane in Cartesian coordinates. 
The gridded H\textsc{i} data are then calibrated for the first sidelobe and the final brightness temperature scale.  
The lower angular resolution H\textsc{i} data from the Leiden--Argentine--Bonn (LAB) Survey (Kalberla et al. 2005) 
are used for the final brightness temperature calibration.

The GALFA--H\textsc{i} first data release (DR1) has been presented by Peek et al. (2011). 
The DR1 data cover 7520 deg$^{2}$ of sky and have been produced using 3046 hours of data 
obtained from 12 individual projects.
RMS noise in a 1 km s$^{-1}$ channel ranges from 140 mK to 60 mK,  with a median of 80 mK. 
The DR1 data can be downloaded at https://purcell.ssl.berkeley.edu/.
The H\textsc{i} data used in this paper are part of DR1 
and have been obtained from TOGS2 (Turn-On GALFA Spectrometer2; PI: C. Heiles \& S. Stanimirovi\'c),
a2004 (H\textsc{i} Survey of the Taurus Molecular Cloud; PI: P. Goldsmith),
and a2174 (H\textsc{i} Survey of the Perseus Molecular Cloud; PI: L. Knee).  
We combined the scans from these three projects and produced an H\textsc{i} cube centered 
at (RA,Dec) = (03$^{\rm h}$29$^{\rm m}$52$^{\rm s}$,$+$30$^{\rm \circ}$34$'$1$''$) in J2000\footnote{In this paper,
all coordinates are quoted in J2000.} with a size of 14.8$^{\rm \circ}$ $\times$ 9$^{\rm \circ}$.

\subsection{Infrared}

We use 60 $\mu$m and 100 $\mu$m images from the Improved Reprocessing of the \textit{IRAS} Survey (IRIS) 
(Miville-Desch$\rm \hat{e}$nes \& Lagache 2005).
IRIS provides an improved zero point calibration and correction for the zodiacal light and striping in the images,
as well as increased angular resolution compared to the original \textit{IRAS} images.
The angular resolution at 60 $\mu$m and 100 $\mu$m is 4$'$ and 4.3$'$, respectively, 
comparable to that of the GALFA--H\textsc{i} data.
The noise level in the IRIS images is approximately 0.07 MJy sr$^{-1}$ 
and 0.48 MJy sr$^{-1}$ at 60 $\mu$m and 100 $\mu$m, respectively. 

\subsection{CO}

We use the CO integrated intensity image provided by Dame et al. (2001). 
Dame et al. (2001) combined CO surveys of the Galactic plane and local molecular clouds 
and produced a composite survey of the entire Galaxy at an angular resolution of 8.4$'$.  
The CO data were obtained by the 1.2-m telescope at the Harvard-Smithsonian Center for Astrophysics (CfA).
The spectra were sampled with an angular spacing of 7.5$'$   
and the final data cube has a uniform RMS noise of 0.25 K per a 0.65 km s$^{-1}$ channel. 
To estimate CO integrated intensity, $I_{\rm CO}$, Dame et al. (2001) integrated CO emission 
from $v = -15$ km s$^{-1}$ to $+15$ km s$^{-1}$.
See Section 2 of Dame et al. (2001) for details of the observations and analysis. 

While we use the CfA CO data for the most of our analysis, $^{13}$CO(1--0) data provided by  
the COMPLETE survey (Ridge et al. 2006a) are used in Section 7.  
As $^{13}$CO(1--0) is a better tracer of dense molecular gas than CO,  
we use the $I_{^{13}\rm CO}$ image with a size of 6.25$^{\circ} \times 3^{\circ}$ to set the boundaries of the dark and star-forming regions in Perseus.
The $^{13}$CO(1--0) data were obtained by the 14-m Five College Radio Astronomy Observatory (FCRAO) telescope
at an angular resolution of 44$''$. 
To study Perseus, Ridge et al. (2006a) integrated $^{13}$CO(1--0) emission from $v = 0$ km s$^{-1}$ to $+20$ km s$^{-1}$ 
with a velocity resolution of 0.07 km s$^{-1}$ and estimated an average RMS noise 
per channel of 0.17 K. 

\subsection{$\boldsymbol{A_{\rm V}}$}

We use the $A_{V}$ image provided by the COMPLETE survey.
This image was constructed from the Two Micron All Sky Survey (2MASS) Point Source Catalog 
using the NICER algorithm (Lombardi \& Alves 2001) at an angular resolution of 5$'$. 
The NICER algorithm estimates reddening along a line of sight by comparing the observed colors of 
background stars to the intrinsic colors (measured in a nearby control field, where the extinction is negligible). 
Note that this method does not assume a grain-size distribution or dust emissivity 
and therefore provides the most reliable estimate of the total gas column density in a molecular cloud  
(e.g., Goodman et al. 2009; GPS09 hereafter).
In this paper, however, we do not directly use the 2MASS $A_{V}$ image to derive 
the total gas column density because it only covers the central region of Perseus,
while we are mainly interested in the cloud outskirts.
Instead, we use the 2MASS $A_{V}$ image to calibrate our $\tau_{100}$ image derived from the IRIS data (see Section 4.4).

\subsection{Treatment of Angular Resolution} 

In this paper, we combine the data sets described above in several ways. 
In each case, we first bring all input data sets to common angular resolution 
determined by the lowest angular resolution data set. 
This is done by performing convolution with an appropriate Gaussian kernel.
For example, to determine the velocity range of H\textsc{i} emission 
based on the correlation between $N$(H\textsc{i}) and the dust column density, 
we smooth the H\textsc{i} data to the angular resolution of the 2MASS $A_{V}$ image, 5$'$ (Section 3.2).  
To derive $N$(H$_{2}$), the IRIS and H\textsc{i} data are convolved with a Gaussian kernel
to have an angular resolution of 4.3$'$.  
Consequently, our final $\Sigma_{\rm H\textsc{i}}$, $\Sigma_{\rm H2}$, and $R_{\rm H2}$ images 
have an angular resolution of 4.3$'$ (Sections 4.5 and 7). 

\section{Methodology: Steps in Deriving \textit{N}(H\textbf{\textsc{i}})}

\subsection{General Properties of H{\textbf{\textsc i}} Spectra}

\begin{figure}
\begin{center}
\includegraphics[scale=0.5]{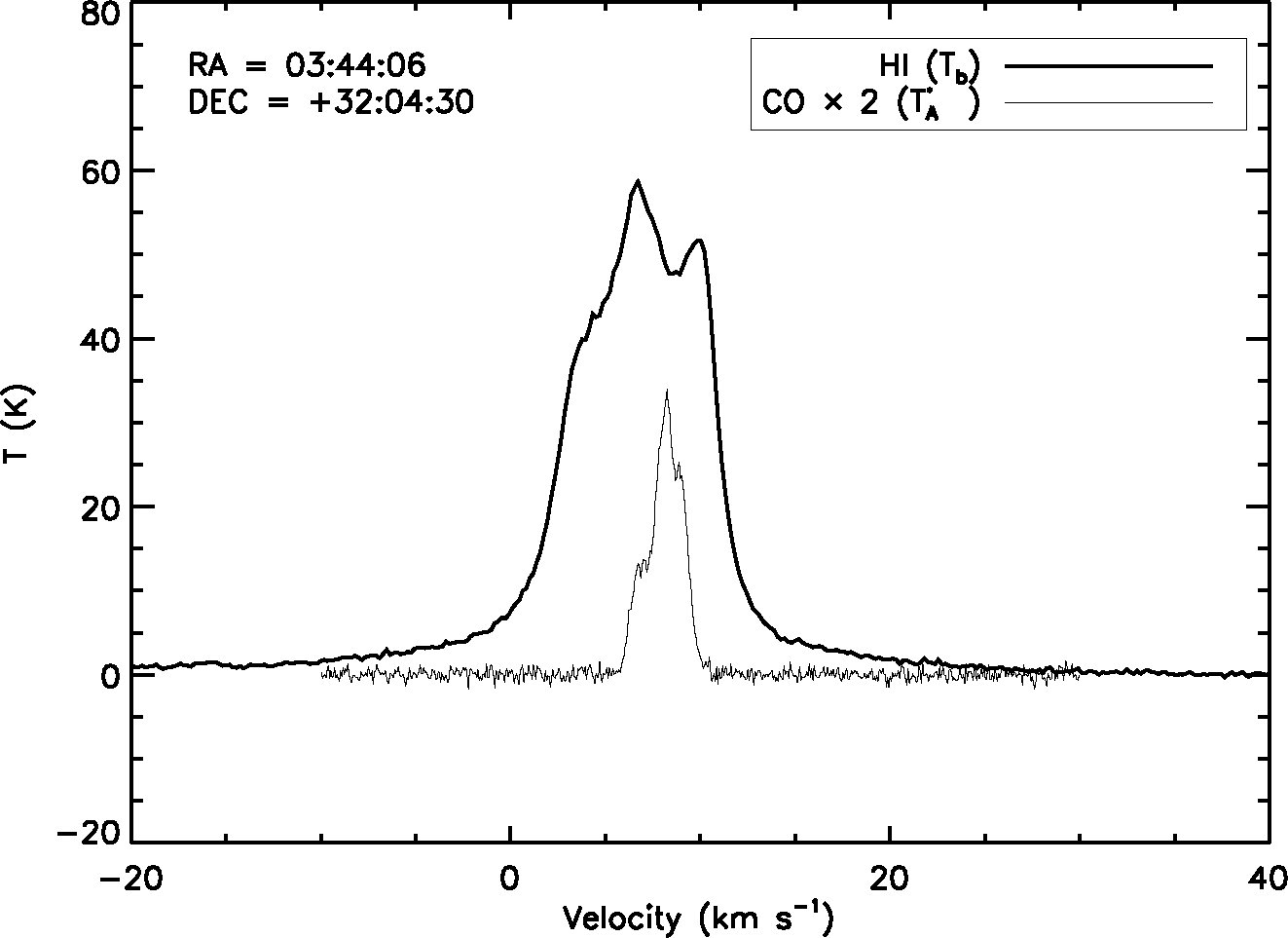}
\caption{\label{f:HI_CO_spectra} H\textsc{i} (thick) and CO (thin) spectra at the position of IC348.
For comparison, CO is amplified twice. 
The main component of H\textsc{i} peaks at $v \sim +5$ km s$^{-1}$
and is offset by 2--3 km s$^{-1}$ from the peak of CO.}
\end{center}
\end{figure}

The main component of the H\textsc{i} emission in Perseus is centered at $v$ = $+4 \sim +8$ km s$^{-1}$.
Compared to the molecular gas traced by CO, the peak of the H\textsc{i} emission is offset by 2--3 km s$^{-1}$. 
As an example, we compare the H\textsc{i} and CO spectra at the position of IC348 in Figure \ref{f:HI_CO_spectra}.
There are (at least) two components in the H\textsc{i} spectrum 
and the stronger one peaks at $v \sim$ $+$5 km s$^{-1}$. 
Interestingly, the location of the dip between the two H\textsc{i} components 
corresponds to the CO peak at $v \sim$ $+$8 km s$^{-1}$.
This dip is likely from H\textsc{i} self-absorption (HISA)
and indicates the presence of very cold neutral medium (spin temperature $T_{\rm s} < 50$ K; e.g., Gibson 2010). 
The correction for HISA requires a knowledge of the relative contributions from cold and warm gas 
to the H\textsc{i} emission, which can be obtained via H\textsc{i} absorption measurements. 
However, only two H\textsc{i} absorption measurements exist for Perseus (e.g., Heiles \& Troland 2003a).
This is clearly too sparse to interpolate the relative contributions from cold and warm gas
throughout the whole molecular cloud. 
Therefore, we do not apply the correction for HISA to $N$(H\textsc{i})
but we investigate the optical depth effect in Section 7.1.1 by applying the first order opacity correction.  

\subsection{Determining the Velocity Range}

\begin{figure}
\begin{center}
\includegraphics[scale=0.5]{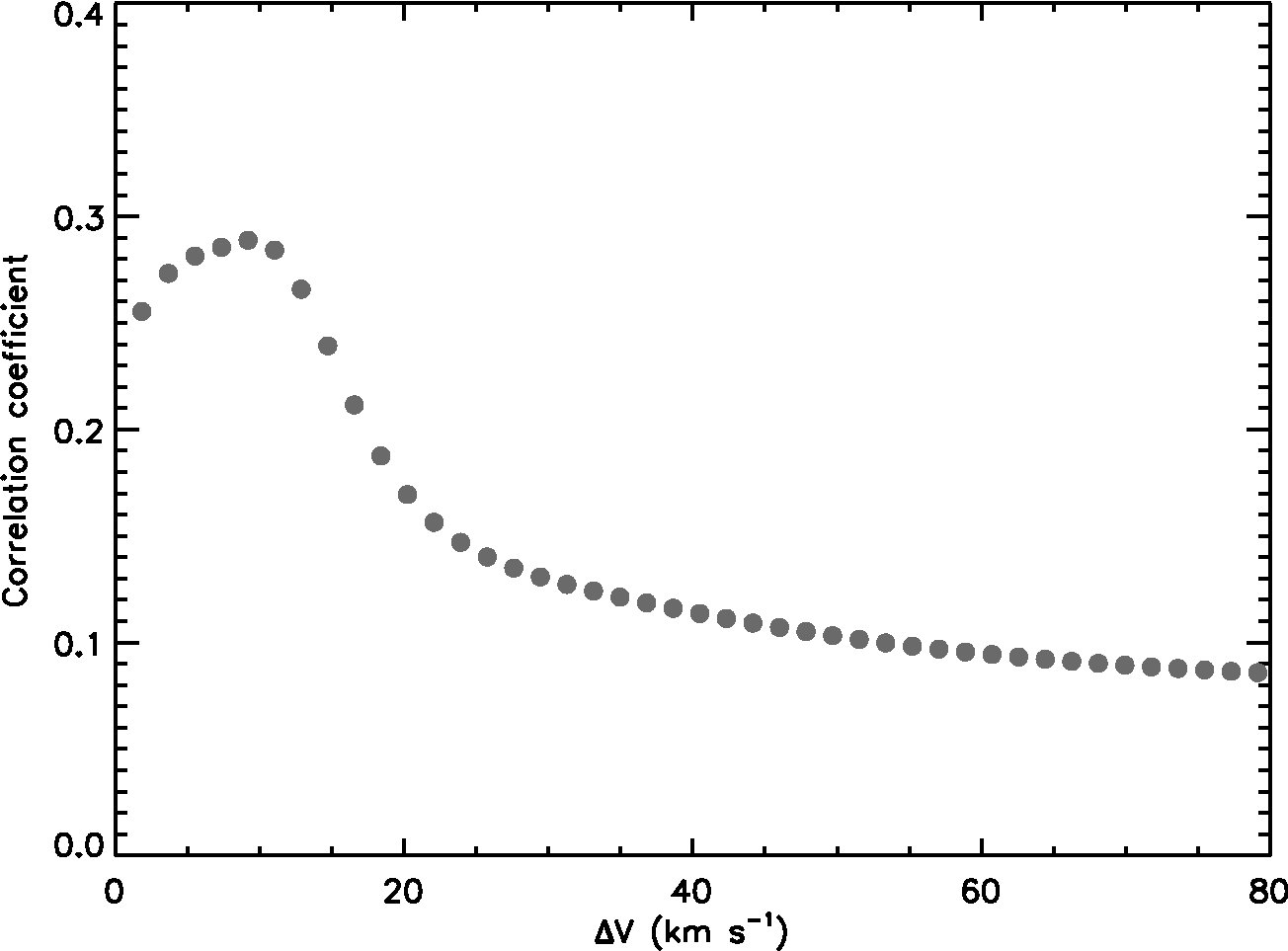}
\caption{\label{f:HI_Av_corr} Correlation coefficient between $N$(H\textsc{i}) and 2MASS $A_{V}$ 
as a function of velocity width $\Delta v$.
Data points with SNR > 5 in the 2MASS $A_{V}$ image are used to calculate the correlation coefficient.}
\end{center}
\end{figure}

\begin{figure}
\begin{center}
\includegraphics[scale=0.35]{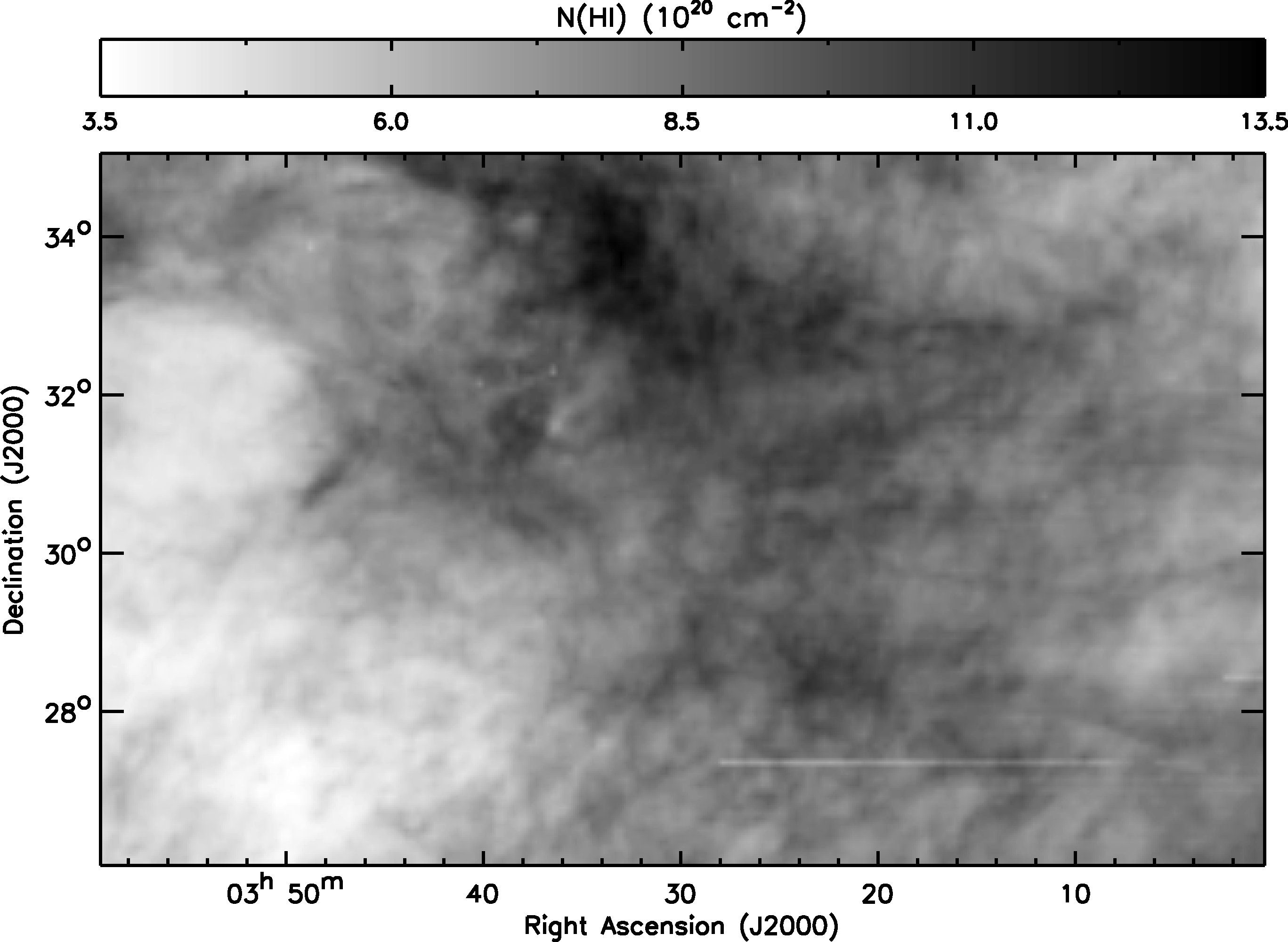}
\caption{\label{f:HI_col} $N$(H\textsc{i}) image at an angular resolution of 4.3$'$.
H\textsc{i} emission is integrated from $v$ = $-5$ km s$^{-1}$ to $+15$ km s$^{-1}$
and $N$(H\textsc{i}) is derived under the assumption of optically thin H\textsc{i} gas.}
\end{center}
\end{figure}

To determine the velocity range of H\textsc{i} emission associated with Perseus, 
we investigate the correlation between $N$(H\textsc{i}) and the dust column density.
The correlation between H\textsc{i} emission and FIR emission has been well-established 
for high Galactic latitudes (e.g., Boulanger et al. 1996).
We use the 2MASS $A_{V}$ image as a tracer of the dust column density 
and smooth the H\textsc{i} data to the angular resolution of the 2MASS $A_{V}$ image.  
We fix the central velocity of H\textsc{i} emission at $v = +$5 km s$^{-1}$ based on Section 3.1 
and integrate H\textsc{i} emission using a velocity width $\Delta v$ increasing from 2 km s$^{-1}$ 
to 80 km s$^{-1}$ with 2 km s$^{-1}$ intervals.  
For each derived $N$(H\textsc{i}) image, we calculate the correlation coefficient between $N$(H\textsc{i}) and 2MASS $A_{V}$.
Only data points with signal-to-noise ratio (SNR) $>5$ in the 2MASS $A_{V}$ image are used for this calculation.
Figure \ref{f:HI_Av_corr} shows the correlation coefficient as a function of $\Delta v$.  
Although the correlation coefficient is never very high, there is a clear peak at $\Delta v \sim 10$ km s$^{-1}$.
The lack of strong correlation between $A_{V}$ and $N$(H\textsc{i}) has been noticed before
and is mostly ascribed to the presence of molecular gas (e.g., Reach et al. 1994; Dame et al. 2001; Leroy et al. 2009).
We take the correlation coefficient of 0.15 (half of the maximum) as a threshold  
and use the corresponding $\Delta v$ of 20 km s$^{-1}$ to derive $N$(H\textsc{i}) 
under the assumption of optically thin H\textsc{i} gas\footnote{For optically thin H\textsc{i} gas, 
$N$(H\textsc{i}) is calculated by $N$(H\textsc{i}) = 1.823 $\times$ $\int T_{\rm b} d v$ (cm$^{-2}$), 
where $T_{\rm b}$ is the brightness temperature.}.  
The derived $N$(H\textsc{i}) ranges from 3.5 $\times$ 10$^{20}$ cm$^{-2}$ 
to 1.3 $\times$ 10$^{21}$ cm$^{-2}$, with a median of 7.9 $\times$ 10$^{20}$ cm$^{-2}$.
We present the $N$(H\textsc{i}) image in Figure \ref{f:HI_col}. 
 
Sancisi et al. (1974) and Imara \& Blitz (2011) estimated $N$(H\textsc{i}) for Perseus independently
and their results are essentially consistent with ours.
Sancisi et al. (1974) obtained H\textsc{i} data for Perseus using the 25-m Dwingeloo telescope
and estimated $N$(H\textsc{i}) by integrating H\textsc{i} emission 
from $v = -10$ km s$^{-1}$ to $+20$ km s$^{-1}$.
In addition, they attempted to subtract the foreground and background $N$(H\textsc{i})
by comparing the H\textsc{i} emission in the direction of Perseus to the H\textsc{i} emission 
measured in a nearby field spatially offset by a few degrees.
They found an average $N$(H\textsc{i}) associated with Perseus $\sim$ 5 $\times$ 10$^{20}$ cm$^{-2}$.
This is close to our median $N$(H\textsc{i}).
On the other hand, Imara \& Blitz (2011) used H\textsc{i} data from the LAB survey 
and estimated $N$(H\textsc{i}) by integrating H\textsc{i} emission 
from $v = -8$ km s$^{-1}$ to $+14$ km s$^{-1}$.
To determine the velocity range for $N$(H\textsc{i}), 
they assumed that H\textsc{i} spectra can be approximated with Gaussian functions
and set $\Delta v$ to be equal to four times the standard deviation of the 
Gaussian component associated with Perseus.
In summary, both Sancisi et al. (1974) and Imara \& Blitz (2011) derived $N$(H\textsc{i})
under the assumption of optically thin H\textsc{i} gas 
and their velocity ranges for $N$(H\textsc{i}) are consistent with ours.

\section{Methodology: Steps in Deriving \textit{N}(H$\boldsymbol{_{2}}$)}

\subsection{Removal of Point Sources from the IRIS Data}

Before further processing, we remove point sources from the 60 $\mu$m and 100 $\mu$m images. 
We identify compact point sources from the \textit{IRAS} Point Source Catalog 
that have $I_{60} > 1$ MJy sr$^{-1}$ and remove them by applying a circular mask 
with a radius equal to twice the angular resolution of the 100 $\mu$m image.   
These masks are shown as blank pixels in Figures \ref{f:IR100}, \ref{f:dust_map}, \ref{f:RH2}, and \ref{f:H2_CO}.  

\subsection{Exclusion of Possible Contaminations}

\begin{figure}
\begin{center}
\includegraphics[scale=0.37]{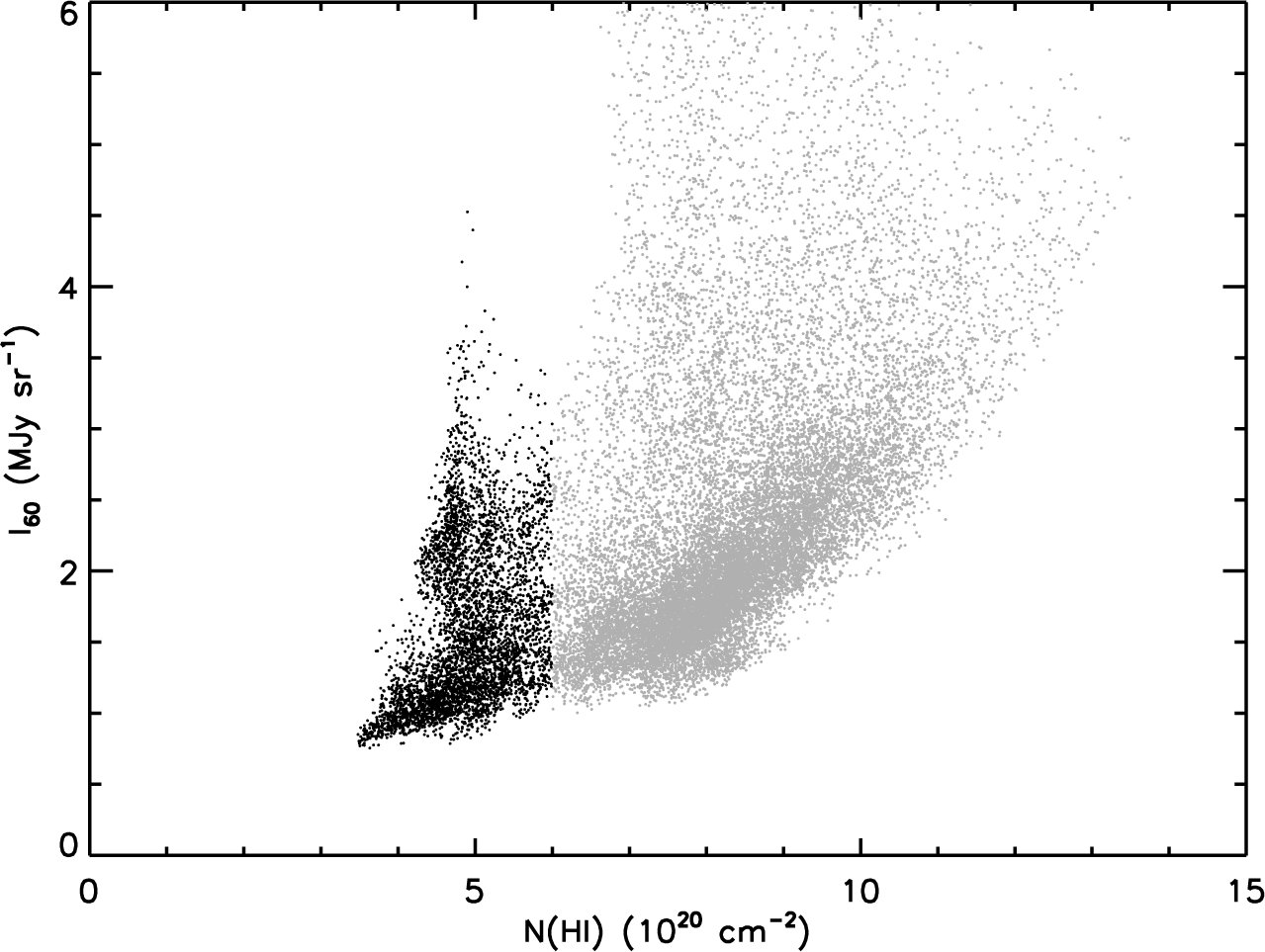}
\includegraphics[scale=0.37]{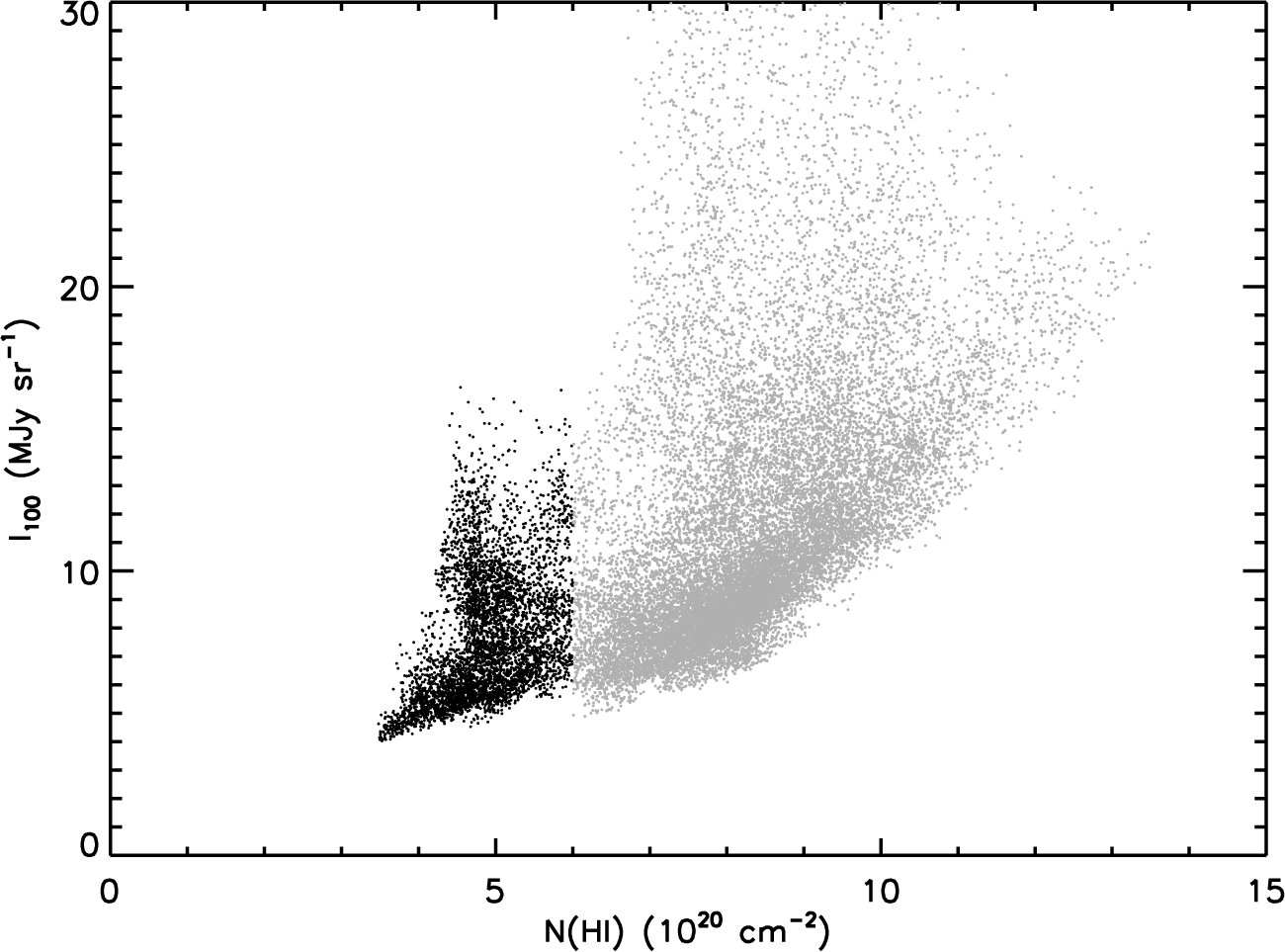}
\caption{\label{f:HI_IR} (Left) $I_{60}$ as a function of $N$(H\textsc{i}). 
(Right) $I_{100}$ as a function of $N$(H\textsc{i}).
The gray data points are assumed to mostly trace Perseus 
and the black data points are likely from the Taurus molecular cloud. 
See Section 4.2.}
\end{center}
\end{figure}

\begin{figure}
\begin{center}
\includegraphics[scale=0.35]{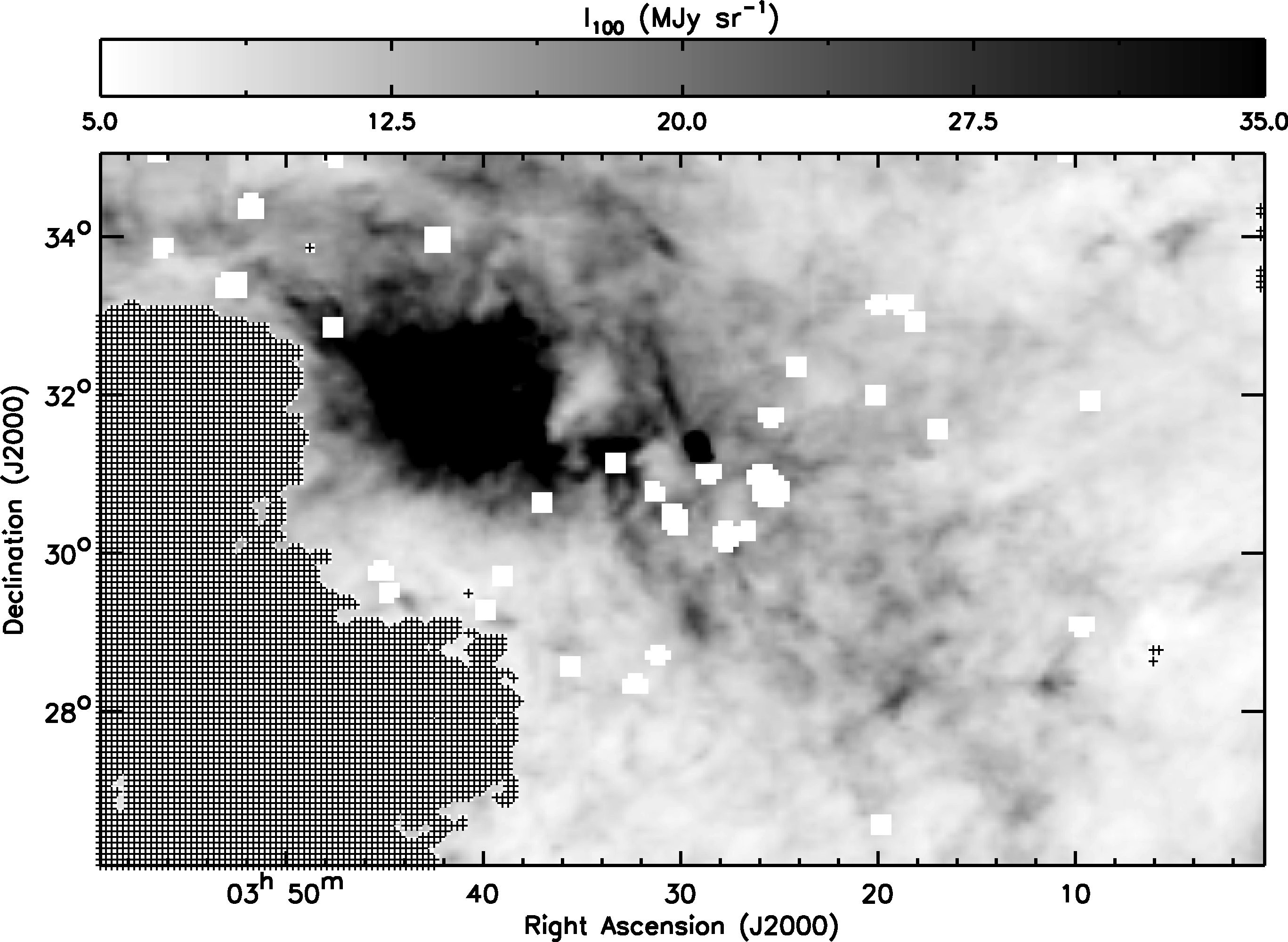}
\caption{\label{f:IR100} IRIS 100 $\mu$m image with the black data points from Figure 5.}
\end{center}
\end{figure}

As Perseus belongs to the Taurus--Auriga--Perseus molecular cloud complex 
and our data cover a large area on the sky ($\sim$15$^{\circ} \times 9^{\circ}$),
there is a concern of including molecular gas not associated with Perseus in our analysis. 
To investigate this, we plot $I_{60}$ and $I_{100}$ as a function of $N$(H\textsc{i}) in Figure \ref{f:HI_IR}. 
As $I_{60}$/$N$(H\textsc{i}) and $I_{100}$/$N$(H\textsc{i}) are approximations of dust-to-gas ratio (DGR), 
an overall linear relation, implying a single DGR, is a reasonable expectation for a single molecular cloud.  
Figure \ref{f:HI_IR} shows that the data points are segregated into two possible distinct features 
at $N$(H\textsc{i}) $\sim$ 6 $\times$ 10$^{20}$ cm$^{-2}$. 
The majority of the data points shows a global linear relation between $N$(H\textsc{i}) and FIR emission.  
These data points are shown in gray and we assume that they mostly trace Perseus.
However, the data points in black have a slightly different slope.
We selected the black data points as having $N$(H\textsc{i}) < 6 $\times$ 10$^{20}$ cm$^{-2}$.  
In Figure \ref{f:IR100}, we show the spatial distribution of the black data points
by overlaying them on the 100 $\mu$m image. 
Clearly, they are all found at the southeast edge of the image in the direction of Taurus.
Taurus is centered at (RA,Dec) = (04$^{\rm h}$30$^{\rm m}$,$+28^{\circ}$) 
and has $v_{\rm CO}$ = $0 \sim +12$ km s$^{-1}$ (e.g., Narayanan et al. 2008).
As a distinct molecular cloud, Taurus would have a different DGR.
To avoid any possible contamination, we exclude (blank) 
the black data points from the IRIS images and further analysis.

In addition to the above consideration,
the warm dust ring that stands out in the dust temperature image (see Section 4.3) 
is an additional source of contamination and we exclude it from further analysis. 

\subsection{$\boldsymbol{T_{\rm dust}}$ and $\boldsymbol{\tau_{100}}$}

For optically thin dust grains at an equilibrium temperature $T_{\rm dust}$, 
the optical depth at 100 $\mu$m, $\tau_{100}$, is determined by: 
\begin{equation}
\tau_{100} = \frac{I_{100}}{B(T_{\rm dust}, \lambda_{100})},
\end{equation}

\noindent where $I_{100}$ is the measured intensity at 100 $\mu$m and  
$B(T_{\rm dust},\lambda_{100})$ is the intensity of a blackbody of temperature $T_{\rm dust}$ at 100 $\mu$m.
To estimate $\tau_{100}$, we first estimate $T_{\rm dust}$ using the ratio $I_{60}/I_{100}$.
Under the assumption that the mass absorption coefficient, $\kappa_{\lambda}$, has a wavelength dependence 
such as $\kappa_{\lambda} \propto \lambda^{-\beta}$, $T_{\rm dust}$ can be estimated from:
\begin{equation} 
\frac{I_{60}}{I_{100}} = \left(\frac{\lambda_{60}}{\lambda_{100}}\right)^{-(3+\beta)} 
\frac{{\rm exp}(hc/\lambda_{100}kT_{\rm dust}) - 1}{{\rm exp}(hc/\lambda_{60}kT_{\rm dust}) - 1},
\end{equation}

\noindent where $h$ is the Planck constant and $k$ is the Boltzmann constant. 
The emissivity spectral index, $\beta$, can vary between 1 (for amorphous carbon) 
and 2 (for metallic and crystalline dielectric material), depending on the grain composition and size.  
Throughout this paper, we adopt $\beta = 2$ based on the discussions in Schnee et al. (2005; S05 hereafter). 
To estimate $T_{\rm dust}$, we smooth the 60 $\mu$m image to the angular resolution of the 100 $\mu$m image.

$I_{60}$ may include a significant contribution from stochastically heated 
very small grains (VSGs) that must be accounted for.  
To do this, we adopt the method used by S05.
First, we smooth the 60 $\mu$m and 100 $\mu$m images to the angular resolution of the $T_{\rm dust}$ map
derived by Schlegel et al. (1998; SFD98 hereafter).
SFD98 used the low resolution DIRBE 100 $\mu$m and 240 $\mu$m data to derive $T_{\rm dust}$;
the contribution from VSGs is expected to be negligible at these wavelengths (e.g., Li \& Draine 2001). 
Second, we calculate $T_{\rm dust}$ based on 100\% of $I_{100}$ 
and a lesser percentage of $I_{60}$, assuming $\beta = 2$. 
The fraction of $I_{60}$ is chosen to minimize the average difference between
our $T_{\rm dust}$ and SFD98's $T_{\rm dust}$.
We find that the best matching fraction is 22\%, meaning that 78\% of $I_{60}$ is due to VSGs. 
This result is consistent with S05, who found 26\%, but lower than the prediction of interstellar dust model 
for the solar neighborhood ($\sim$40\%; D\'esert et al. 1990).

We present the $T_{\rm dust}$ image in Figure \ref{f:dust_map}.  
Our $T_{\rm dust}$ image is generally consistent with the $T_{\rm dust}$ image derived by S05, 
but shows systematically slightly lower $T_{\rm dust}$. 
The difference between our $T_{\rm dust}$ and S05's $T_{\rm dust}$, 
$T_{\rm dust} - T_{\rm dust,S05}$, has a median of $-1$ K. 
This discrepancy likely arises from the fact that S05 adopted a $T_{\rm dust}$-dependent formula 
for $\beta$ and used a different fitting algorithm.
Their $T_{\rm dust}$-dependent formula for $\beta$ considered the observed inverse relation
between $T_{\rm dust}$ and $\beta$ and had two free parameters (see Equation 2 of S05).
To simultaneously fit these two free parameters, S05 used the IDL routine \textsc{amoeba}, 
which adopts the downhill method of Nelder \& Meade (1965),
while we estimated the best matching parameter using the least-squares method.

$T_{\rm dust}$ across Perseus ranges from $\sim$16 K to $\sim$22 K.
The lowest $T_{\rm dust}$ regions spatially correlate with the highest $A_{V}$,
in agreement with the expectation for dense and cold molecular gas. 
Several localized regions with high $T_{\rm dust}$ stand out; 
two are the star-forming regions, IC348 and NGC1333, 
while the third is a ring-like feature located at (RA,Dec) = 
(03$^{\rm h}$39$^{\rm m}$30$^{\rm s}$,$+$32$^{\rm \circ}$) with a radius of $\sim$56$'$.  
Ridge et al. (2006b) investigated this ring in detail 
using \textit{IRAS}, \textit{MSX}, 2MASS $A_{V}$, and CO data  
and concluded that the ring, the product of an H\textsc{ii} region 
driven by the B-type star HD 278942, is likely located behind IC348. 
However, they also argued that the ring is interacting with Perseus,  
based on the spatial correlation between 8 $\mu$m emission and 2MASS $A_{V}$.   
The ring is visible in the 2MASS $A_{V}$ image but at a much lower contrast, 
implying that the ring is not as significant column density feature as one might infer from the IRIS data.
Therefore, we exclude it from further analysis by applying a circular mask of radius of 56$'$. 
This mask size is consistent with what Ridge et al. (2006b) used to separate the ring 
from the Perseus cloud component, $\sim$50$'$. 

At this point, we refined the zero point of the $\tau_{100}$ image 
by assuming that the dust column density traced by $\tau_{100}$ is proportional to $N$(H\textsc{i})
for the regions dominated by diffuse atomic gas.
Under this assumption, $\tau_{100}$ would approach zero as $N$(H\textsc{i}) approaches zero   
and any offset from this relation is likely due to the uncertainties in the zero points of the IRIS images.
As a result of this exercise, we added 1.8 $\times$ 10$^{-4}$ to the $\tau_{100}$ image.
The observed offset in $\tau_{100}$ implies a small additional uncertainty in $T_{\rm dust}$, 
which is of order of 0.03--0.6 K.

\subsection{$\boldsymbol{A_{V}}$}

Once we estimate $\tau_{100}$ using the VSG--corrected $T_{\rm dust}$,
we convert $\tau_{100}$ to $A_{V}$ by:
\begin{equation}
A_{V} = X \tau_{100}, 
\end{equation}
\noindent where $X$ is the conversion factor that relates the dust column density to the extinction in the $V$-band. 
The conversion factor $X$ = 720 is estimated by finding the best agreement between our derived $A_{V}$ image  
and the 2MASS $A_{V}$ image for the region where the two images overlap.
Our $A_{V}$ and 2MASS $A_{V}$ have a 1:1 relation but with an average scatter of $\sim$0.5 mag.
This scatter around the 1:1 relation is mostly attributed to the 
variation in $T_{\rm dust}$ along a line of sight (e.g., GPS09). 

We note that the bootstrapping of our $A_{V}$ image derived from the IRIS data 
with the $A_{V}$ image derived from the 2MASS extinction data is motivated by GPS09. 
For Perseus, GPS09 inter-compared three methods for measuring total gas column density; 
extinction mapping at near-infrared (NIR) wavelengths, thermal emission mapping at FIR wavelengths, 
and CO emission mapping.
They found that the dust-based measures (NIR extinction and FIR emission)
are superior to the gas-based measure (CO emission) and particularly suggested 
the NIR extinction as the best probe of total gas column density.
Any molecular transition, such as CO emission, traces a limited range of volume densities
due to the critical density for excitation, opacity, and depletion. 
The dust-based measures have a wider dynamic range, 
but the NIR extinction measure saturates at high extinctions where background sources cannot be observed  
and the FIR emission measure is limited at low extinction values. 
In addition, the FIR emission measure suffers from its intrisic uncertainty, 
the variation of $T_{\rm dust}$ along a line of sight. 
Overall, the NIR extinction measure is the best probe up to its saturation level. 
Therefore, our calibration of $\tau_{100}$ to the 2MASS $A_{V}$ image would provide 
a better estimate of the total gas column density, 
compared to the total gas column density derived solely from the IRIS data.

\subsection{Dust-to-Gas Ratio and $\boldsymbol{N(\rm H_{2})}$} 
Finally, we calculate $N$(H$_{2}$) by:  
\begin{equation}
N({\rm H_{2}}) = \frac{1}{2}\left(\frac{A_{V}}{\rm DGR} - N({\rm H\textsc{i}})\right), 
\end{equation}
\noindent where DGR is defined by DGR = $A_{V} / N({\rm H})$ 
and $N({\rm H}) = N({\rm H{\textsc{i}}}) + 2N({\rm H_{2}})$.

To measure DGR in Perseus, we plot $A_{V}$ as a function of $N({\rm H\textsc{i}})$ 
over the whole cloud in Figure \ref{f:HI_Av}.
As we expect $N(\rm H)$ $\sim$ $N(\rm H{\textsc{i}})$ for the regions dominated by diffuse atomic gas
(which correspond to the majority of data points in Figure \ref{f:HI_Av}),
the fitted slope $A_{V} / N(\rm H\textsc{i}) = 1.1 \times 10^{-21}$ mag cm$^{2}$ is a good measure of DGR.
Our estimate is $\sim$2 times higher than the typical Galactic DGR, 5.3 $\times$ 10$^{-22}$ mag cm$^{2}$ 
(e.g., Bohlin et al. 1978).
This is not concerning as notable variations have been found for the Galactic DGR. 
For example, Kim \& Martin (1996) compiled $A_{V}$/$N$(H) values for the Galaxy 
derived from UV absorption measurements from literature 
and found that $A_{V}$/$N$(H) can be higher than the typical Galactic value by a factor of three 
(see their Figure 4).

We also compare our DGR estimate with values obtained from two independent methods: \\ 
(1) \textit{Copernicus/FUSE absorption measurements}: 
Toward HD 22951, HD 23180, and BD$+31^{\circ}643$, 
\textit{Copernicus/FUSE} measured H$_{2}$ directly in absorption 
(Savage et al. 1977; Rachford et al. 2002).
HD 22951 and HD 23180 are B1 stars and BD$+31^{\circ}643$ is a B5 star.
Towards BD$+31^{\circ}643$, Snow et al. (1994) measured $A_{V} = 2.68$ mag 
and combined this with the observed $N(\rm H\textsc{i})$ and $N(\rm H_{2})$, 
resulting in $A_{V}/N(\rm H) = 5.5 \times 10^{-22}$ mag cm$^{2}$. 
However, extinction parameters are not known for HD 22951 and HD 23180. 
Considering that HD 22951, HD 23180, and BD$+31^{\circ}643$ are close to each other on the sky,  
if we apply $R_{V} = 3.19$ measured for BD$+31^{\circ}643$ to both HD 22951 and HD 23180, 
we obtain $A_{V}/N(\rm H)$ = (4.6--6) $\times ~10^{-22}$ mag cm$^{2}$.
While these estimates are close to the typical Galactic DGR, 
we note that they are likely lower limits on the DGR of Perseus,  
considering that most of the lines of sight surveyed by the \textit{Copernicus/FUSE} 
are composed of two or more diffuse clouds (e.g., Rachford et al. 2002, 2009).
Rachford et al. (2002) compared the observed high rotational transition column densities of H$_{2}$ 
to the model of H$_{2}$ formation, destruction, and ro-vibrational excitation developed by Browning et al. (2003) 
and found that it is impossible to reproduce the observed column densities under a single cloud assumption. 
Instead, models with multiple clouds along the line of sight, and/or multiple pathways of incident UV radiation, 
match much better the observed column densities.
Stars preferentially form in regions with high DGR, 
where dust grains provide both high shielding against photodissociation and sites for H$_{2}$ formation.  
Therefore, if several diffuse clouds exist along the line of sight toward Perseus, 
the component associated with Perseus likely has a higher DGR.  
In this case, the DGR measured along the whole line of sight is lower than the actual DGR of Perseus.\\ 


\noindent (2) \textit{Upper limit on the DGR}: 
Equation (4) itself provides an upper limit on the DGR.  
If the DGR is too high, a large number of data points will have a negative value of $N$(H$_{2}$).  
For example, if DGR > 1.5 $\times$ 10$^{-21}$ mag cm$^{2}$, 
more than 50\% of the pixels with $A_{V}$ > 1 mag will have a negative value of $N$(H$_{2}$). 
Therefore, it is most likely that the DGR in Perseus is less than 1.5 $\times$ 10$^{-21}$ mag cm$^{2}$. \\

\noindent \textit{Conclusion}:  
Based on two independent constraints on the DGR,
we conclude that our estimate of $A_{V} / N(\rm H\textsc{i}) = 1.1 \times 10^{-21}$ mag cm$^{2}$ 
is midway between the lower and upper limits and therefore represents a very reasonable estimate. 

\subsection{Uncertainty in \textit{N}(H$\boldsymbol{_{2}}$)}

We estimate the uncertainty in $N(\rm H_{2})$ for each pixel by performing a series of Monte Carlo simulations.
In these simulations, we take into account the uncertainties in $N(\rm H\textsc{i})$ and $A_{V}$ 
and how they propagate into $N(\rm H_{2})$.  

For the uncertainty in $N(\rm H\textsc{i})$, we generate 1000 $N$(H\textsc{i}) images 
by using 1000 velocity widths randomly drawn from a Gaussian distribution 
that peaks at 20 km s$^{-1}$ with 1$\sigma$ of 4 km s$^{-1}$ (based on Section 3.2).
The median 1$\sigma$ of $N(\rm H\textsc{i})$ due to the systematic uncertainty 
in using a particular velocity width is 5.6 $\times$ 10$^{19}$ cm$^{-2}$. 

For the uncertainty in $A_{V}$, we assess the range of $A_{V}$ 
by varying the input parameters and deriving $A_{V}$ each time.
We add/subtract 1$\sigma$ noise to/from the 60 $\mu$m and 100 $\mu$m images
(0.07 MJy sr$^{-1}$ at 60 $\mu$m and 0.48 MJy sr$^{-1}$ at 100 $\mu$m) and use $\beta = 1$ or 2. 
Therefore, we derive the $A_{V}$ images with 8 different combinations of the input parameters 
and for each case the contribution from VSGs to $I_{60}$ and $X$ for $A_{V} = X \tau_{100}$ 
are determined by finding the best matching parameters.
To include the uncertainties of the zero point,
we also derive the $\tau_{100}$ images with and without the zero point calibration.  
We find that the contribution from VSGs to $I_{60}$ varies from 0.12 to 0.24 with a median of 0.17,  
while $X$ varies from 630 to 810 with a median of 728. 
Finally, we compare the 16 $A_{V}$ images and find the maximum and minimum estimates of $A_{V}$ for each pixel.

The uncertainty in $N(\rm H\textsc{i})$ and the maximum/minimum estimates of $A_{V}$ are incorporated 
in a Monte Carlo simulation to produce 1000 $N(\rm H_{2})$ images. 
We use the distribution of the simulated $N(\rm H_{2})$ to estimate 
the uncertainty in $N(\rm H_{2})$ on a pixel-by-pixel basis.
The median 1$\sigma$ of $N$(H$_{2}$) is 3.6 $\times$ 10$^{19}$ cm$^{-2}$.  

In addition to the above considerations,
several systematic effects are also likely to affect $N(\rm H_{2})$;
(1) the assumption of a single dust population and temperature along a line of sight, 
(2) the assumption of a single DGR, and 
(3) the assumption that H$_{2}$ is the dominant source of excess dust emission. 
As it is difficult to estimate the magnitude of these effects, 
we do not include them in our Monte Carlo simulations.  

\section{General Results}

\subsection{$\boldsymbol{T_{\rm dust}}$}

$T_{\rm dust}$ across Perseus ranges from $\sim$16 K to $\sim$22 K with a median of 17 K. 
This is in agreement with SFD98 who found $T_{\rm dust} \sim$ 17 K for the same region
and S05 who found a slightly higher $T_{\rm dust} \sim$ 18 K for the main body of Perseus. 
Similarly, Stepnik et al. (2003) derived $T_{\rm dust}$ $\sim$ 17 K for the outskirts of Taurus 
using the DIRBE data at 140 $\mu$m and 240 $\mu$m.

\begin{figure}
\begin{center}
\includegraphics[scale=0.35]{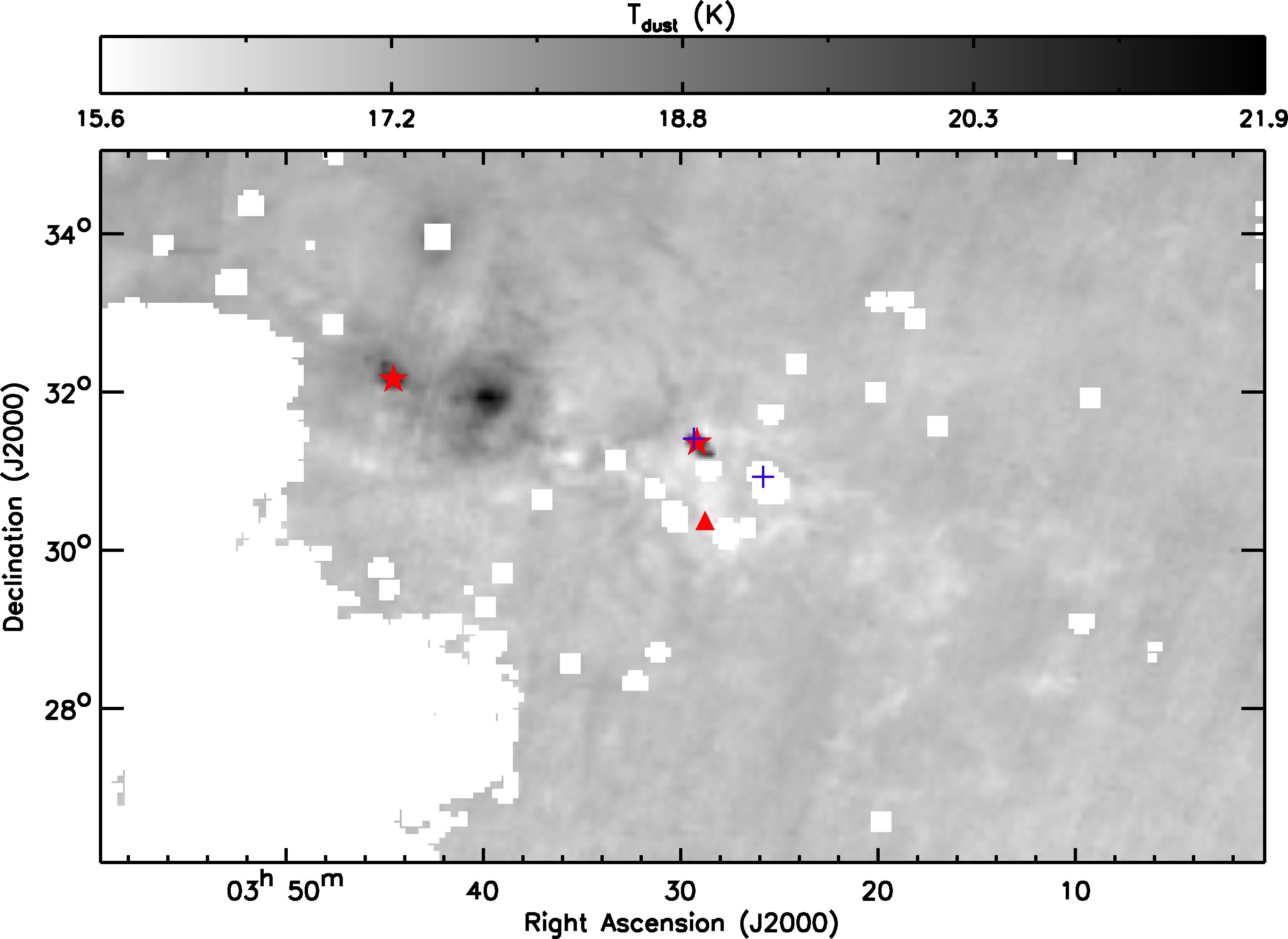}
\caption{\label{f:dust_map} $T_{\rm dust}$ image derived from the IRIS 60 $\mu$m and 100 $\mu$m images 
after accounting for the contribution from VSGs.
Several B-type stars as potential dust heating sources are marked;  
B8 V-type stars (BD$+30^{\circ}$540 and BD$+30^{\circ}$549) are shown as blue crosses (\v{C}ernis 1990);
B5 V-type stars (BD$+31^{\circ}$643 and SVS3) are shown as red stars (\v{C}ernis 1990; Luhman et al. 2003);
B4 IV--V-type star (BD$+29^{\circ}$566) is shown as a red triangle (\v{C}ernis 1990).}
\end{center}
\end{figure}

\begin{figure}
\begin{center}
\includegraphics[scale=0.6]{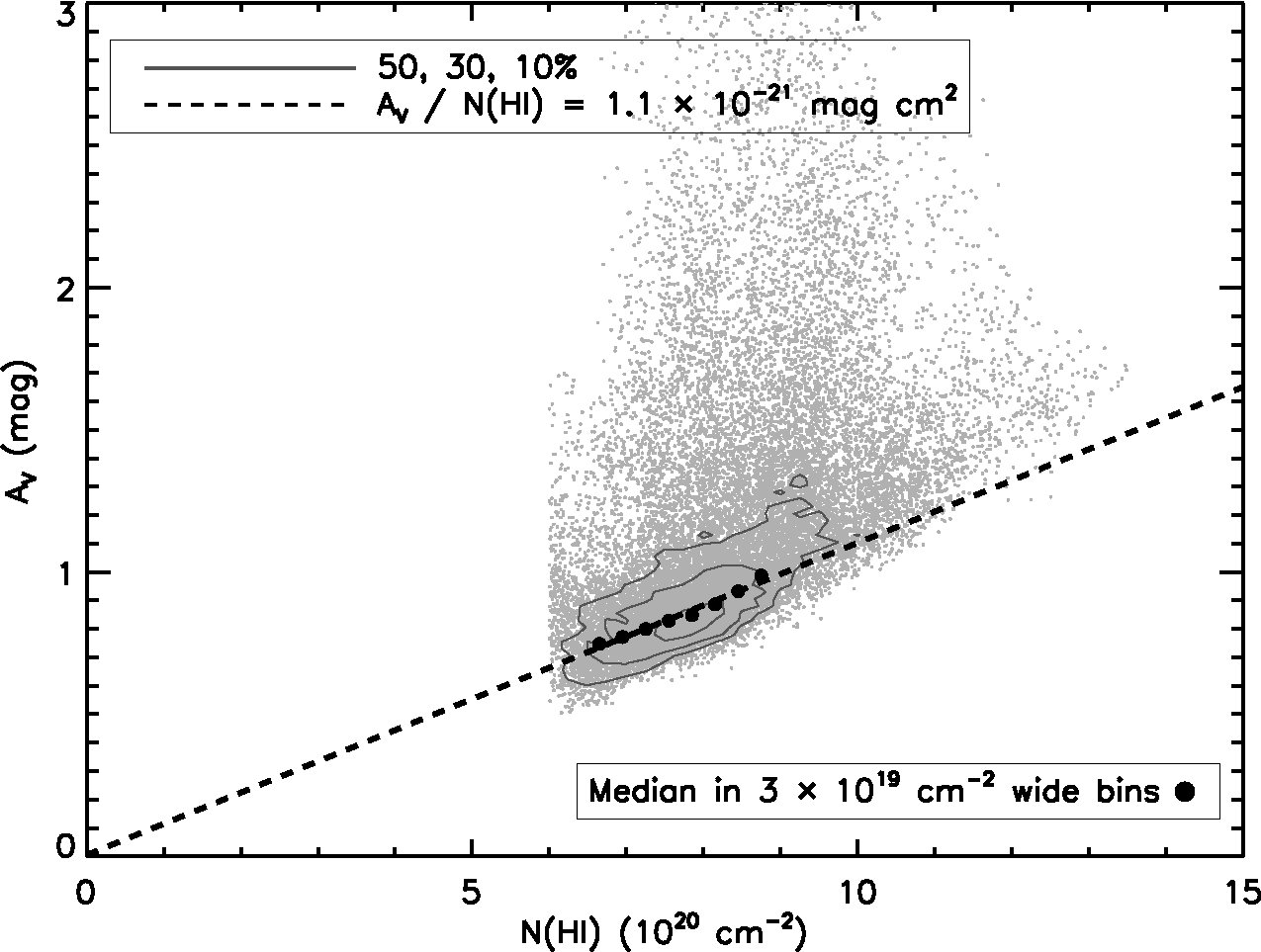}
\caption{\label{f:HI_Av} $A_{V}$ as a function of $N(\rm H\textsc{i})$.
The contour levels correspond to 50\%, 30\%, and 10\% of the total data points.  
The black circles are median $A_{V}$ values in 3 $\times$ 10$^{19}$ cm$^{-2}$ wide bins.  
The linear least-squares fit for the median $A_{V}$ values is shown as a black dashed line.}
\end{center}
\end{figure}

As shown in Figure~\ref{f:dust_map}, the lowest $T_{\rm dust}$ regions spatially correlate
with the A$_V$ and CO images, in agreement with the expectation for dense and cold clouds. 
The prominent minima in $T_{\rm dust}$ correspond to B5, B1E, B1, L1448, L1451, and L1455. 
These dark regions contain a number of IR sources and Herbig-Haro objects 
that are considered to be associated with pre-main sequence stars (e.g., Bally et al. 2008),
but the luminosities of these objects are so low that their influence on the overall energy budget 
of the clouds would be negligible.
Therefore, $T_{\rm dust}$ within these dark regions is determined by the external ISRF and dust grain properties.
Low $T_{\rm dust}$ in B5, B1E, B1, L1448, L1451, and L1455 compared to the surrounding regions 
suggests that the inside of these regions gas is dense enough 
to have sufficient shielding against the external ISRF, leading to efficient cooling.
For example, B5 has limb-brightening in CO, 
indicating a cold cloud interior surrounded by a photo-heated exterior (e.g., Beichman et al. 1988). 

\begin{landscape}
\begin{figure}
\includegraphics[scale=0.46]{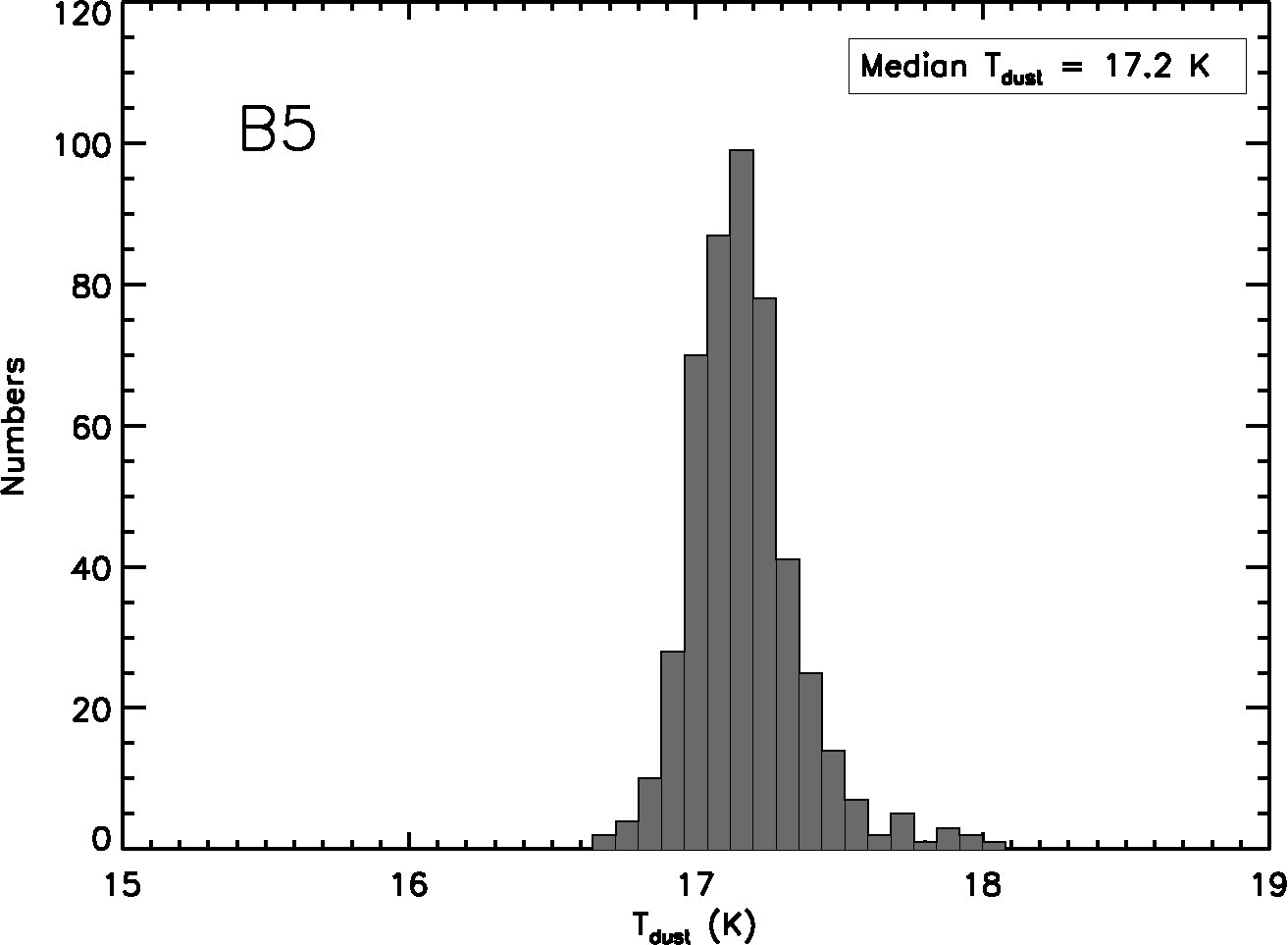}
\includegraphics[scale=0.46]{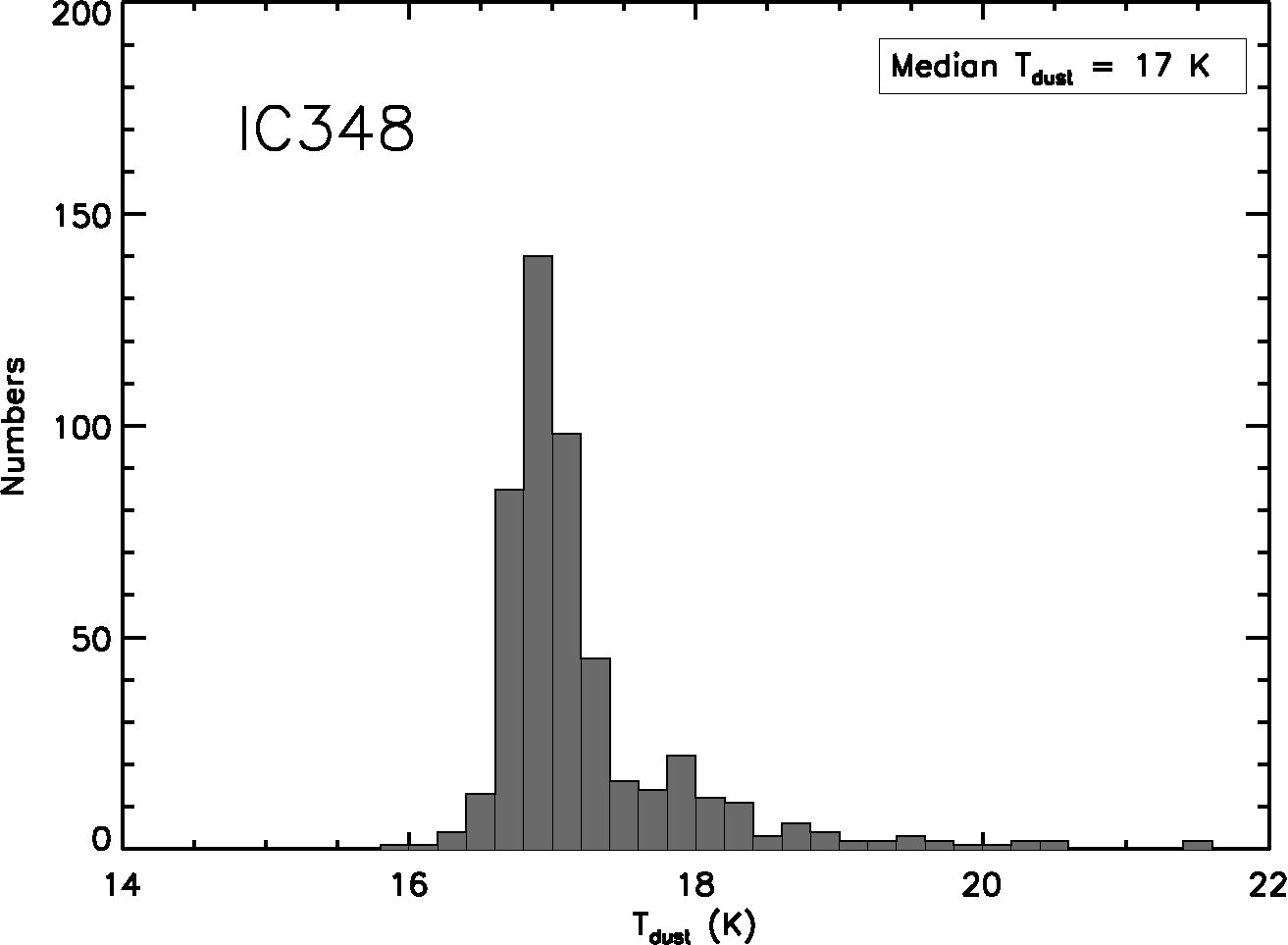}
\includegraphics[scale=0.46]{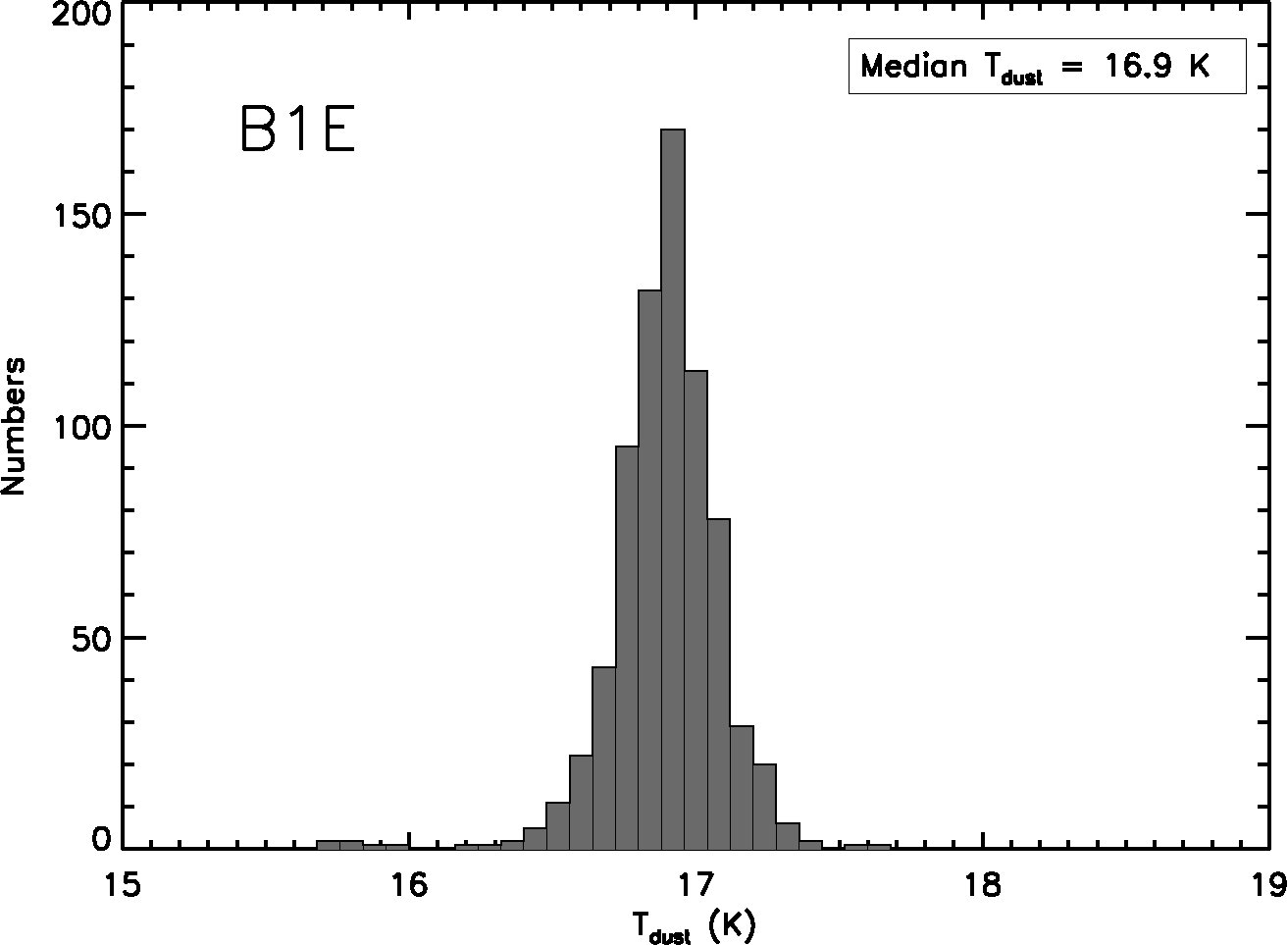}
\caption{\label{f:dust_hist} Histograms of $T_{\rm dust}$.
The median $T_{\rm dust}$ is given in the top right corner of each plot.
(Left top) B5. (Right top) IC348. (Left bottom) B1E.}
\end{figure}
\end{landscape}

Figure~\ref{f:dust_map} also shows several localized regions with high $T_{\rm dust}$.  
One of the prominent features is the warm dust ring at (RA,Dec) = 
(03$^{\rm h}$39$^{\rm m}$30$^{\rm s}$,$+32^{\circ}$), already discussed in Section 4.3. 
The other high $T_{\rm dust}$ features include IC348 and NGC1333.  
To quantify how much hotter these star-forming regions are,
we plot the histograms of $T_{\rm dust}$ in Figure~\ref{f:dust_hist}. 
For the histograms, we use the data points in the rectangular boxes shown in Figure \ref{f:RH2}.
Each box has either a dark or a star-forming region and extends up to the edge of the image. 
The width of each box is set based on the COMPLETE $^{13}$CO(1--0) data to include the whole region of interest.
While all regions have a similar median $T_{\rm dust}$ of 17 K, 
the star-forming regions have data points with higher $T_{\rm dust}$.
For example, for IC348 and NGC1333, 11\% and 2\% of data points have $T_{\rm dust}$ > 18 K,
while for B5, B1E, and B1, all data points have $T_{\rm dust}$ < 18 K.
The data points with $T_{\rm dust}$ > 18 K are found in the central parts of IC348 and NGC1333,
where B-type stars are located (blue crosses and red stars in Figure~\ref{f:dust_map}), 
and this implies that the internal radiation field may play a significant role 
in heating dust grains inside IC348 and NGC1333.
This possibility is discussed further in the next section. 

\subsection{ISRF: External or Internal?}

We now investigate possible heating sources that can result in the measured $T_{\rm dust}$.
As we mentioned in Section 1, KMT09's model assumes a uniform external ISRF.
Understanding the origin of ISRF is therefore important for comparison between observations and the model.
There are two possible sources of ISRF that are responsible for the measured $T_{\rm dust}$ across Perseus;  
(1) the UV flux from the B-type stars in Perseus (local)
and (2) the UV radiation field determined by the overall stellar distribution in the Galaxy (global). 

(1) We consider two B5 V-type stars in Perseus (red stars in Figure~\ref{f:dust_map}) 
as possible sources of the local ISRF.
All other stars have spectral types later than B5\footnote{Two B8 V-type stars in NGC1333 
(blue crosses in Figure~\ref{f:dust_map}) contribute negligibly to the UV flux.
We do not consider the B4 IV--V-type star (red triangle in Figure~\ref{f:dust_map}) 
due to its uncertain stellar classification  
(e.g., Lesh 1969; Klochkova \& Kopylov 1985; \v{C}ernis 1990). 
Nevertheless, inclusion of this star does not change our result.}. 
We assume that the spectral energy distribution of a B-type star is approximately a black body curve 
(e.g., Spaans et al. 1994) and adopt $T_{\rm eff}$ = 15600 K for a B5 V-type star (Morton \& Adams 1968).  
For $\lambda$ = 912--2066 \AA, two B5 V-type stars emit 2.12 $\times$ 10$^{12}$ erg cm$^{-2}$ s$^{-1}$.
If a distance of 15 pc from each star, which corresponds to $\sim$1/3 of the longitudinal size of our sky 
coverage, is assumed, the UV flux from the two B5 V-type stars is reduced to 
3.04 $\times$ 10$^{-5}$ erg cm$^{-2}$ s$^{-1}$ (1/$r^{2}$ geometrical dilution). 
For dust grains in thermal equilibrium, $T_{\rm dust}$ is determined by (Lequeux 2005):  
\begin{equation}
G = 4.6 \times 10^{-11} \left(\frac{a}{0.1~\mu\rm{m}}\right)T_{\rm dust}^{6}~\rm{erg~cm^{-2}~s^{-1}}, 
\end{equation}
\noindent where $G$ is the UV flux and $a$ is the size of dust grains.
An absorption efficiency  $Q_{\rm a}$ = 1, appropriate for large dust grains that dominate the FIR emission 
and have a size comparable to UV wavelengths, 
and a dust emissivity spectral index $\beta$ = 2 are assumed for Equation (5). 
Therefore, dust grains with a radius $a$ = 0.1 $\mu$m in the UV radiation field of 
3.04 $\times$ 10$^{-5}$ erg cm$^{-2}$ s$^{-1}$ would have $T_{\rm dust} \sim$ 9.3 K.
Note that we do not consider the decrement of UV flux by dust attenuation.
Therefore, if the B-type stars are the only sources of the ISRF,
we would expect $T_{\rm dust} \la 9.3$ K in the outskirts of Perseus.  

(2) We consider the global Galactic ISRF of Porter \& Strong (2005).
Porter \& Strong (2005) adopted the Galactic stellar and dust distribution models 
and calculated the radiation field from stars at $\lambda$ = 0.1--1000 $\mu$m.
The ISRF was found to be the most intense in the inner Galaxy, 
to decrease toward the Galactic poles, but to remain steady up to $\sim$5 kpc from the Galactic plane.
For $R_{\rm gal}$ (Galactocentric radius) = 0 and $z$ (Galactocentric height) = 0, 
the radiation field at 0.1 $\mu$m is $\sim$0.08 eV cm$^{-3}$ 
and for $R_{\rm gal}$ = 0 and $z$ = 5 kpc, the radiation field at 0.1 $\mu$m is $\sim$0.01 eV cm$^{-3}$ 
(see Figure 2 of Porter \& Strong 2005). 
If we assume that this $\sim$87.5\% decrement of the UV radiation field at $R_{\rm gal}$ = 0 from $z$ = 0 to 5 kpc 
would be applicable for $R_{\rm gal} \sim$ 8 kpc, the UV radiation field at $z \sim -100$ pc 
(the location of Perseus; $R_{\rm gal} \sim$ 8 kpc and $z \sim -100$ pc) would be between 
2.7 $\times$ 10$^{-3}$ (the UV radiation field in the solar neighborhood at $\lambda$ = 912--2066 \AA; $G_{0}$) 
and 3.4 $\times$ 10$^{-4}$ erg cm$^{-2}$ s$^{-1}$ (12.5\% of $G_{0}$). 
Applying Equation (5), this corresponds to 14 K < $T_{\rm dust}$ < 20 K. 
As our median $T_{\rm dust} \sim$ 17 K is indeed in this range, 
the UV ISRF determined by the overall stellar distribution in the Galaxy could be 
the dominant heating source for dust grains across Perseus.

Therefore, we conclude that Perseus is embedded in the uniform Galactic ISRF 
that heats dust grains up to $T_{\rm dust} \sim$ 17 K, 
except for the central parts of IC348 and NGC1333 where the radiation from the B-type stars is dominant.

\section{The H\textsc{i}--$\boldsymbol {\rm H_{2}}$ Transition in Perseus: Theoretical Perspective}

KMT08 modeled H$_{2}$ formation and photodissociation for a spherical cloud 
that is bathed in a uniform ISRF.
They assumed that the H\textsc{i}--H$_{2}$ transition is infinitely sharp
and found that the location of the transition inside an
atomic-molecular complex is solely determined by two dimensionless numbers:
\begin{equation}
\label{e:ksi}
\chi = \frac{f_{\rm diss} \sigma_{\rm d} c E^{\ast}_{\rm 0}}{n_{\rm CNM} \mathcal{R}}
\end{equation}

\begin{equation}
\tau_{\rm R} = n_{\rm CNM} \sigma_{\rm d} R. 
\end{equation}

\noindent Here $f_{\rm diss}$ $\sim$ 0.1 is the fraction of absorptions of LW photons
that produce H$_{2}$ dissociation rather than decay back to a bound state,
$\sigma_{\rm d}$ is the dust absorption cross section per H nucleus in the LW band,
$E^{\ast}_{\rm 0}$ is the number density of LW photons,
$n_{\rm CNM}$ is the number density of CNM in the H\textsc{i} shielding layer,
$\mathcal{R}$ is the H$_{2}$ formation rate coefficient on dust grains,
and $R$ is the cloud radius.
$\chi$ is the ratio of the rate at which LW photons are absorbed by dust grains
to the rate at which they are absorbed by H$_{2}$.
We can consider $\chi$ as a dimensionless measure of the strength of the ISRF.
For example, in strong radiation fields, H$_{2}$ molecules are easily photodissociated
and therefore the absorption of LW photons is dominated by dust grains (large $\chi$).
On the contrary, in weak radiation fields, H$_{2}$ molecules survive
and due to their large resonant cross section the absorption of LW photons
is dominated by H$_{2}$ molecules (small $\chi$).
$\tau_{\rm R}$ is simply the dust optical depth that the cloud would have if its density is equal to $n_{\rm CNM}$.

Using $\chi$ and $\tau_{\rm R}$, KMT08 derived $x_{\rm H2}$, 
the fraction of the cloud radius where the H\textsc{i}--H$_{2}$ transition occurs:
\begin{equation}
\begin{split}
x_{\rm H2} & = \left[1 - \frac{3 \psi}{4(\tau_{\rm R} + 0.2 \psi)}\right]^{1/3} \\
           & \approx 1 - \frac{\psi}{4\tau_{\rm R} - 0.7 \psi} \\
\end{split}
\end{equation}
\noindent where
\begin{equation*}
\psi = \chi \frac{2.5 + \chi}{2.5 + \chi e}. 
\end{equation*}

\noindent Then, the dust optical depth through the H\textsc{i} shielding layer, 
$\tau_{\rm H\textsc{i}}$, can be determined by: 
\begin{equation}
\begin{split}
\tau_{\rm H\textsc{i}} & = n_{\rm CNM} \sigma_{\rm d} R (1 - x_{\rm H2})  \\
                      & \approx \frac{\tau_{\rm R} \psi}{4\tau_{\rm R} - 0.7 \psi}. \\
\end{split}
\end{equation}

\noindent The H\textsc{i} shielding surface density for H$_{2}$ formation, 
$\Sigma_{\rm H\textsc{i},s}$, is simply: 
\begin{equation}
\Sigma_{\rm H\textsc{i},s} = \tau_{\rm H\textsc{i}}\left(\frac{\mu_{\rm H}}{\sigma_{\rm d}}\right), 
\end{equation}

\noindent where $\mu_{\rm H}$ is the mass per H nucleus.

To investigate what fraction of the gas in an atomic-molecular complex is molecular,  
KMT09 simplified the above equations as follows.
First, as both $\sigma_{\rm d}$ and $\mathcal{R}$ are the measures of dust surface area, 
their ratio is nearly unity and drops out of Equation (6). 
Second, $n_{\rm CNM}$ is determined from the condition of pressure balance 
between the cold neutral medium (CNM) and warm neutral medium (WNM) phases:
\begin{equation}
\begin{split}
n_{\rm CNM} & = \phi_{\rm CNM} n_{\rm min} \\
            & = \phi_{\rm CNM} \frac{31 G'_{\rm 0}}{1 + 3.1 Z'^{0.365}}. \\
\end{split}
\end{equation}

\noindent Here $G'_{\rm 0}$ is the incident UV radiation field and $Z'$ is the metallicity.
The primes indicate that quantities are normalized to the values in the solar neighborhood.
There is a maximum temperature at which CNM can be in pressure balance with WNM
and the corresponding density is the minimum CNM density, $n_{\rm min}$ (Wolfire et al. 2003).
Typically, $\phi_{\rm CNM}$ $\sim$ 1--10.
As $n_{\rm CNM}$ $\propto$ $G'_{\rm 0}$ $\propto$ $E^{\ast}_{\rm 0}$,
$G'_{\rm 0}$ and $E^{\ast}_{\rm 0}$ drop out of Equation (6) 
and $\chi$ becomes purely a function of $Z'$ and $\phi_{\rm CNM}$:
\begin{equation}
\chi = 2.3 ~ \frac{1 + 3.1 Z'^{0.365}}{\phi_{\rm CNM}}. 
\end{equation}

To estimate $\tau_{\rm R}$, KMT09 started from the mean column density, $\Sigma_{\rm comp}$, 
of a spherical cloud that consists of a molecular core of number density $n_{\rm mol}$
and a H\textsc{i} shielding layer of number density $n_{\rm CNM}$:
\begin{equation}
\Sigma_{\rm comp} = \frac{4}{3} \mu_{\rm H} n_{\rm CNM} R \left[1 + (\phi_{\rm mol} - 1) x_{\rm H2}^{3}\right]. 
\end{equation}

\noindent Here $\phi_{\rm mol}$ is the ratio of $n_{\rm mol}$ to $n_{\rm CNM}$.
Once $\tau_{\rm c}$, the dust optical depth that the cloud would have
if its H\textsc{i} and H$_{2}$ are uniformly mixed, is defined by:
\begin{equation}
\tau_{\rm c} = \frac{3}{4} \left(\frac{\Sigma_{\rm comp} \sigma_{\rm d}}{\mu_{\rm H}}\right),
\end{equation}

\noindent $\tau_{\rm R}$ can be expressed as a function of $\tau_{\rm c}$, $\phi_{\rm mol}$, and $x_{\rm H2}$:
\begin{equation}
\begin{split}
\tau_{\rm R} & = \frac{3}{4}\left(\frac{\Sigma_{\rm comp} \sigma_{\rm d}}{\mu_{\rm H}}\right) 
               \frac{1}{\left[1 + (\phi_{\rm mol} - 1) x_{\rm H2}^{3}\right]} \\
             & = \tau_{\rm c} \frac{1}{\left[1 + (\phi_{\rm mol} - 1) x_{\rm H2}^{3}\right]}. 
\end{split}
\end{equation}

As a consequence of these simplified functions,
the H$_{2}$ mass fraction, $f_{\rm H2}$, can be expressed with the following analytic function:
\begin{equation}
\begin{split}
f_{\rm H2} & = \frac{M_{\rm H2}}{M} \\
           & = 1 - \frac{3 \psi}{4 \tau_{\rm c}} \left[1 + \frac{0.8 \psi \phi_{\rm mol}}{4 \tau_{\rm c} + 
               3 (\phi_{\rm mol} - 1) \psi}\right]^{-1}, \\
\end{split}
\end{equation}

\noindent where $M_{\rm H2}$ is the H$_{2}$ mass and $M$ is the total cloud mass.
Finally, the H$_{2}$-to-H\textsc{i} ratio, $R_{\rm H2}$, can be expressed as:
\begin{equation}
\begin{split}
R_{\rm H2} & = f_{\rm H2} / f_{\rm H\textsc{i}} \\
           & = f_{\rm H2} / (1 - f_{\rm H2}) \\
           & = \frac{4 \tau_{\rm c}}{3 \psi} \left[1 + \frac{0.8 \psi \phi_{\rm mol}}{4 \tau_{\rm c} + 
               3(\phi_{\rm mol} - 1) \psi}\right] - 1. \\
\end{split}
\end{equation}
With $\psi$ being a function of $Z'$ and $\phi_{\rm CNM}$, 
and $\tau_{\rm c}$ being a function of the cloud gas column density, 
$R_{\rm H2}$ is determined by the total gas column density,
$Z'$, $\phi_{\rm CNM}$, and $\phi_{\rm mol}$, and is independent of the strength of the ISRF. 
While the total gas column density and $Z'$ are direct observables, 
$\phi_{\rm CNM}$ and $\phi_{\rm mol}$ are not. 
KMT09 adopted $\phi_{\rm CNM}$ = 3 and $\phi_{\rm mol}$ = 10 as fiducial values. 
Note that the H$_{2}$-to-H\textsc{i} ratio can observationally be expressed as 
$R_{\rm H2}=\Sigma_{\rm H2} / \Sigma_{\rm H\textsc{i}}$.

Another important model prediction is that $\Sigma_{\rm H\textsc{i},s}$ has a dependence on $Z'$ 
and is almost independent of the strength of the ISRF.  
Based on Equation (9) and (10), $\tau_{\rm H\textsc{i}}$ slightly increases with $Z'$ via $\tau_{\rm R}$ and $\psi$ 
and $\sigma_{\rm d}$ increases with $Z'$ via $\sigma_{\rm d}$ = $\sigma_{d}' Z'$.
Therefore, $\Sigma_{\rm H\textsc{i},s}$ decreases with $Z'$ to a power less than unity 
as $\Sigma_{\rm H\textsc{i},s} \propto \tau_{\rm H\textsc{i}}/\sigma_{\rm d}$.
For solar metallicity, KMT09 predict $\Sigma_{\rm H\textsc{i},s}$ $\sim$ 10 M$_{\odot}$ pc$^{-2}$.

\section{The H\textsc{i}--$\boldsymbol {\rm H_{2}}$ Transition in Perseus: Observational Perspective}

From our derived $N$(H\textsc{i}) and $N$(H$_{2}$) images, 
we calculate $\Sigma_{\rm H\textsc{i}}$ and $\Sigma_{\rm H2}$ by: 

\begin{equation*}
\Sigma_{\rm H\textsc{i}}~(\rm M_{\odot}~pc^{-2}) = 
\frac{\textit{N}(H\textsc{i})}{1.25 \times 10^{20}~(cm^{-2})}
\end{equation*}
\begin{equation}
\Sigma_{\rm H2}~(\rm M_{\odot}~pc^{-2}) = 
\frac{\textit{N}(H_{2})}{6.25 \times 10^{19}~(cm^{-2})} .
\end{equation}

\noindent We find that $\Sigma_{\rm H\textsc{i}}$ is relatively constant across Perseus 
with a range of 5--11 M$_{\odot}$ pc$^{-2}$,
while $\Sigma_{\rm H2}$ has a wider range of 0--73 M$_{\odot}$ pc$^{-2}$.
This results in a small dynamic range of $\Sigma_{\rm H\textsc{i}}+\Sigma_{\rm H2}$, 
from 8 M$_{\odot}$ pc$^{-2}$ to 30 M$_{\odot}$ pc$^{-2}$ across most of the Perseus cloud.  

We then derive $R_{\rm H2}$ = $\Sigma_{\rm H2}$/$\Sigma_{\rm H\textsc{i}}$, 
which allows us to directly test KMT09's predictions. 
The $R_{\rm H2}$ image is presented in Figure \ref{f:RH2}.
We find that the derived $R_{\rm H2}$ ranges from 0 to 10, with a median of 0.1. 
This suggests that diffuse molecular gas is pervasive in Perseus. 
For example, the data points with $R_{\rm H2}$ > 1, which can be considered to be molecular-dominated, 
comprise only 9\% of the total data points. 
We note that spatial variations in the $R_{\rm H2}$ distribution are 
mainly determined by the $\Sigma_{\rm H2}$ distribution, 
due to the small dynamic range of $\Sigma_{\rm H\textsc{i}}$. 
The high $R_{\rm H2}$ features spatially correlate with the $A_{V}$ and CO images 
and correspond to the dark and star-forming regions in Perseus.   

\subsection{$\boldsymbol{\Sigma_{\rm H\textsc{i}}}$ vs
$\boldsymbol{\Sigma_{\rm H\textsc{i}}+\Sigma_{\rm H2}}$ for the Dark and Star-forming Regions}

To investigate KMT09's predictions, we focus on several dark and star-forming regions in Perseus 
and measure $R_{\rm H2}$ radially to probe the cloud envelopes with the lowest $R_{\rm H2}$. 
In Figure \ref{f:sigmaHI_sigmaTotal}, we plot $\Sigma_{\rm H\textsc{i}}$ 
as a function of $\Sigma_{\rm H\textsc{i}}+\Sigma_{\rm H2}$ for B5, IC348, B1E, B1, and NGC1333.  
For each plot we use the data points in the rectangular boxes shown in Figure \ref{f:RH2}.  
The overlaid curves are KMT09's predictions for particular combinations of input parameters, 
$Z'$, $\phi_{\rm CNM}$, and $\phi_{\rm mol}$.
These parameters will be determined and discussed in Section 7.2.1.

We find an almost constant $\Sigma_{\rm H\textsc{i}}$ $\sim$ 6--8 M$_{\odot}$ pc$^{-2}$ for each region.
This is consistent with the small range of $\Sigma_{\rm H\textsc{i}}$ found across the whole Perseus cloud.  
However, the median value of $\Sigma_{\rm H\textsc{i}}$ slightly changes between the regions.  
For example, B5 has the lowest median $\Sigma_{\rm H\textsc{i}}\sim5.8$ M$_{\odot}$ pc$^{-2}$
and NGC1333 has the highest median $\Sigma_{\rm H\textsc{i}} \sim\ 8.2$ M$_{\odot}$ pc$^{-2}$. 
A similar result for Perseus was pointed out by Sancisi et al. (1974) 
who found that $\Sigma_{\rm H\textsc{i}}$ remains approximately constant 
at $\sim$8 M$_{\odot}$ pc$^{-2}$ for a wide range of interstellar extinction.
Several extragalactic studies have also noticed that $\Sigma_{\rm H\textsc{i}}$ 
does not exceed $\sim$10 M$_{\odot}$ pc$^{-2}$ for nearby galaxies 
(e.g., Wong \& Blitz 2002; Blitz \& Rosolowsky 2004, 2006; Bigiel et al. 2008; Wong et al. 2009).  

In KMT09's model, H$_{2}$ formation requires a certain amount of $\Sigma_{\rm H\textsc{i},s}$ 
to shield H$_{2}$ against photodissociation.
Once this minimum $\Sigma_{\rm H\textsc{i}}$ is achieved, 
H$_{2}$ forms out of H\textsc{i} and $\Sigma_{\rm H\textsc{i}}$ remains constant.  
KMT09's model predicts $\Sigma_{\rm H\textsc{i},s}$ $\sim$ 10 M$_{\odot}$ pc$^{-2}$
for solar metallicity and this is indeed consistent with what we find in Figure \ref{f:sigmaHI_sigmaTotal}.
For all five regions, we find saturation of $\Sigma_{\rm H\textsc{i}} \sim$ 6--8 M$_{\odot}$ pc$^{-2}$. 
However, in the case of B5 and NGC1333, we do not probe the regions of purely atomic gas, 
which correspond to the linear portion of the model curves in Figure \ref{f:sigmaHI_sigmaTotal}. 
For IC348, B1E, and B1, we see hints of a possible turnover between atomic- and molecular-dominated zones
at $\Sigma_{\rm H\textsc{i}}+\Sigma_{\rm H2}$ = 5--8 M$_{\odot}$ pc$^{-2}$,
yet our observations do not sample well the purely atomic zones.
This inability to probe purely atomic regions is not due to the sensitivity of 
our $\Sigma_{\rm H\textsc{i}}$ and $\Sigma_{\rm H2}$ images. 
In Figure \ref{f:sigmaHI_sigmaTotal}, we show the median 3$\sigma$ values of 
$\Sigma_{\rm H\textsc{i}}$ and $\Sigma_{\rm H\textsc{i}}+\Sigma_{\rm H2}$ for the whole Perseus cloud as the black dashed lines. 
All data points are clearly above the 3$\sigma$ values 
and this suggests that the lack of turnover detection is due to the highly extended atomic envelopes. 
For example, in the case of B5 and NGC1333,
we are probing radial profiles up to $\sim$15--20 pc from the centers of the regions. 
Therefore, the purely atomic envelopes are likely located more than $\sim$20 pc from the molecular peaks. 

\subsubsection{Effect of High Optical Depth}

An alternative or additional explanation for the relatively constant $\Sigma_{\rm H\textsc{i}}$ we measured 
could be a high optical depth of H\textsc{i} in Perseus.
If H\textsc{i} gas is optically thick, the measured brightness temperature $T_{\rm b}$ levels off
at a temperature close to the kinetic temperature $T_{\rm k}$.
Therefore, under the assumption of optically thin H\textsc{i} gas 
(which we used to calculate our $\Sigma_{\rm H\textsc{i}}$), 
$\Sigma_{\rm H\textsc{i}}$ would saturate as $\Sigma_{\rm H\textsc{i}}$ $\propto$ $T_{\rm b}$ $\sim$ $T_{\rm k}$ 
and therefore be underestimated.
There are a few lines of evidence that indicate the presence of cold, optically thick CNM in Perseus.
For example, HISA features were found toward the centers of several dark and star-forming regions (e.g., Ridge et al. 2006a). 
In addition, Heiles \& Troland (2003b) analyzed the H\textsc{i} emission/absorption spectra 
obtained toward two radio sources behind Perseus, 3C93.1 and NRAO140,  
and found CNM components in Perseus (based on radial velocity information) 
that have optical depths $\tau$ $\sim$ 1 and $\sim$7.

The best way to estimate the effect of optical depth is to measure CNM in absorption
in the direction of many background radio sources behind Perseus. 
This provides the total column density of atomic hydrogen gas ($\rm CNM+WNM$) along each line of sight, 
which can be compared with the column density we measured in Section 3.2 
to derive the optical depth corrections for the whole cloud.  
As only two such measurements exist so far, we are in the process of obtaining additional H\textsc{i} absorption measurements.  
Here we investigate only the magnitude of the optical depth effect.

If we assume $T_{\rm s}$ = 70 K as measured by Heiles \& Troland (2003b),  
the corrected $N$(H\textsc{i}) will need to be calculated using the following equation:
\begin{equation}
N(\textrm{H}\textsc{i})~(\textrm{cm}^{-2}) = 1.823 \times 10^{18}~T_{\textrm{s}} \int \textrm{ln}
\left(\frac{T_{\textrm{s}}}{T_{\textrm{s}} - T_{\textrm{b}}}\right) dv.
\end{equation}

\noindent If we apply this equation to our $N$(H\textsc{i}) image, 
we find that $\Sigma_{\rm H\textsc{i}}$ changes by a factor of 1.2--2. 
Another approach involves using the statistics of CNM and WNM from Heiles \& Troland (2003b).
Their Table 1 summarizes the CNM and WNM column densities for all sources in their survey. 
We can then estimate the fraction of CNM column density (relative to the total column density) 
as a function of WNM column density.
Using only the table entries for a region centered on Perseus and $\sim$1200 deg$^{2}$ in size,
we find that the CNM fraction is close to 50\% for WNM column density less than 1.2 $\times$ 10$^{21}$ cm$^{-2}$. 
This again suggests that our $N$(H\textsc{i}) or $\Sigma_{\rm H\textsc{i}}$ could be underestimated by a factor of two.

Our excercise shows that a high optical depth along each line of sight could result in 
a factor of two spread in $\Sigma_{\rm H\textsc{i}}$ for Perseus. 
This is not substantial and therefore the almost constant $\Sigma_{\rm H\textsc{i}}$ we measured 
is likely a real physical feature, possibly driven by the conversion of H\textsc{i} into H$_{2}$.
However, the magnitude of the optical depth effect we estimated is approximate 
and measuring the true $\Sigma_{\rm H\textsc{i}}$ distribution across Perseus needs a more sophisticated investigation.

\subsection{$\boldsymbol{R_{\rm H2}}$ vs
$\boldsymbol{\Sigma_{\rm H\textsc{i}}+\Sigma_{\rm H2}}$ for the Dark and Star-forming Regions}

We now investigate KMT09's prediction of $R_{\rm H2}$ vs $\Sigma_{\rm H\textsc{i}}+\Sigma_{\rm H2}$
for the dark and star-forming regions in Perseus. 
In Figure \ref{f:RH2_sigmaTotal}, we plot $R_{\rm H2}$ 
as a function of $\Sigma_{\rm H\textsc{i}}+\Sigma_{\rm H2}$ for B5, IC348, B1E, B1, and NGC1333. 
For each plot we use the data points in the rectangular boxes shown in Figure \ref{f:RH2}. 

We find that the relation between $R_{\rm H2}$ and $\Sigma_{\rm H\textsc{i}}+\Sigma_{\rm H2}$ 
is remarkably consistent for all dark and star-forming regions. 
Each plot in Figure \ref{f:RH2_sigmaTotal} shows a very sharp rise of $R_{\rm H2}$ 
at low $\Sigma_{\rm H\textsc{i}}+\Sigma_{\rm H2}$, 
a turnover around $R_{\rm H2}$ $\sim$ 1, and a slow increase toward higher $R_{\rm H2}$. 
For the two star-forming regions, $R_{\rm H2}$ reaches a value of $\sim$10,
while for the three dark regions, $R_{\rm H2}$ is less than 5.

\subsubsection{Best-fit Curves}

In Figure \ref{f:RH2_sigmaTotal}, we overlay KMT09's predictions as red solid curves and find excellent agreement.
The overlaid curves were derived using Equation (17) with the assumption 
of $\phi_{\rm mol} = 10$ and $Z' = 1$ for all dark and star-forming regions. 
The fiducial value $\phi_{\rm mol} =10$ was adopted following the discussion in KMT09. 
We note that $\phi_{\rm mol}$ does not make a significant change in $R_{\rm H2}$.   
For example, between $\phi_{\rm mol} = 10$ and 50, 
$R_{\rm H2}$ at $\Sigma_{\rm H\textsc{i}}+\Sigma_{\rm H2}$ = 100 M$_{\odot}$ pc$^{-2}$ only varies by a factor of 1.1.  
Our assumption of a single $Z' = 1$ is based on the result of Gonz\'alez Hern\'andez et al. (2009)
and our DGR investigation in Section 4.5.  
Gonz\'alez Hern\'andez et al. (2009) studied the chemical composition of \v{C}ernis 52, 
an A3 V-type star in the direction of Perseus and derived a metallicity of 
[Fe/H] = $-0.01 \pm 0.15$ (corresponding to 0.7--1.4 Z$_{\odot}$).
They determined a distance to \v{C}ernis 52 of 231$^{+135}_{-85}$ pc 
and in conjuction with the radial velocity information 
concluded that \v{C}ernis 52 is likely a member of IC348.
As a single DGR fits most of the diffuse regions in Perseus,  
there is probably no significant variation of $Z'$ across the cloud. 
Therefore, we adopt $Z' = 1$ for all dark and 
star-forming regions in Perseus throughout this paper 
and constrain only $\phi_{\rm CNM}$ from our fitting\footnote{
We have attempted to determine the best-fit curve 
for $R_{\rm H2}$ vs $\Sigma_{\rm H\textsc{i}}+\Sigma_{\rm H2}$ 
by using both $Z$ and $\phi_{\rm CNM}$ as free parameters 
but have found that the results are consistent with the case of $Z' = 1$ 
and varying $\phi_{\rm CNM}$ within 1$\sigma$.}.

We performed Monte Carlo simulations to determine the best-fit curves 
with consideration of uncertainties in both $R_{\rm H2}$ and $\Sigma_{\rm H\textsc{i}}+\Sigma_{\rm H2}$.  
We added random offsets to $R_{\rm H2}$ and $\Sigma_{\rm H\textsc{i}}+\Sigma_{\rm H2}$ 
and repeated this process 1000 times. 
For each variable, offsets were drawn from a Gaussian distribution 
with the standard deviation equal to the measured noise. 
For each realization, we determined the best-fit curve by assuming $\phi_{\rm mol} = 10$ and $Z' = 1$  
and finding $\phi_{\rm CNM}$ that minimizes the sum of the squares of the residuals.
IDL routine \textsc{mpfitfun} (Markwardt 2009) was used for the fitting. 
Finally, we estimate the median $\phi_{\rm CNM}$ among the simulated 1000 $\phi_{\rm CNM}$ 
and use it as the best-fit parameter.
We summarize this best-fit $\phi_{\rm CNM}$ for each region in Table 1.

\subsubsection{Fitting Results}

We find that $\phi_{\rm CNM}$ ranges from $\sim$6 to $\sim$10.
As $\phi_{\rm CNM}$ determines the typical CNM number density via $n_{\rm CNM} = \phi_{\rm CNM} n_{\rm min}$,  
$\phi_{\rm CNM}$ = 6--10 translates into $T_{\rm CNM}$ = 60--75 K 
under the assumption of thermal pressure allowing the coexistence of CNM and WNM (see Equation 19 of KMT09). 
This temperature range is indeed consistent with typical CNM properties in the solar neighborhood. 
Heiles \& Troland (2003b) examined H\textsc{i} emission/absorption spectra 
obtained toward 79 radio sources and found a median $T_{s}$ of $\sim$70 K for CNM.

While $\phi_{\rm CNM}$ appears to systematically decrease toward the west of Perseus, 
we find no significant difference in $\phi_{\rm CNM}$ between the dark and star-forming regions.  
This implies that H\textsc{i} envelopes surrounding these regions have similar $n_{\rm CNM}$.
We estimate $n_{\rm CNM}$ = 20--30 cm$^{-3}$ using $G_{0}'$ = $G / G_{0}$ = 0.4 
derived from Equation (5) with $T_{\rm dust}$ = 17 K.
The similar $n_{\rm CNM}$ further suggests a similar normalized radiation field $\chi$ of 1--1.5. 
As $\chi$ measures the relative importance of dust shielding and H$_{2}$ self-shielding,  
$\chi\sim 1$ implies that dust shielding and H$_{2}$ self-shielding are equally important for H$_{2}$ formation in Perseus. 

\subsubsection{Column Density at The H\textsc{i}--H$_{2}$ Transition}

KMT09 assume an infinitely sharp H\textsc{i}--H$_{2}$ transition
and in this case $R_{\rm H2}$ would be zero at the H\textsc{i} shielding column density.  
However, in reality, the transition cannot be infinitely sharp 
and KMT09 estimate $R_{\rm H2}$ $\sim$ 0.25 at the H\textsc{i} shielding column density
by comparing the theoretical H\textsc{i} shielding column densities 
with the detailed numerical radiative transfer calculations.
For solar metallicity, KMT09 predict (as shown with the red solid curves) that $R_{\rm H2}$ $\sim$ 0.25 occurs
at $\Sigma_{\rm H\textsc{i}}+\Sigma_{\rm H2}$ = 7--10 M$_{\odot}$ pc$^{-2}$. 
This is indeed consistent with what we see in Figure \ref{f:RH2_sigmaTotal}; 
$R_{\rm H2}$ $\sim$ 0.25 at $\Sigma_{\rm H\textsc{i}}+\Sigma_{\rm H2}$ = 6--12 M$_{\odot}$ pc$^{-2}$ 
or $N{\rm (H\textsc{i})}+2N(\rm{H_{2}})$ = (8--14) $\times$ 10$^{20}$ cm$^{-2}$. 
This transition column density is a factor of 2--3 larger than 
what previous studies found in the Galaxy via UV absorption measurements (e.g., Savage et al. 1977; Gillmon et al. 2006).  
However, considering that the previous studies defined the transition column density at $R_{\rm H2}$ $\sim$ 0.1  
and $\Sigma_{\rm H2}$ linearly increases with $\Sigma_{\rm H\textsc{i}}+\Sigma_{\rm H2}$, 
our result is consistent with the previous studies.

\subsection{$\boldsymbol{R_{\rm H2}}$ Radial Profiles}

To investigate how $R_{\rm H2}$ radially changes from a molecular peak,
we plot $R_{\rm H2}$ as a function of distance from the center of 
each dark and star-forming region in Figure \ref{f:RH2_radial}. 
We use the data points in the rectangular boxes (see Figure \ref{f:RH2})
and calculate the mean $R_{\rm H2}$ in three-pixel size bins 
(bin size = 12.9$'$, corresponding to 1.1 pc at the distance of 300 pc) to remove small-scale fluctuations.

For all five regions, we find two power-law functions ($R_{\rm H2} \propto r^{-\alpha}$) 
in their radial profiles up to $r \sim 150'$.  
Beyond $r \sim 150'$, the radial profiles show slight fluctuations due to the presence of small-scale structure 
in the $R_{\rm H2}$ image and/or become flat.  
We perform a linear least-squares fit to each radial profile up to $r \sim 150'$
and summarize the results in Table 2 with other properties of the radial profiles.
We list the slopes of power-law functions ($\alpha_1$ and $\alpha_2$),
the radius where the transition from the shallow to steep power-law functions occurs ($r_{\rm b}$), 
$R_{\rm H2}$ at $r_{\rm b}$ ($R_{\rm b, H2}$), 
the radius where the H\textsc{i}--H$_{2}$ transition occurs ($r_{\rm t}$), 
the radius where hydrogen is predominantly in the form of H\textsc{i} ($r_{\rm H\textsc{i}}$), 
and the thickness of the H\textsc{i}--H$_{2}$ transition region ($d_{\rm t}$). 
We define $r_{\rm t}$ as a radius at $R_{\rm H2}$ $\sim$ 0.25, $r_{\rm H\textsc{i}}$ 
as a radius at $R_{\rm H2}$ $\sim$ 0.1, and $d_{\rm t}$ as $r_{\rm t} - r_{\rm b}$.

We find no significant difference between the dark and star-forming regions 
in terms of their $R_{\rm H2}$ radial distribution. 
The shallow portion of the radial profiles has a slope of $\alpha_{2}$ = 0.4--0.7,
while the steep portion has a slope of $\alpha_{1}$ = 1.4--3.7.
The transition from the shallow to the steep power-law function occurs at 
$r_{\rm b}$ $\sim$ 30--90$'$ or $\sim$3--8 pc.  
At this $r_{\rm b}$, $R_{\rm H2}$ ranges from 0.5 to 1.2. 
In addition, we find that H$_{2}$ extends up to $r \sim$ 200$'$ or $\sim$17 pc 
from the centers of the dark and star-forming regions.
This suggests that molecular clouds are much more extended than what is suggested by molecular tracers such as CO. 
The H\textsc{i}--H$_{2}$ transition, however, seems to be relatively sharp.
We estimate $d_{\rm t}$ = $r_{\rm t} - r_{\rm b}$, which measures the size of diffuse H$_{2}$ layer
and therefore could be considered as the thickness of the transition region,
and find $d_{\rm t}$ $\sim$ 30--60$'$, corresponding to $\sim$3--5 pc. 
We then estimate the ratio of the thickness of the H\textsc{i}--H$_{2}$ transition region to the region size, 
$d_{\rm t}/R$ < $d_{\rm t} / r_{\rm H\textsc{i}}$ = 0.2--0.4.
As we do not know the exact size of the dark and star-forming regions, 
$d_{\rm t} / r_{\rm H\textsc{i}}$ provides only an upper limit on $d_{\rm t}/R$.  

The other interesting thing to note is that the $R_{\rm H2}$ radial distribution becomes 
smoother toward the west of Perseus. 
Namely, $\Delta \alpha = |\alpha_{1} - \alpha_{2}|$ becomes smaller from B5 to NGC1333. 
This may be related to $n_{\rm CNM}$ decreasing from B5 to NGC1333 
(inferred from $\phi_{\rm CNM}$ decreasing from B5 to NGC1333 with the same $n_{\rm min}$; see Section 7.2.1),  
even though it is not clear how $n_{\rm CNM}$ affects the smoothness of $R_{\rm H2}$ radial distribution. 
In KMT09's model, $n_{\rm CNM}$ determines the location of the H\textsc{i}--H$_{2}$ transition 
and due to the assumption of the infinitely sharp transition, there is no such relation 
between $n_{\rm CNM}$ and the smoothness of the $R_{\rm H2}$ radial distribution.
This systematic variation of $\Delta \alpha$ across Perseus could be resulted from 
the change of internal density structure and/or the complicated viewing geometry.    

\section{Comparison of H$\boldsymbol{_{2}}$ and CO Distributions}

In Figure~\ref{f:H2_CO}, we compare the H$_{2}$ and CO distributions.
The black and gray contours show the CfA $I_{\rm CO}$ data and our $\Sigma_{\rm H2}$ data, respectively. 
To ease the comparison, we plot only 3$\sigma$ contours for both data sets.  
Figure~\ref{f:H2_CO} shows that H$_{2}$ is found wherever significant CO exists. 
However, the reverse is not true; a considerable amount of H$_{2}$ exists beyond the 3$\sigma$ contour of $I_{\rm CO}$. 
While H$_{2}$ is generally more extended than CO, there are significant spatial variations.  
For example, CO and H$_{2}$ trace each other remarkably well in the southwest part of Perseus,  
while H$_{2}$ is significantly more extended elsewhere (particularly around B5, B1E, B1, IC348, and NGC1333). 

\subsection{Relative Distribution of $\boldsymbol{\rm H_{2}}$ and CO}

To quantify the relative distribution of H$_{2}$ and CO, we plot $\Sigma_{\rm H2}$ and $I_{\rm CO}$ 
as a function of distance from the center of each dark and star-forming region in Figure \ref{f:H2_CO_radial}.
We first smooth the $\Sigma_{\rm H2}$ image to the angular resolution of the CfA $I_{\rm CO}$ image 
and use the data points in the rectangular boxes shown in Figure \ref{f:RH2}.
We then calculate the mean $\Sigma_{\rm H2}$ and $I_{\rm CO}$ in two-pixel size bins 
(bin size = 16.8$'$, corresponding to 1.5 pc at the distance of 300 pc).
Finally, each radial profile is normalized by its peak value 
and therefore Figure \ref{f:H2_CO_radial} shows the relative distribution of H$_{2}$ and CO.
As spatial averaging improves SNR, the radial profiles of $\Sigma_{\rm H2}$ and $I_{\rm CO}$ allow us 
to investigate the relative distribution of H$_{2}$ and CO more reliably than the visual comparison of 3$\sigma$ contours. 

We find that all five regions have extended H$_{2}$ compared to CO at all radii. 
For IC348, B1E, and NGC1333, H$_{2}$ radial profiles show slight fluctuations and/or become flat at $r > 150'$,
which has been seen in Figure \ref{f:RH2_radial}. 
We measure the radius where each H$_{2}$ and CO radial profile drops off to 1/4 of its peak value 
($r_{\rm H2}$ and $r_{\rm CO}$) and calculate the ratio of $r_{\rm H2}$ to $r_{\rm CO}$.
We find that the ratio, $r_{\rm H2}/r_{\rm CO}$, ranges from 1.2 to 2, with a median of 1.4. 
This suggests that on average H$_{2}$ is $\sim$1.4 times more extended than CO in Perseus, 
implying the existence of a substantial amount of ``CO-dark'' gas.

\subsection{Comparison to Wolfire et al. (2010): ``CO-dark'' Gas}

The ``CO-dark'' gas refers to interstellar gas that exists in the form of H$_{2}$ 
along with C\textsc{i} and C\textsc{ii}, but little or no CO.
Such gas has been predicted from theoretical models of PDRs, 
which showed that the C\textsc{ii}--CO transition occurs deeper into an interstellar cloud 
than the H\textsc{i}--H$_{2}$ transition due to the more effective self-shielding for H$_{2}$ than CO
(e.g., van Dishoeck \& Black 1988; Tielens \& Hollenbach 1985). 

Recently, Wolfire et al. (2010; WHM10 hereafter) modeled the ``CO-dark'' gas 
and explored how the fraction of H$_{2}$ gas mass in this ``CO-dark'' component 
depends on the mass of a cloud ($M$), the UV radiation field ($G_{0}'$), 
the metallicity ($Z'$), and the mean $V$-band extinction ($\overline{A}_{V}$). 
They used a PDR code to calculate the H$_{2}$, C\textsc{i}, and C\textsc{ii} abundances 
in a spherical cloud and found that the fraction of ``CO-dark'' gas, $f_{\rm DG}$, 
is $\sim$0.3 for solar metallicity over a wide range of $M$ and $G_{0}'$. 
Here $f_{\rm DG}$ is defined as $f_{\rm DG}$ = ($M(r'_{\rm H2}) - M(r'_{\rm CO})$)/$M(r'_{\rm H2})$, 
where $r'_{\rm H2}$ and $r'_{\rm CO}$ is the radius of the H$_{2}$ and CO part of the cloud, respectively. 
Formally, $r'_{\rm H2}$ is defined as the radius where 2$n_{\rm H2} = n_{\rm H\textsc{i}}$ 
and $r'_{\rm CO}$ is defined as the radius where the optical depth from the cloud surface to $r'_{\rm CO}$ is 1.
More simply, when expressed in terms of $\Delta A_{V, \rm DG} = A_{V}(r'_{\rm CO}) - A_{V}(r'_{\rm H2})$, 
$f_{\rm DG}$ depends only on the ratio $\Delta A_{V, \rm DG} / \overline{A}_{\rm V}$.
This is because $\Delta A_{V, \rm DG}$ is a measure of the ``CO-dark'' gas mass 
and $\overline{A}_{\rm V}$ is a measure of the H$_{2}$ gas mass. 
Interestingly, WHM10 found that $\Delta A_{V, \rm DG}$ is a weak function of $G_{0}'$ and is almost constant. 
Therefore, $f_{\rm DG}$ is strongly dependent only on $\overline{A}_{\rm V}$.  
For example, for a cloud with CO mass of $10^6$ M$_{\odot}$ and solar metallicity, 
$f_{\rm DG} \sim 0.8$ is expected at $\overline{A}_{V} =2$ mag, 
while $f_{\rm DG} \sim 0.2$ is expected at $\overline{A}_{V} =20$ mag.  

To compare our observations to WHM10's prediction, 
we calculate the H$_{2}$ mass within the 3$\sigma$ boundary of $I_{\rm CO}$ ($M(r_{\rm 3\sigma,CO})$) 
and compare it with the H$_{2}$ mass within the 3$\sigma$ boundary of $\Sigma_{\rm H2}$ ($M(r_{\rm 3\sigma,H2})$).
We find $f_{\rm DG}$ = ($M(r_{\rm 3\sigma,H2}) - M(r_{\rm 3\sigma,CO})$)/$M(r_{\rm 3\sigma,H2}) \sim$ 0.3. 
Note that our definition of $f_{\rm DG}$ is different from that of WHM10. 
This is because $r'_{\rm CO}$ cannot be directly measured from observations. 
Nevertheless, $M(r'_{\rm H2})$ and $M(r'_{\rm CO})$ of WHM10 represent 
the total H$_{2}$ and CO masses and therefore our $f_{\rm DG}$ is conceptually similar with what WHM10 used.
We estimate $\overline{A}_{V} \sim$ 1.5 mag for the data points within $r_{\rm 3\sigma,H2}$. 
As WHM10 do not provide a prediction of $f_{\rm DG}$ at $\overline{A}_{V}$ < 2 mag, 
we extrapolate it from their Figure 11; $f_{\rm DG} \sim$ 0.8 is expected at $\overline{A}_{V} \sim$ 1.5 mag.
Note that the expected $f_{\rm DG} \sim$ 0.8 at $\overline{A}_{V} \sim$ 1.5 mag is for a cloud 
that has a much higher mass ($M \sim 10^{6}$ M$_{\odot}$) and a stronger radiation field ($G_{0}' \sim 10$) than Perseus.  
Considering the predicted insensitivity of $f_{\rm DG}$ to the cloud mass and the strength of radiation field, 
however, we measure a factor of three lower value, $f_{\rm DG} \sim$ 0.3.

To investigate this discrepancy between our $f_{\rm DG}$ and WHM10's prediction, 
we first examine the dependence of $f_{\rm DG}$ on the DGR. 
As shown in Section 4.5, our estimate of $\Sigma_{\rm H2}$ 
and consequently $f_{\rm DG}$ depends on the DGR via Equation (4).
In the same section, we discussed the lower and upper limits on the DGR.
If we use the lower limit on the DGR instead of our adopted DGR,
$\overline{A}_{V}$ decreases to 1.1 mag and $f_{\rm DG}$ increases to 0.6. 
Our estimate is a factor of two lower than WHM10's prediction 
at $\overline{A}_{V}$ $\sim$ 1.1 mag, $f_{\rm DG}$ $\sim$ 0.9. 
Similarly, if we use the upper limit on the DGR,
$\overline{A}_{V}$ increases to 1.9 mag and $f_{\rm DG}$ decreases to 0.2.
Our estimate is now a factor of four lower than WHM10's prediction 
at $\overline{A}_{V}$ $\sim$ 1.9 mag, $f_{\rm DG}$ $\sim$ 0.8.
Note that ``CO-dark'' gas exists in Perseus even for these extreme cases 
and the discrepancy with the model persists.

Second, we test if $f_{\rm DG}$ is a strong function of $\overline{A}_{V}$ as WHM10 predicted.  
For this purpose, we divide the $\Sigma_{\rm H2}$ image into four regions (see Figure \ref{f:H2_CO})  
and estimate $f_{\rm DG}$ and $\overline{A}_{V}$ for each region. 
We use the $\Sigma_{\rm H2}$ image derived from our adopted DGR, 1.1 $\times$ 10$^{-21}$ mag cm$^{2}$.
In Table 3, we list the estimated $f_{\rm DG}$ and $\overline{A}_{V}$ for the four regions. 
We find $f_{\rm DG}$ $\sim$ 0.3 for regions 1 and 2 and $f_{\rm DG}$ = 0.1--0.2 for regions 3 and 4. 
All regions have similar mean extinction, $\overline{A}_{V}$ = 1.4--1.7 mag. 
While we have only four data points covering a small range of $\overline{A}_{V}$,  
we measure a relatively large range of $f_{\rm DG}$, from $\sim$0.1 to $\sim$0.3.
The large variation we find in $f_{\rm DG}$ is not consistent with WHM10, 
but echos the observations of other diffuse clouds in the Galaxy. 
For example, the Polaris cloud has $\overline{A}_{V} \sim$ 0.3 mag (Heithausen \& Thaddeus 1990) 
and $f_{\rm DG} \sim$ 0.36 (Abdo et al. 2010).
On the other hand, the Pegasus cloud has $\overline{A}_{V} \sim$ 0.6 mag (Yamamoto et al. 2006)
and $f_{\rm DG} \sim$ 0.6 (Grenier et al. 2005).

This region-to-region variation we find in $f_{\rm DG}$ could be due to the DGR systematically changing across Perseus. 
For example, a factor of two lower DGR would allow regions 3 and 4 to have similar $f_{\rm DG}$ as regions 1 and 2. 
However, we do not find any evidence for such systematic change of the DGR 
when examining the correlation between $N$(H\textsc{i}) and FIR emission (see Section 4.2).
As an alternative, the complicated viewing geometry of the H$_{2}$ and CO distributions 
could result in the regional variation in $f_{\rm DG}$.

\section{Discussion}

The excellent agreement between our data and KMT09's predictions (Sections 7.1 and 7.2) suggests 
that KMT09's model captures the fundamental physics of H$_{2}$ formation reasonably well even on sub-pc scales.
This is certainly encouraging, but slightly surprising considering how simple KMT09's model is.
Essentially, for a spherical cloud bathed in a uniform and isotropic ISRF,
the most important factor for H$_{2}$ formation is the H\textsc{i} shielding surface density $\Sigma_{\rm H\textsc{i},s}$.
Once enough shielding against photodissociation is achieved by $\Sigma_{\rm H\textsc{i},s}$,
all additional $\Sigma_{\rm H\textsc{i}}$ is fully converted into $\Sigma_{\rm H2}$, 
resulting in a uniform $\Sigma_{\rm H\textsc{i}}$ distribution   
and therefore a linear increase of $\Sigma_{\rm H2}$ with $\Sigma_{\rm H\textsc{i}}+\Sigma_{\rm H2}$.  

Before going into a more detailed discussion of several important aspects of KMT09's model,
we would like to remind the reader that two main predictions summarized in Section 6 are not independent.
The predicted behavior of $R_{\rm H2}$ as a function of $\Sigma_{\rm H\textsc{i}}+\Sigma_{\rm H2}$ 
is a consequence of saturation of $\Sigma_{\rm H\textsc{i}}$. 
Therefore, the excellent agreement between our data and KMT09's predictions in Figure 12
is entirely driven by the constant $\Sigma_{\rm H\textsc{i}}$.
While we are not able to probe the purely atomic envelope of Perseus,
the constant $\Sigma_{\rm H\textsc{i}}$ of 6--8 M$_{\odot}$ pc$^{-2}$ we find 
is in agreement with KMT09's prediction of $\Sigma_{\rm H\textsc{i},s}$ for solar metallicity. 

In addition, we caution that our data do not allow us to fully test 
the dependence of $R_{\rm H2}$ on metallicity and ISRF. 
KMT09's model predicts that $R_{\rm H2}$ depends on metallicity, 
while being almost insensitive to the strength of ISRF. 
While our findings are in agreement with KMT09's predictions 
for solar metallicity and $\sim$2/5 of the Galactic ISRF, 
the lack of metallicity and ISRF variations across Perseus does not permit more detailed model testing. 
Future studies of high-latitude molecular clouds and GMCs in external galaxies with sub-solar metallicity 
will provide crucial tests of KMT09's model. 
We now turn to several crucial model assumptions and discuss their possible limitations.

\subsection{Equilibrium vs Non-equilibrium $\boldsymbol{\rm H_{2}}$ Formation}

The most fundamental principle underlying KMT09's model is H$_{2}$ formation in equilibrium.  
In equilibrium, H$_{2}$ formation and photodissociation are balanced and H$_{2}$ abundance is locally constant. 
Several authors have investigated the time evolution of H$_{2}$
and found a typical timescale of H$_{2}$ formation in equilibrium, $\tau_{\rm H2}$ = 10--30 Myr
(e.g., Goldsmith et al. 2007; Liszt 2007)

As $\tau_{\rm H2}$ is comparable to or longer than the estimated lifetime of GMCs, 
$\tau_{\rm life} \sim 10$ Myr (McKee \& Ostriker 2007 and references therein), 
the role of interstellar turbulence in accelerating H$_{2}$ formation has received 
significant attention in recent years 
(e.g., Hartmann et al. 2001; Bergin et al. 2004; Glover \& Mac Low 2007; Glover et al. 2010). 
The main idea behind the models of H$_{2}$ formation in turbulent media 
is that shocks induced by cloud--cloud collisions, spiral density waves, 
supernovae explosions, etc., compress diffuse atomic gas, 
allowing H$_{2}$ to form in the postshock gas where high enough shielding is achieved against photodissociation. 
While earlier studies have been limited to 1D, Glover \& Mac Low (2007) and Glover et al. (2010)
performed 3D magnetohydrodynamical (MHD) simulations of turbulence with detailed chemical modeling. 
They showed that a significant amount of H$_{2}$ could form within 1--2 Myr, 
due to the turbulence-driven clumping that accelerates H$_{2}$ formation. 
However, the H$_{2}$ abundance keeps increasing over time and never reaches equilibrium by 20--30 Myr.
Eventually, the gas will become fully molecular 
unless the H$_{2}$ formation is terminated by destruction of GMCs via stellar feedback, 
e.g., photoevaporation by H\textsc{ii} regions and protostellar outflows.
This sharply contrasts with KMT09's model. 

Our results are consistent with the model of H$_{2}$ formation in equilibrium 
and suggest that turbulence may play a secondary role in the H$_{2}$ formation. 
Clearly, further studies are needed to compare abundances of H$_{2}$ 
and other molecular species with the predictions from both stationary and turbulent models.  

\subsection{The Importance of WNM in the Envelopes of GMCs} 

We find that $\Sigma_{\rm H\textsc{i}}$ is uniform across Perseus with a range of 6--8 M$_{\odot}$ pc$^{-2}$, 
consistent with KMT09's prediction of $\Sigma_{\rm H\textsc{i},s}$ for solar metallicity. 
However, as CNM is far more effective at shielding against photodissociation than WNM, 
the predicted $\Sigma_{\rm H\textsc{i},s}$ comes purely from CNM,  
while our measured $\Sigma_{\rm H\textsc{i}}$ includes the contributions from both CNM and WNM. 
As we discussed in Section 7.1.1, our measured $\Sigma_{\rm H\textsc{i}}$ could be underestimated by a factor of two for high optical depth gas (CNM). 
While $\Sigma_{\rm H\textsc{i}}$ corrected for high optical depth would likely still be comparable to KMT09's prediction, 
the importance of WNM in the envelopes of atomic-molecular complexes clearly  
needs to be taken into account in the models of H$_{2}$ formation. 
This is also supported by Bolatto et al. (2011) who found a relatively good agreement 
between KMT09's prediction and the observed $R_{\rm H2}$ vs $\Sigma_{\rm H\textsc{i}} + \Sigma_{\rm H2}$ 
in the Small Magellanic Cloud, where WNM dominates the total H\textsc{i} (Dickey et al. 2000). 
At the same time, it is essential to observationally constrain the relative contribution from CNM and WNM to $\Sigma_{\rm H\textsc{i}}$. 
We are in the process of obtaining H\textsc{i} absorption measurements for Perseus, 
which will allow us to separate the contributions from cold and warm gas.

\subsection{The Importance of Internal Radiation}

As we showed in Section 5.2, in the case of IC348 and NGC1333, 
the embedded B-type stars are the dominant sources of the radiation field.  
KMT09's model does not take into account the role of internal radiation in determining $R_{\rm H2}$, 
with the assumption that massive stars producing significant amounts of dissociating photons 
do not stay inside their parent molecular clouds for a long time. 
While we clearly see the importance of the B-type stars for dust heating, 
we find no significant difference in our analysis of $R_{\rm H2}$ for the dark and star-forming regions. 
This supports the idea that the internal radiation is not important for H$_{2}$ formation. 
However, as we show below, this is valid for Perseus  
because it does not contain any stars earlier than the B5 V-type. 
 
When O- or B-type stars suddenly turn on in a molecular cloud, 
UV photons from the stars photodissociate H$_{2}$ and create H\textsc{i} dissociation regions around H\textsc{ii} regions. 
This variation of the UV radiation field has a much shorter timescale ($\sim$10$^{3}$ yr) 
than H$_{2}$ formation ($\sim$10$^{7}$ yr) and therefore totally changes 
the thermal and chemical structures of the molecular cloud (e.g., Hollenbach \& Natta 1995; Bertoldi \& Draine 1996).  
However, the sphere of photodissociation influence depends greatly on the stellar type.

Roger \& Dewdney (1992) studied the properties of H\textsc{i} dissociation regions 
around O- and B-type main sequence stars.
Their source list includes SVS3, the B5 V-type star in NGC1333 (red star in Figure \ref{f:dust_map}).
An H\textsc{i} region associated with SVS3 was measured 
to have $M_{\rm H\textsc{i}}$ = 0.017 M$_{\odot}$ and $R_{\rm H\textsc{i}}$ = 0.03 pc (Rodr\'iguez et al. 1990).
This is consistent with an H\textsc{i} dissociation region created by a star of $T_{\rm eff} \sim 16600$ K 
in a dense medium of $n_{\rm H\textsc{i}} \sim 6000$ cm$^{-3}$. 
The size of the H\textsc{i} dissociation region around this particular B5 star is only $\sim$1/10 of our spatial resolution. 
Therefore, the H\textsc{i} dissociation regions around the individual B-type stars in Perseus are below 
our spatial resolution and therefore are negligible for our consideration of $R_{\rm H2}$.  

However, the internal radiation is still important for dust heating inside IC348 and NGC1333.  
This can be understood by the fact that not every UV photon produces photodissociation of H$_{2}$. 
While UV and visible photons heat dust grains, 
only the UV photons in the narrower LW band (912--1120 \AA) photodissociate H$_{2}$. 
In addition, only $\sim$10\% of the LW photons actually photodissociate H$_{2}$ ($f_{\rm diss} \sim 0.1$).  
Therefore, the UV photons from the B-type stars in IC348 and NGC1333 could be dominant for dust heating,  
while relatively ineffective for photodissociation of H$_{2}$.

If any stars earlier than B3 V resided in Perseus, on the other hand, 
the role of the internal radiation in determining $R_{\rm H2}$ would be no longer negligible. 
For example, Roger \& Dewdney (1992) found that an H\textsc{i} region associated with IRAS 23545$+$6508 
has $M_{\rm H\textsc{i}}$ = 1.4 M$_{\odot}$ and $R_{\rm H\textsc{i}}$ = 0.4 pc 
and these properties are consistent with an H\textsc{i} dissociation region 
created by a B3 V type star in a medium of $n_{\rm H\textsc{i}} \sim 100$ cm$^{-3}$. 
The size of the H\textsc{i} dissociation region is now comparable to our spatial resolution 
and therefore $R_{\rm H2}$ around the stars earlier than B3 V would be significantly modified. 
Therefore, depending on the stellar type of stars residing inside GMCs, 
the internal radiation should not be ignored.

\section{Summary} 

In this paper, we study the H\textsc{i}--H$_{2}$ transition 
across the Perseus molecular cloud on sub-pc scales ($\sim$0.4 pc). 
We use the highest resolution H\textsc{i} data for Perseus from the GALFA--H\textsc{i} survey, 
which allow us to probe the outskirts as well as the main body of the cloud.
The velocity range of the H\textsc{i} emission associated with Perseus is determined  
based on the correlation between $N$(H\textsc{i}) and the dust column density traced by the 2MASS $A_{V}$ image. 
We then derive the $N$(H\textsc{i}) image under the assumption of optically thin H\textsc{i} gas.
The dust column density is estimated using the IRIS 60 $\mu$m and 100 $\mu$m images
and is calibrated with the 2MASS $A_{V}$ image. 
In combination with the $N$(H\textsc{i}) image and the derived DGR, 
we use the dust column density to derive the $N$(H$_{2}$) image. 

The $N$(H\textsc{i}) and $N$(H$_{2}$) images and the existing CO data for Perseus 
allow us to test two recent models of H$_{2}$ and CO formation in equilibrium.
For several dark and star-forming regions (B5, B1E, B1, IC348, and NGC1333), 
we investigate $\Sigma_{\rm H\textsc{i}}$ and $R_{\rm H2}$ as a function of 
$\Sigma_{\rm H\textsc{i}}+\Sigma_{\rm H2}$ and compare the results to KMT09's predictions. 
We also compare the H$_{2}$ and CO distributions 
and estimate the fraction of H$_{2}$ mass in the ``CO-dark'' component, $f_{\rm DG}$.  
We investigate $f_{\rm DG}$ as a function of $\overline{A}_{V}$  
and compare the result to WHM10's prediction.
Several of the main results can be summarized as follows:  

1. For both dark and star-forming regions in Perseus, 
we find an almost uniform $\Sigma_{\rm H\textsc{i}}$ $\sim$ 6--8 M$_{\odot}$ pc$^{-2}$.
This is consistent with KMT09's prediction of the H\textsc{i} shielding surface density 
for solar metallicity, $\Sigma_{\rm H\textsc{i},s} \sim 10$ M$_{\odot}$ pc$^{-2}$. 

2. The purely atomic envelopes are located more than $\sim$20 pc from the centers of the dark and star-forming regions.  
We probe the $\Sigma_{\rm H\textsc{i}}$ and $\Sigma_{\rm H\textsc{i}}+\Sigma_{\rm H2}$ distributions 
up to $\sim$20--30 pc from the centers of the regions
and barely detect the turnover between atomic- and molecular-dominated zones for IC348, B1E, and B1.  

3. The relation between $R_{\rm H2}$ and $\Sigma_{\rm H\textsc{i}}+\Sigma_{\rm H2}$ on a log-linear scale 
is remarkably consistent for all dark and star-forming regions, 
having a very sharp rise of $R_{\rm H2}$ at low $\Sigma_{\rm H\textsc{i}}+\Sigma_{\rm H2}$,  
a turnover around $R_{\rm H2}$ $\sim$ 1, and a slow increase toward higher $R_{\rm H2}$.  

4. The dark and star-forming regions have similar $\phi_{\rm CNM}$ = 6--10, which implies that  
these regions are embedded in the atomic envelopes with $n_{\rm CNM}$ = 20--30 cm$^{-3}$.
In addition, $\phi_{\rm CNM}$ = 6--10 indicates a similar normalized radiation field $\chi \sim 1$, 
suggesting that dust shielding and H$_{2}$ self-shielding are equally important for H$_{2}$ formation. 

5. We measure the H\textsc{i}--H$_{2}$ transition gas column density of 
$N{\rm(H\textsc{i})} + 2N(\rm{H_{2}})$ = (8--14) $\times$ 10$^{20}$ cm$^{-2}$ 
or $\Sigma_{\rm H\textsc{i}} + \Sigma_{\rm H2}$ = 6--12 M$_{\odot}$ pc$^{-2}$ at $R_{\rm H2}$ $\sim$ 0.25.
This is consistent with the previous estimates for the Galaxy. 

6. We estimate the fraction of ``CO-dark'' gas, $f_{\rm DG} \sim 0.3$, for Perseus, 
a factor of three lower than WHM10's prediction. 
This implies that a significant amount of H$_{2}$ gas is not traced by Galactic CO surveys. 

7. We investigate the dependence of $f_{\rm DG}$ on $\overline{A}_{V}$
and find a relatively large variation of $f_{\rm DG}$ $\sim$ 0.1--0.3 for similar $\overline{A}_{V} \sim 2$ mag.   
This is not consistent with WHM10's prediction that $f_{\rm DG}$ is a strong function of $\overline{A}_{V}$. 

While we generally find good agreement with KMT09's predictions, 
several outstanding questions remain. 
Our estimated $\Sigma_{\rm H\textsc{i}}$ $\sim$ 6--8 M$_{\odot}$ pc$^{-2}$ 
is likely dominated by WNM, while KMT09's model predicts shielding by CNM alone.
Clearly, the WNM in the atomic envelopes of GMCs needs to be considered 
in the models of H$_{2}$ formation as a source of shielding against photodissociation.
The importance of gravitationally bound vs diffuse gas is another important aspect,
as explored in Ostriker et al. (2010).
In addition, KMT09's model does not take into account the role of internal radiation for H$_{2}$ formation. 
This is valid for Perseus as it does not contain early-type stars to photodissociate 
a significant amount of H$_{2}$ that could be detected at our spatial resolution. 
However, if GMCs are young enough to harbor early-type stars 
that are able to change the thermal and chemical structures of the GMCs, 
the internal radiation cannot be ignored.     
Most importantly, our Perseus data on sub-pc scales are consistent with
KMT09's model for H$_{2}$ formation in equilibrium
and suggest that turbulence may not be of primary importance for H$_{2}$ formation.
Yet, the timescale for equilibrium H$_{2}$ formation is longer than the estimated lifetime of most GMCs.
This problem remains to be resolved. 
Future detailed comparisons of observations with non-equilibrium, turbulence-driven H$_{2}$ formation
models will be important to investigate the role of turbulence for formation of H$_{2}$ and various molecular species.

We thank the anonymous referee for suggestions that improved this work. 
We also thank Ron Allen, Joss Bland-Hawthorn, Alberto Bolatto, Bruce Elmegreen, Fabian Heitsch, 
Mordecai Mac Low, Chris McKee, Eve Ostriker, Rene Plume, and Mark Wolfire for stimulating discussions. 
M.-Y.L., S.S., M.E.P., C.H., E.J.K., J.E.G.P., A.B., J.G., N.P., and D.S. acknowledge 
support from NSF grants AST-0707597, 0917810, 0707679, and 0709347.
M.-Y.L. and S.S. thank the Research Corporation for Science Advancement for their support. 
K.A.D. acknowledges funding from the European Community's Seventh Framework Program 
under grant agreement no PIIF-GA-2008-221289.
J.E.G.P. was supported by HST-HF-51295.01A, provided by NASA through a Hubble Fellowship grant from STScI, 
which is operated by AURA under NASA contract NAS5-26555.
The Arecibo Observatory is part of the National Astronomy and Ionosphere
Center, which is operated by Cornell University under a
cooperative agreement with the National Science Foundation.
The National Radio Astronomy Observatory is a facility of the National Science Foundation 
operated under cooperative agreement by Associated Universities, Inc. 
We credit the use of the KARMA visualization software (Gooch 1996).

\clearpage

\begin{table}
\begin{center}
{\bf TABLE 1} \\
\textsc{Fitting Results for $R_{\rm H2}$ vs $\Sigma_{\rm H\textsc{i}}+\Sigma_{\rm H2}$} \\
\vskip 0.2cm
\begin{tabular}{c c}\hline \hline
Region & $\phi_{\rm CNM}^{\rm a}$ \\ \hline
B5 & $9.61 \pm 1.32$ \\
IC348 & $8.37 \pm 0.88$ \\
B1E & $8.32 \pm 1.22$ \\
B1 & $7.95 \pm 1.06$ \\
NGC1333 & $6.26 \pm 0.82$ \\
\hline
\end{tabular}
\end{center}
{\small $\rm {}^{a}$ Uncertainties in $\phi_{\rm CNM}$ are estimated 
from the distribution of simulated 1000 $\phi_{\rm CNM}$.}
\end{table}

\begin{table}
\begin{center}
{\bf TABLE 2} \\
\textsc{Properties of $R_{\rm H2}$ Radial Profiles} \\
\vskip 0.2cm
\begin{tabular}{c c c c c c c c c c}\hline \hline
Region & $\alpha^{\rm a}$ & $\delta^{\rm a}$ & $\alpha 1$ & $\alpha 2$ & $r_{\rm b}^{\rm b}$ & $R_{\rm b,H2}^{\rm c}$ & 
$r_{\rm t}^{\rm d}$ & $r_{\rm H\textsc{i}}^{\rm e}$ & $d_{\rm t}^{\rm f}$ \\
 & (J2000) & (J2000) &  &  & ($'$) & & ($'$) & ($'$) & ($'$) \\ \hline
B5 & 03 47 39 & $+$32 51 28 & $3.69 \pm 0.15$ & $0.44 \pm 0.01$ & $\sim$90 & $\sim$0.7 & $\sim$120 & $\sim$160 & $\sim$30 (< 0.19) \\
IC348 & 03 44 13 & $+$32 04 13 & $3.22 \pm 0.06$ & $0.57 \pm 0.01$ & $\sim$70 & $\sim$1.2 & $\sim$120 & $\sim$150 & $\sim$50 (< 0.33) \\
B1E & 03 36 45 & $+$31 12 40 & $1.67 \pm 0.02$ & $0.42 \pm 0.01$ & $\sim$30 & $\sim$1.2 & $\sim$90 & $\sim$150 & $\sim$60 (< 0.40) \\
B1 & 03 33 18 & $+$31 08 23 & $2.21 \pm 0.05$ & $0.70 \pm 0.01$ & $\sim$60 & $\sim$0.8 & $\sim$110 & $\sim$170 & $\sim$50 (< 0.29) \\
NGC1333 & 03 29 17 & $+$31 21 16 & $1.43 \pm 0.04$ & $0.61 \pm 0.01$ & $\sim$60 & $\sim$0.5  & $\sim$100 & $\sim$190 & $\sim$40 (< 0.21) \\ 
\hline
\end{tabular}
\end{center}
{\small $\rm {}^{a}$ Central position of each region.
For B5 and B1, the centers correspond to the $I_{^{13}\rm CO}$ peaks as their $R_{\rm H2}$ peaks are masked. 
For IC348, B1E, and NGC1333, the centers correspond to the $R_{\rm H2}$ peaks.} \\
{\small $\rm {}^{b}$ Radius where the transition from the steep to shallow power-law functions occurs.} \\
{\small $\rm {}^{c}$ $R_{\rm H2}$ at $r_{\rm b}$.} \\
{\small $\rm {}^{d}$ Radius where the H\textsc{i}--H$_{2}$ transition occurs. This radius is defined at $R_{\rm H2}$ $\sim$ 0.25.} \\ 
{\small $\rm {}^{e}$ Radius where gas is mainly in H\textsc{i}. This radius is defined at $R_{\rm H2}$ $\sim$ 0.1.} \\ 
{\small $\rm {}^{f}$ Thickness of the H\textsc{i}--H$_{2}$ transition region. This is measured as $r_{\rm t} - r_{\rm b}$.
The values in parantheses are $d_{\rm t}/r_{\rm H\textsc{i}}$.} \\
{\small $\rm {}^{b, c, d, e}$ Estimated by eye.} \\
{\small $\rm {}^{b, d, e, f}$ 1$'$ corresponds to 0.09 pc at the distance of 300 pc.}
\end{table}

\begin{table}
\begin{center}
{\bf TABLE 3} \\
\textsc{$f_{\rm DG}$ and $\overline{A}_{V}$ for Perseus} \\
\vskip 0.2cm
\begin{tabular}{c c c}\hline \hline
Region & $f_{\rm DG}^{\rm a}$ & $\overline{A}_{V}$ (mag)$^{\rm a}$ \\ \hline
1 & 0.32 & 1.71 \\
2 & 0.31 & 1.73 \\
3 & 0.18 & 1.58 \\
4 & 0.10 & 1.43 \\
\hline
\end{tabular}
\end{center}
{\small $\rm {}^{a}$ 1$\sigma$ uncertainty is less than 1\% of each quantity.} 
\end{table}
\clearpage

\clearpage

\clearpage

\begin{figure}
\begin{center}
\includegraphics[scale=0.35]{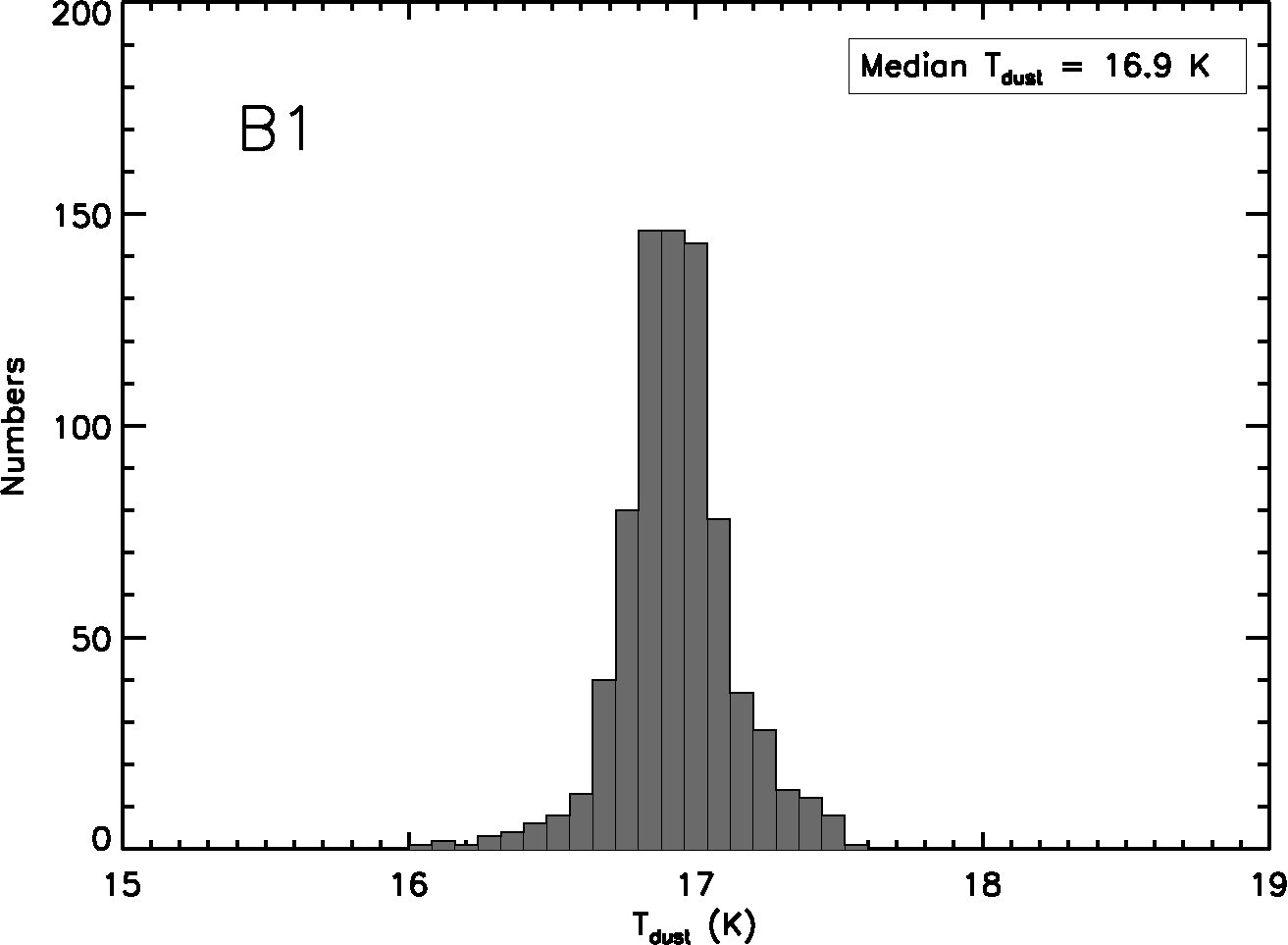}
\includegraphics[scale=0.35]{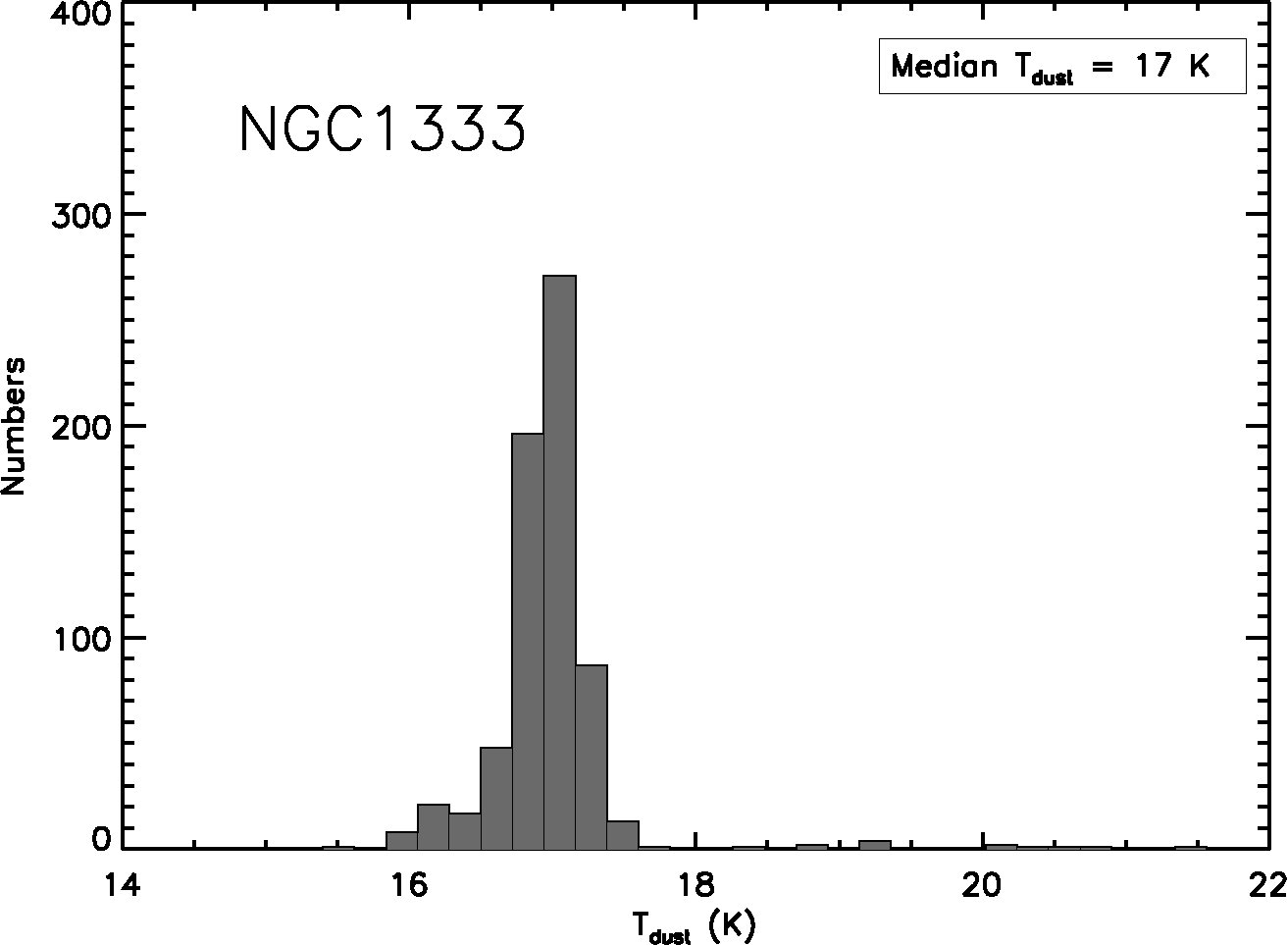}
\vskip 0.3cm
\begin{minipage}{16cm}
Fig. \ref{f:dust_hist}.--- (Continued) (Left) B1. (Right) NGC1333.
\end{minipage}
\end{center}
\end{figure}

\begin{figure}
\begin{center}
\includegraphics[scale=0.35]{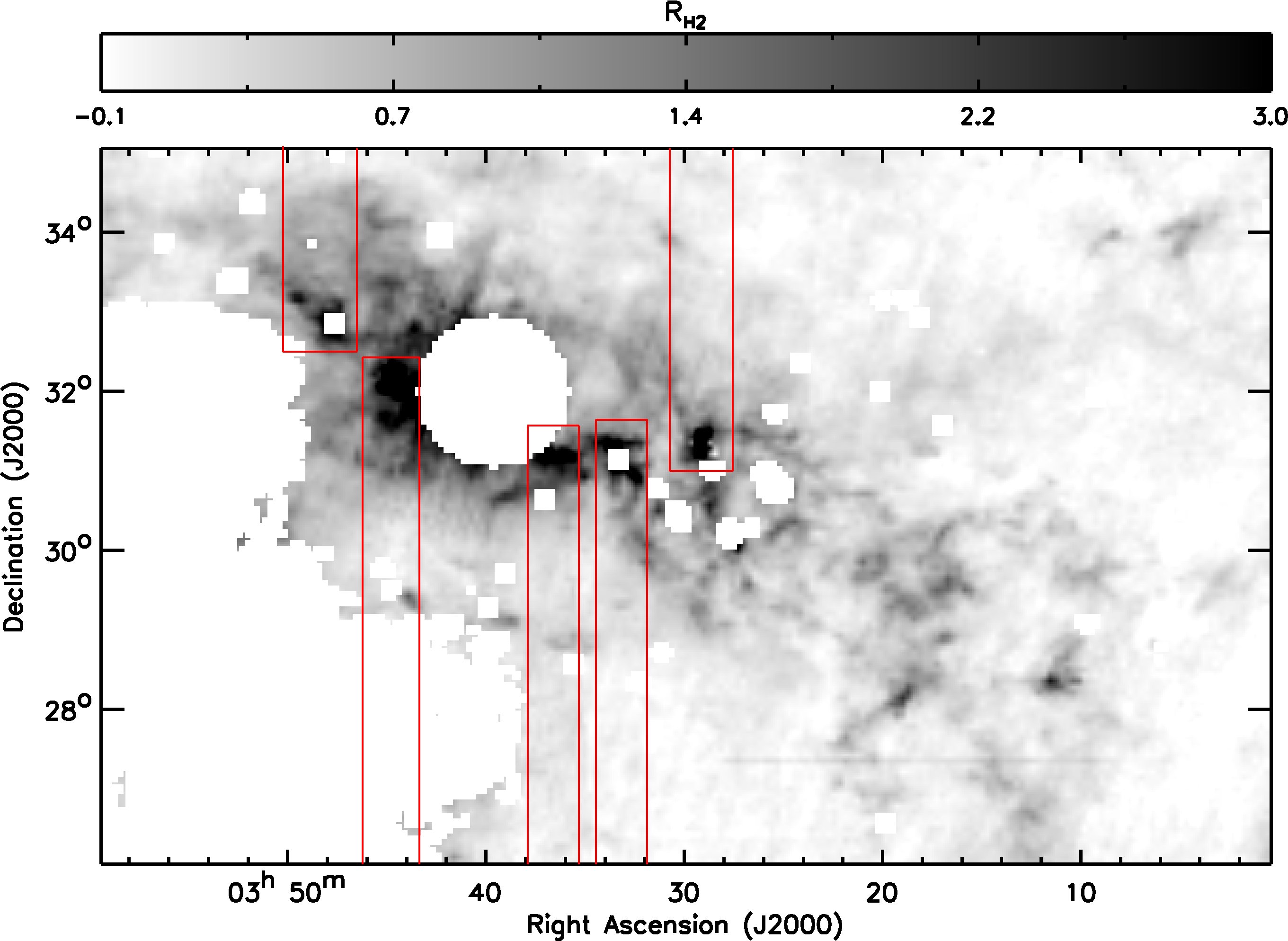}
\caption{\label{f:RH2} $R_{\rm H2}$ image.
The red rectangular boxes show the boundaries of each dark and star-forming region set by the COMPLETE $^{13}$CO(1--0) data.} 
\end{center}
\end{figure}
\clearpage

\begin{landscape}
\begin{figure}
\includegraphics[scale=0.46]{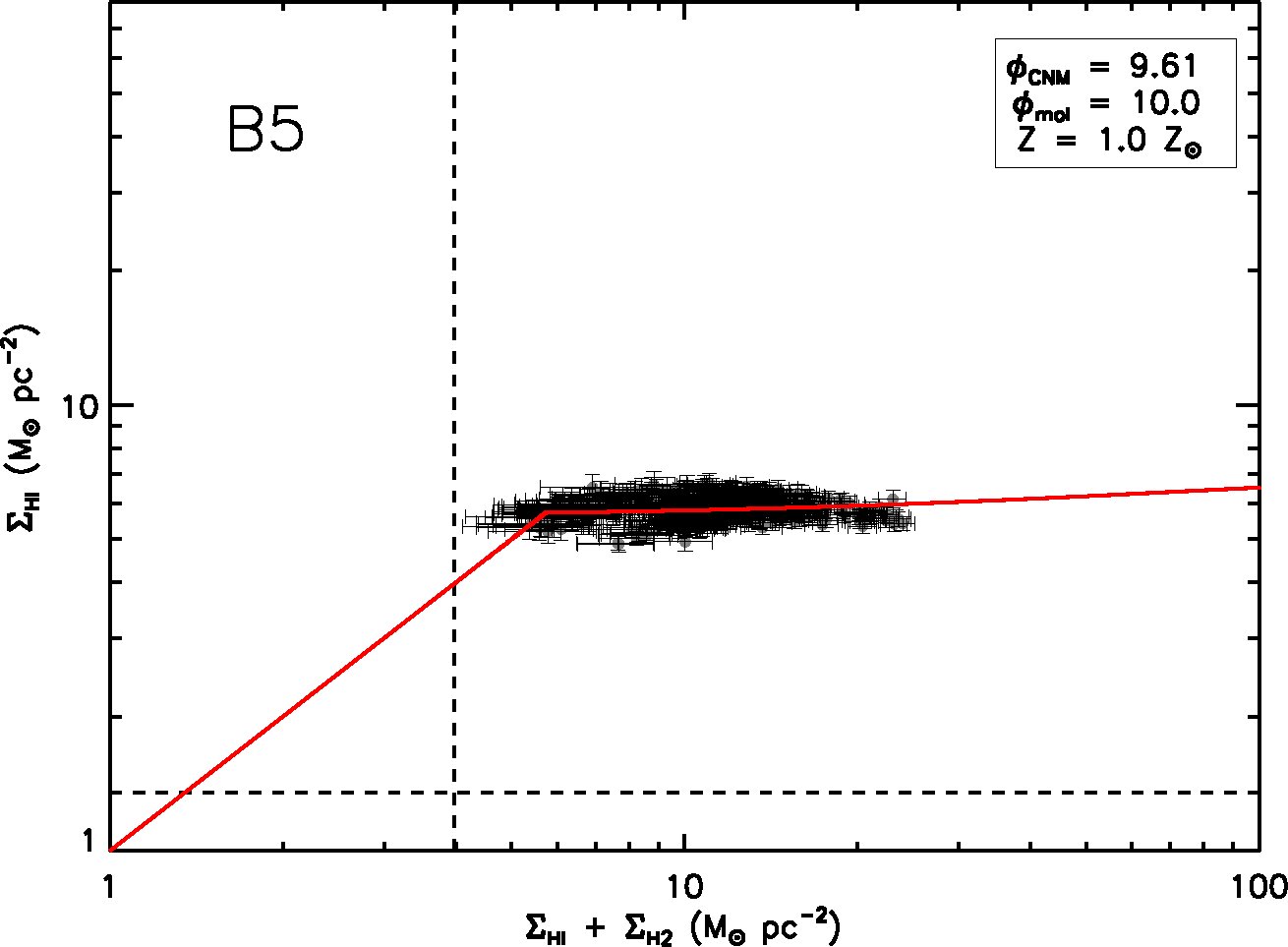}
\includegraphics[scale=0.46]{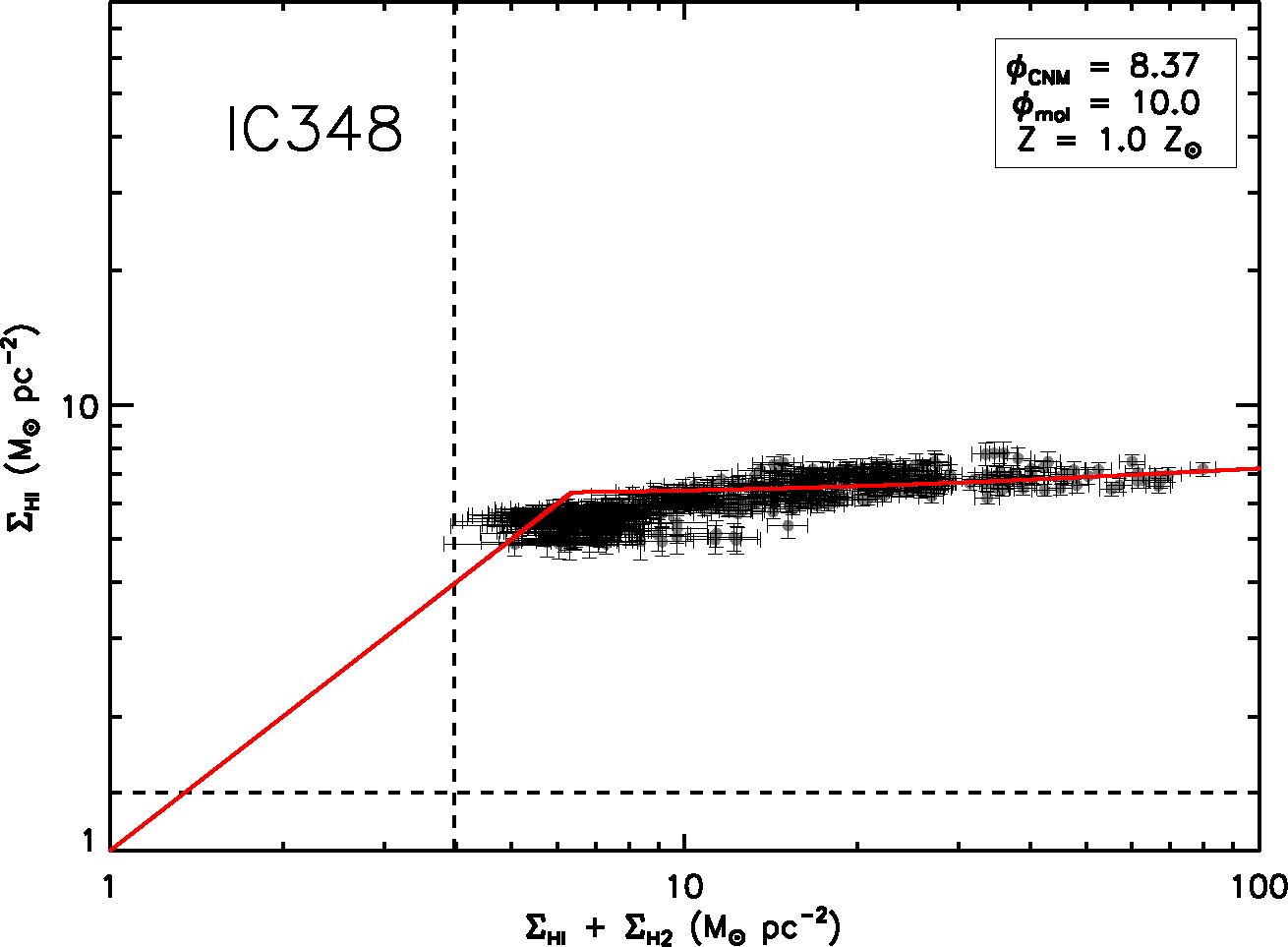}
\includegraphics[scale=0.46]{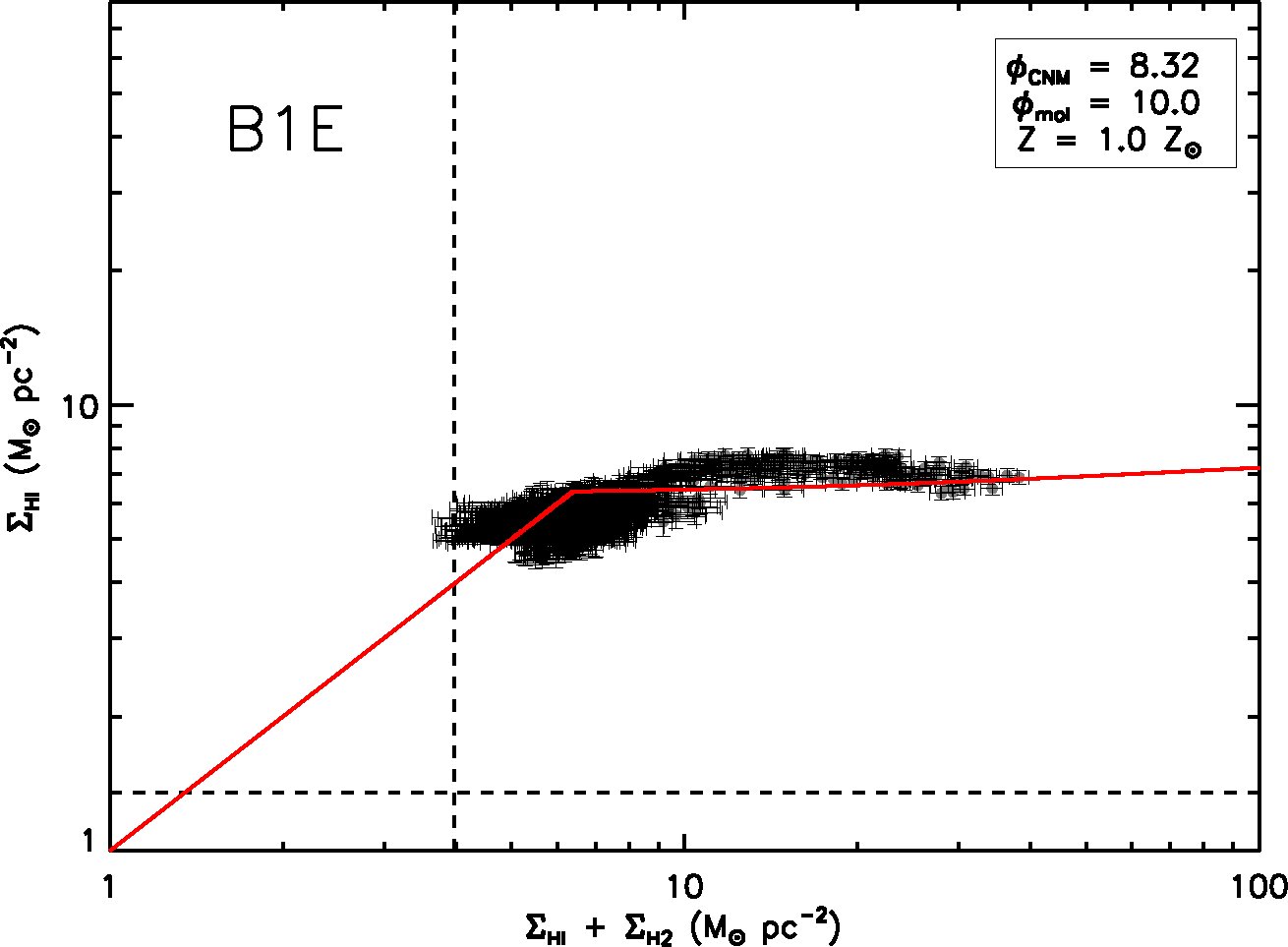}
\caption{\label{f:sigmaHI_sigmaTotal} $\Sigma_{\rm H\textsc{i}}$ as a function of $\Sigma_{\rm H\textsc{i}}+\Sigma_{\rm H2}$.
All data points in the rectangular boxes (see Figure \ref{f:RH2}) are used for plotting. 
The median 3$\sigma$ values of $\Sigma_{\rm H\textsc{i}}$ and $\Sigma_{\rm H\textsc{i}}+\Sigma_{\rm H2}$ 
for the whole Perseus cloud are shown as the black dashed lines. 
The red curves are the best-fit model curves determined in Section 7.2.1.
The best-fit parameters are summarized in the top right corner of each plot.  
(Left top) B5. (Right top) IC348. (Left bottom) B1E.}
\end{figure}
\end{landscape}

\begin{figure}
\begin{center}
\includegraphics[scale=0.35]{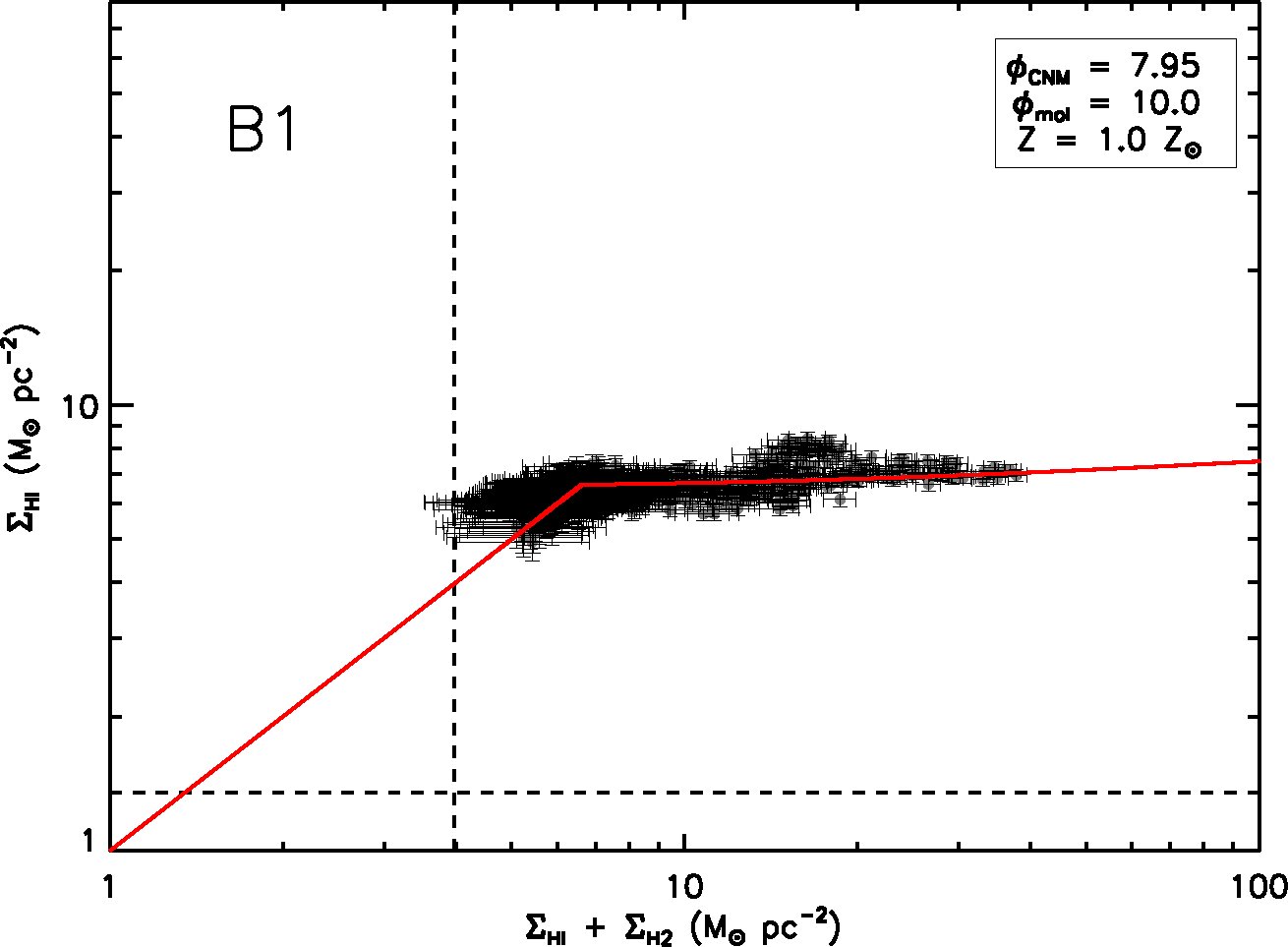}
\includegraphics[scale=0.35]{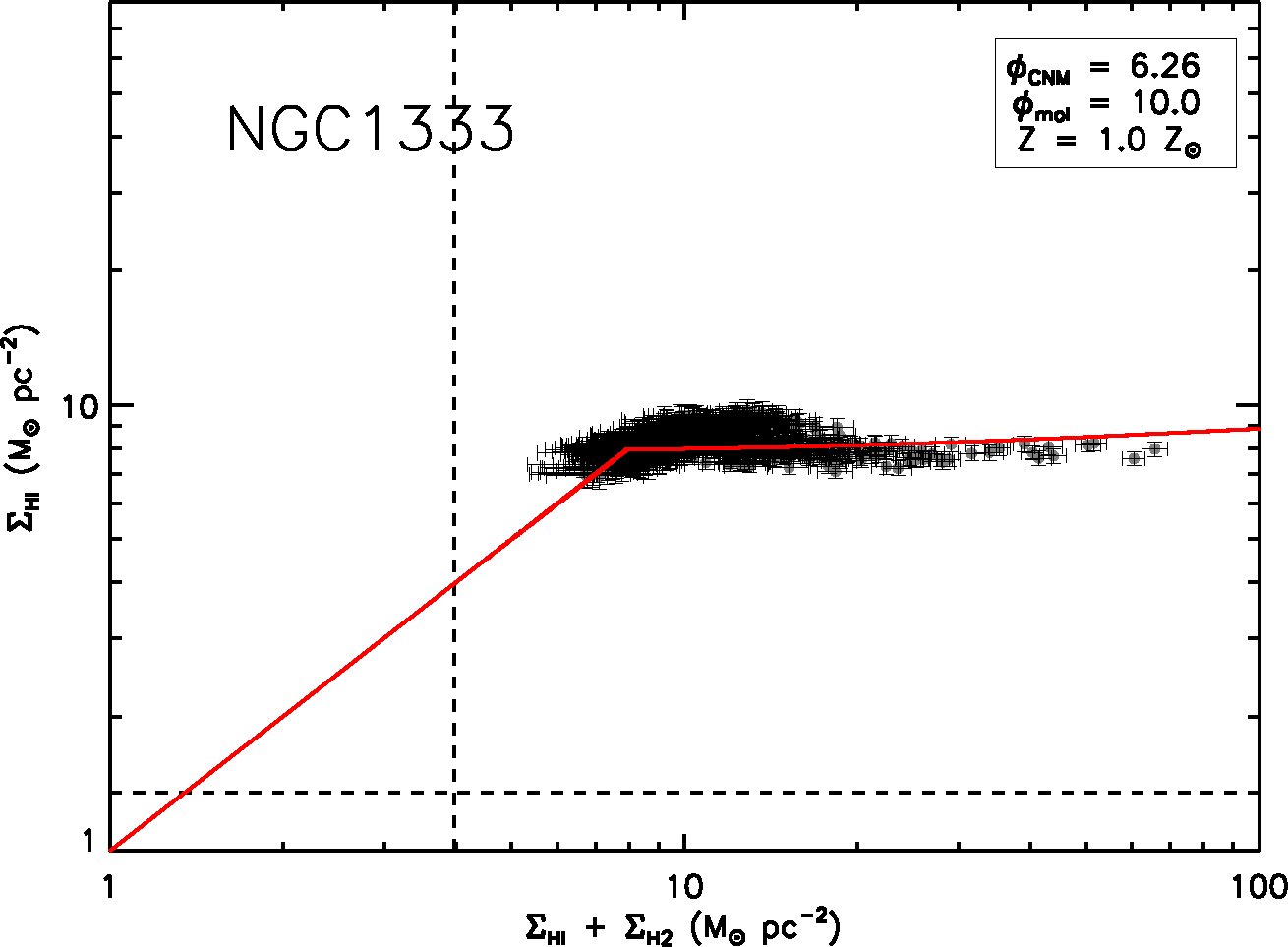}
\vskip 0.3cm
\begin{minipage}{16cm}
Fig. \ref{f:sigmaHI_sigmaTotal}.--- (Continued) (Left) B1. (Right) NGC1333.
\end{minipage}
\end{center}
\end{figure}

\begin{landscape}
\begin{figure}
\includegraphics[scale=0.46]{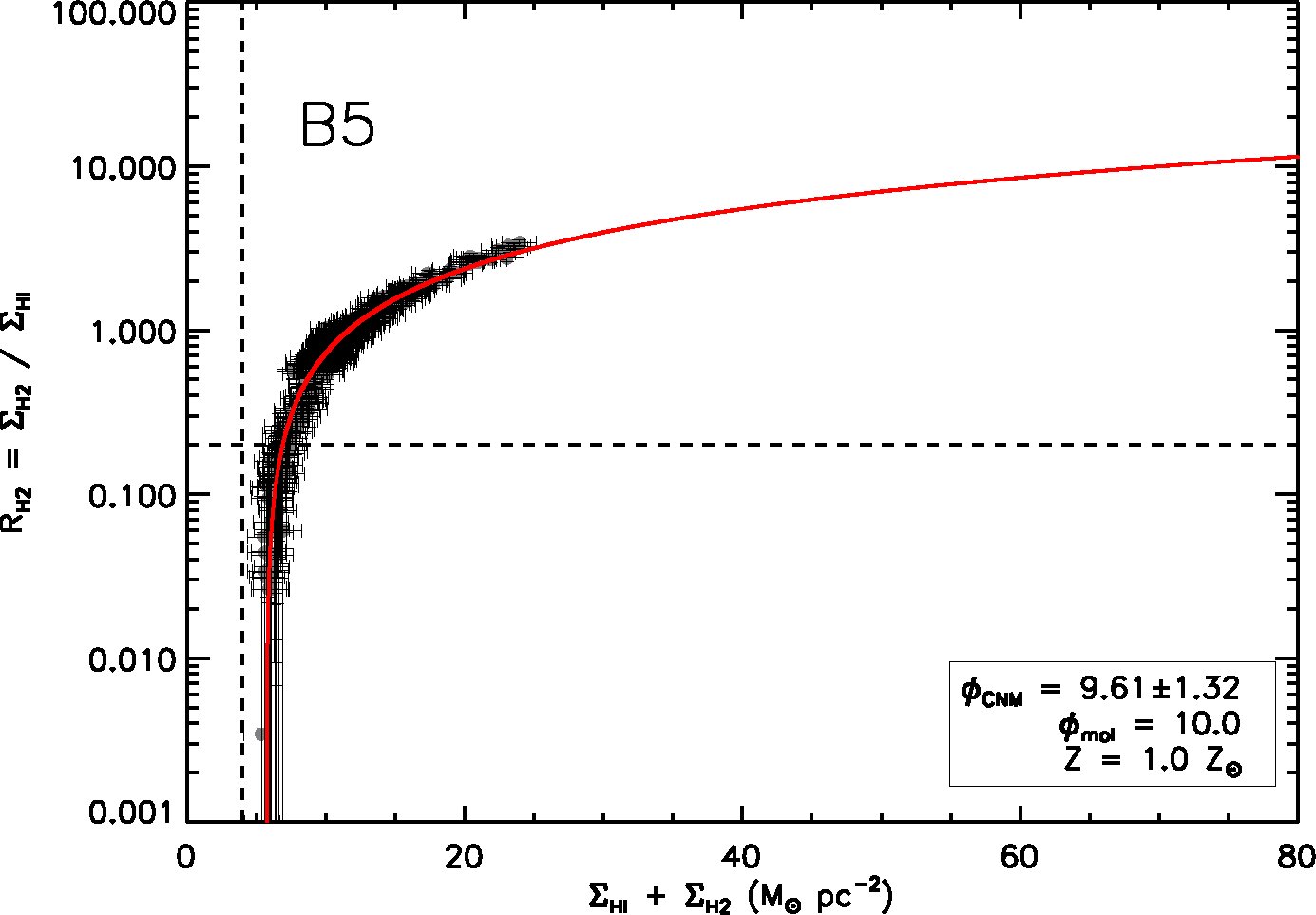}
\includegraphics[scale=0.46]{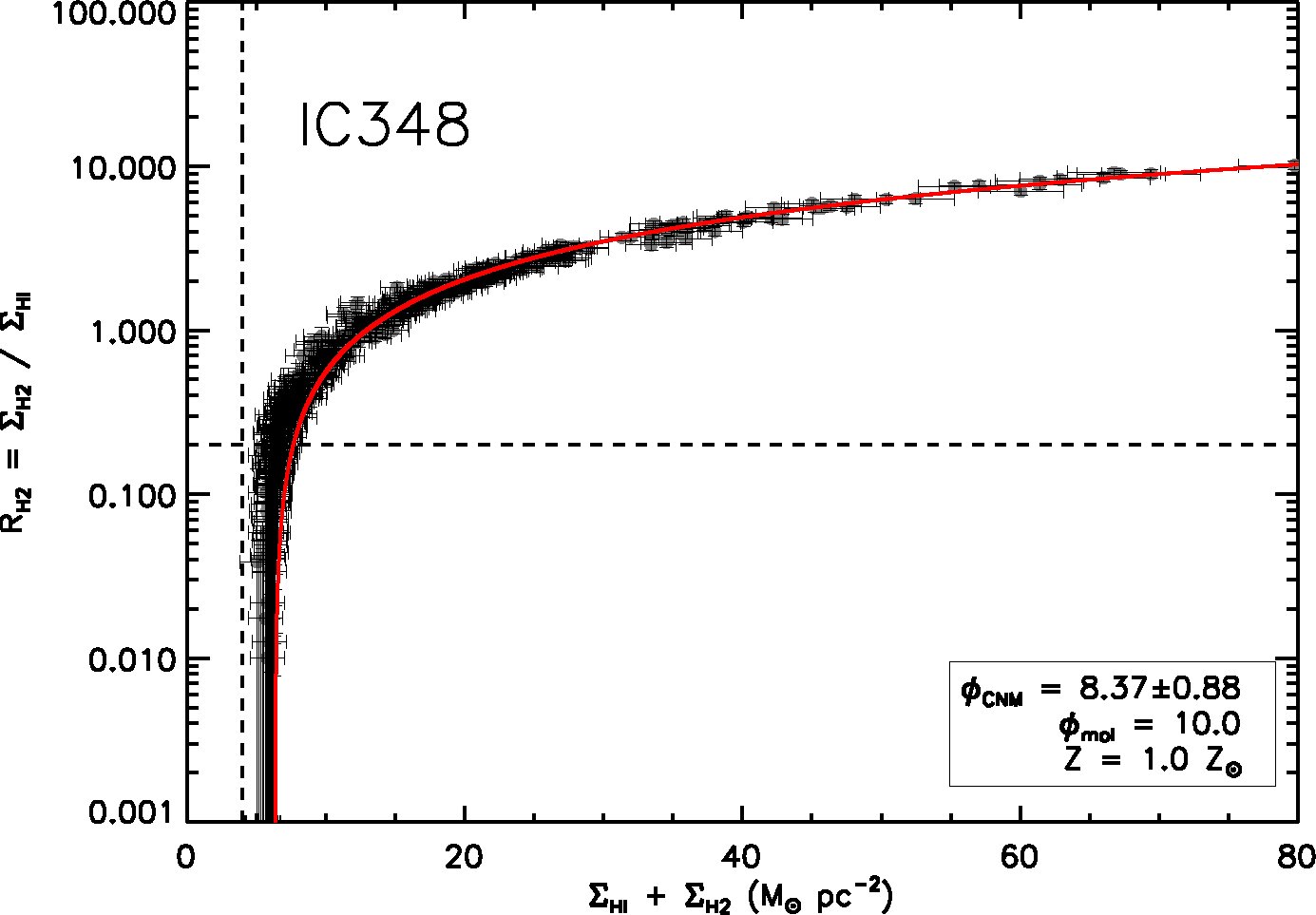}
\includegraphics[scale=0.46]{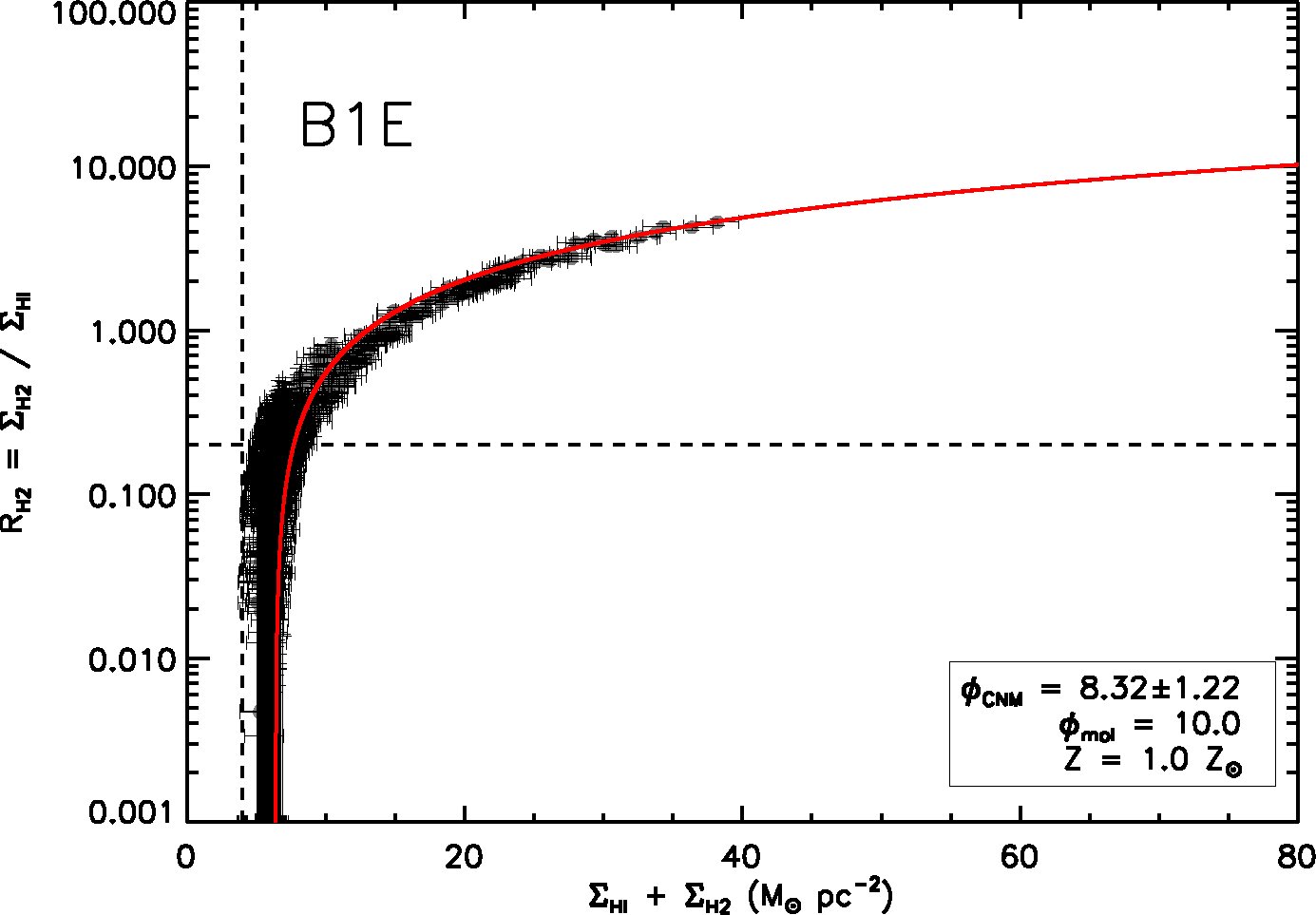}
\caption{\label{f:RH2_sigmaTotal} $R_{\rm H2}$ as a function of $\Sigma_{\rm H\textsc{i}}+\Sigma_{\rm H2}$. 
All data points in the rectangular boxes (see Figure \ref{f:RH2}) are used for plotting.
The median 3$\sigma$ values of $R_{\rm H2}$ and $\Sigma_{\rm H\textsc{i}}+\Sigma_{\rm H2}$ 
for the whole Perseus cloud are shown as the black dashed lines. 
The best-fit model curves determined in Section 7.2.1. are shown in red.
The best-fit parameters are summarized in the lower right corner of each plot.
(Left top) B5. (Right top) IC348. (Left bottom) B1E.}
\end{figure}
\end{landscape}

\begin{figure}
\begin{center}
\includegraphics[scale=0.34]{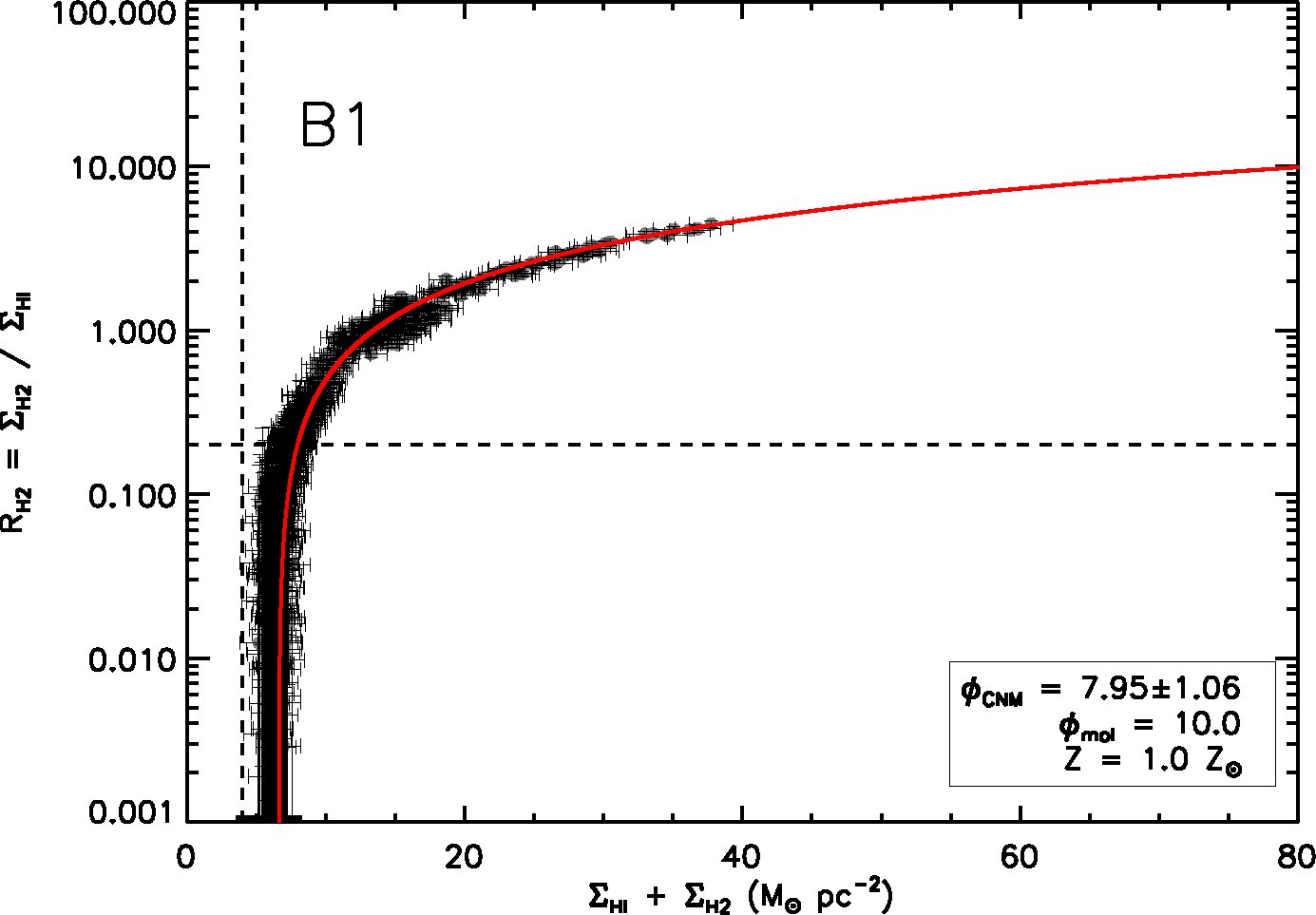}
\includegraphics[scale=0.34]{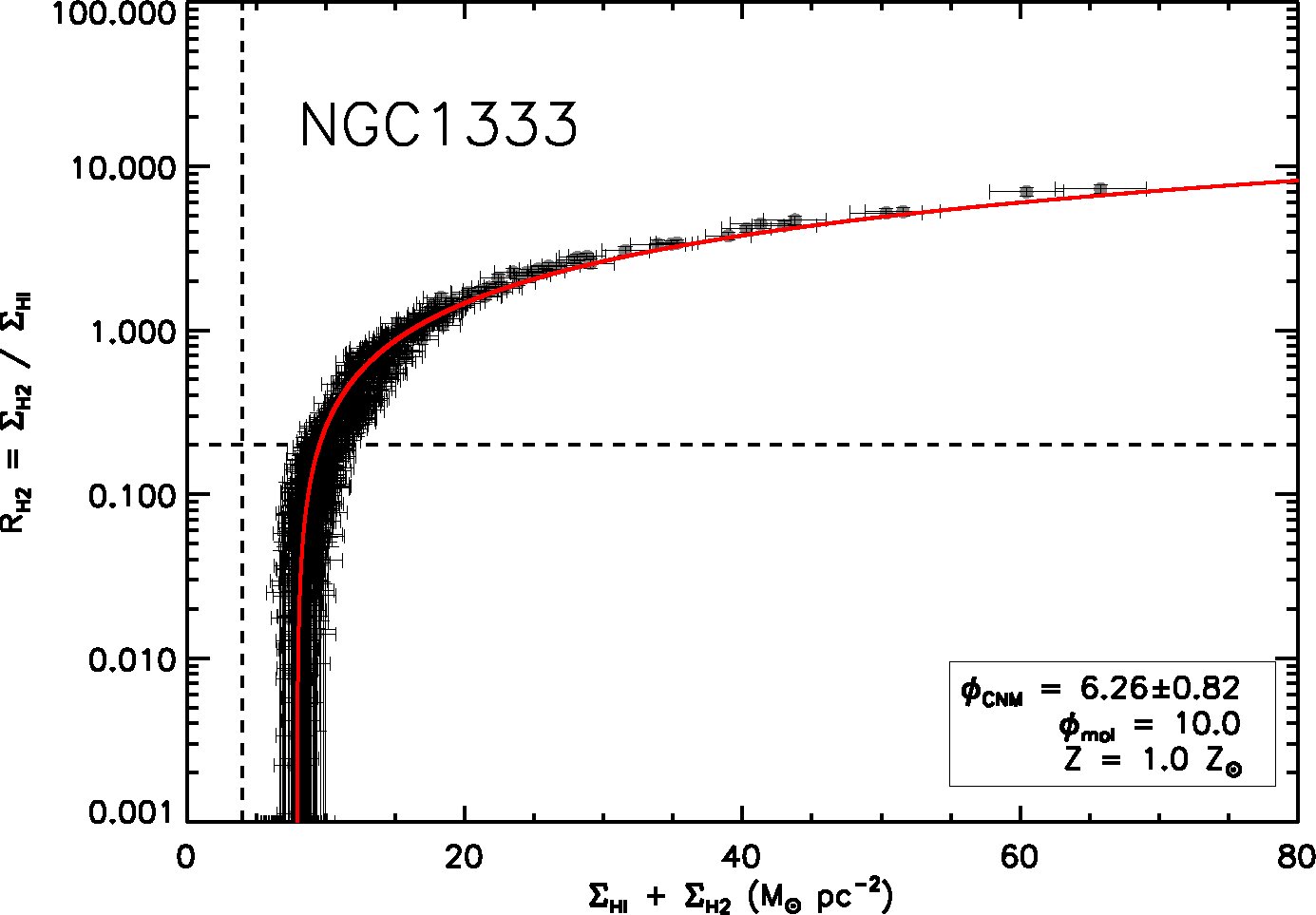}
\vskip 0.3cm
\begin{minipage}{16cm}
Fig. \ref{f:RH2_sigmaTotal}.--- (Continued) (Left) B1. (Right) NGC1333.
\end{minipage}
\end{center}
\end{figure}

\begin{landscape}
\begin{figure}
\includegraphics[scale=0.46]{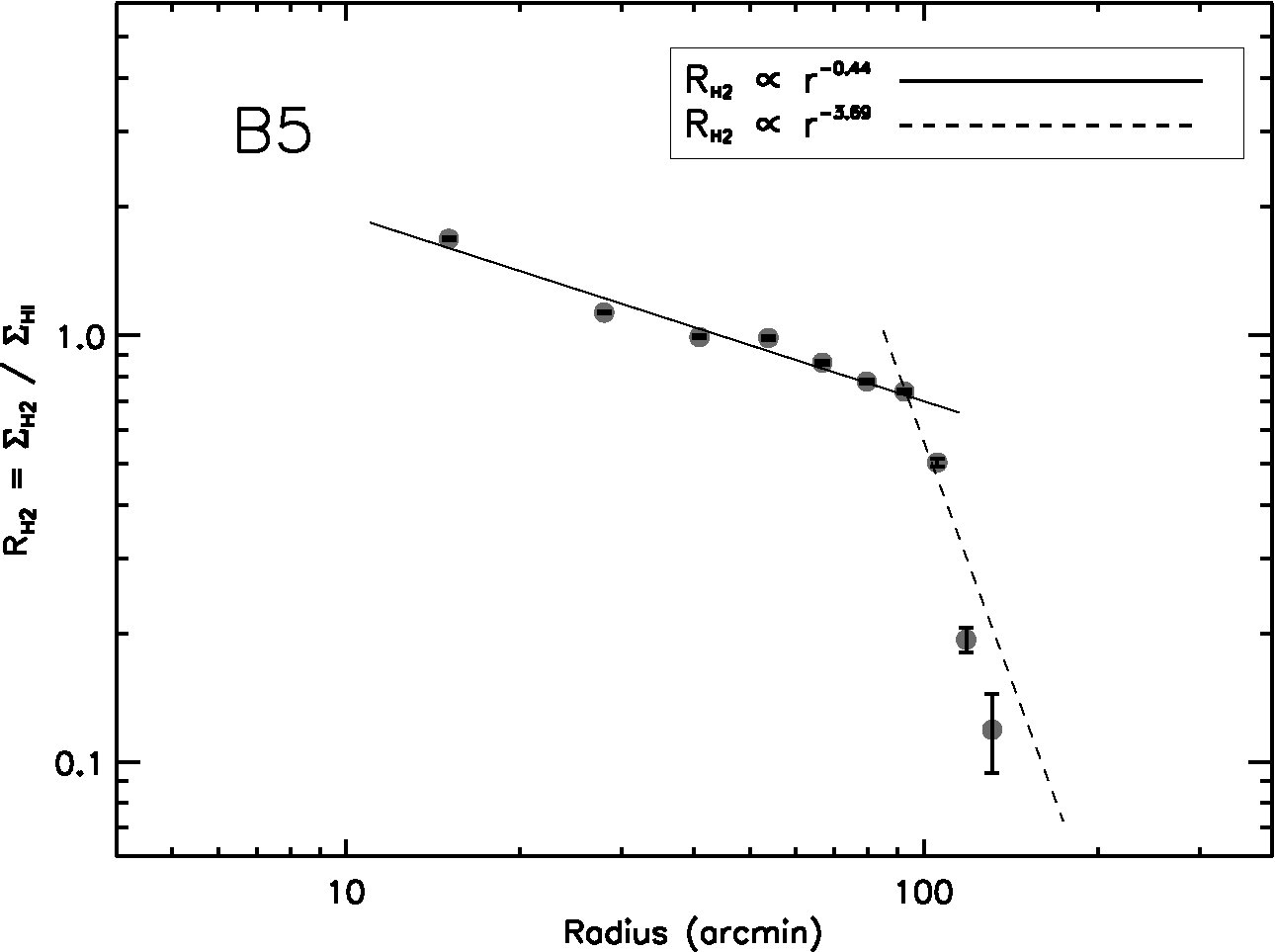}
\includegraphics[scale=0.46]{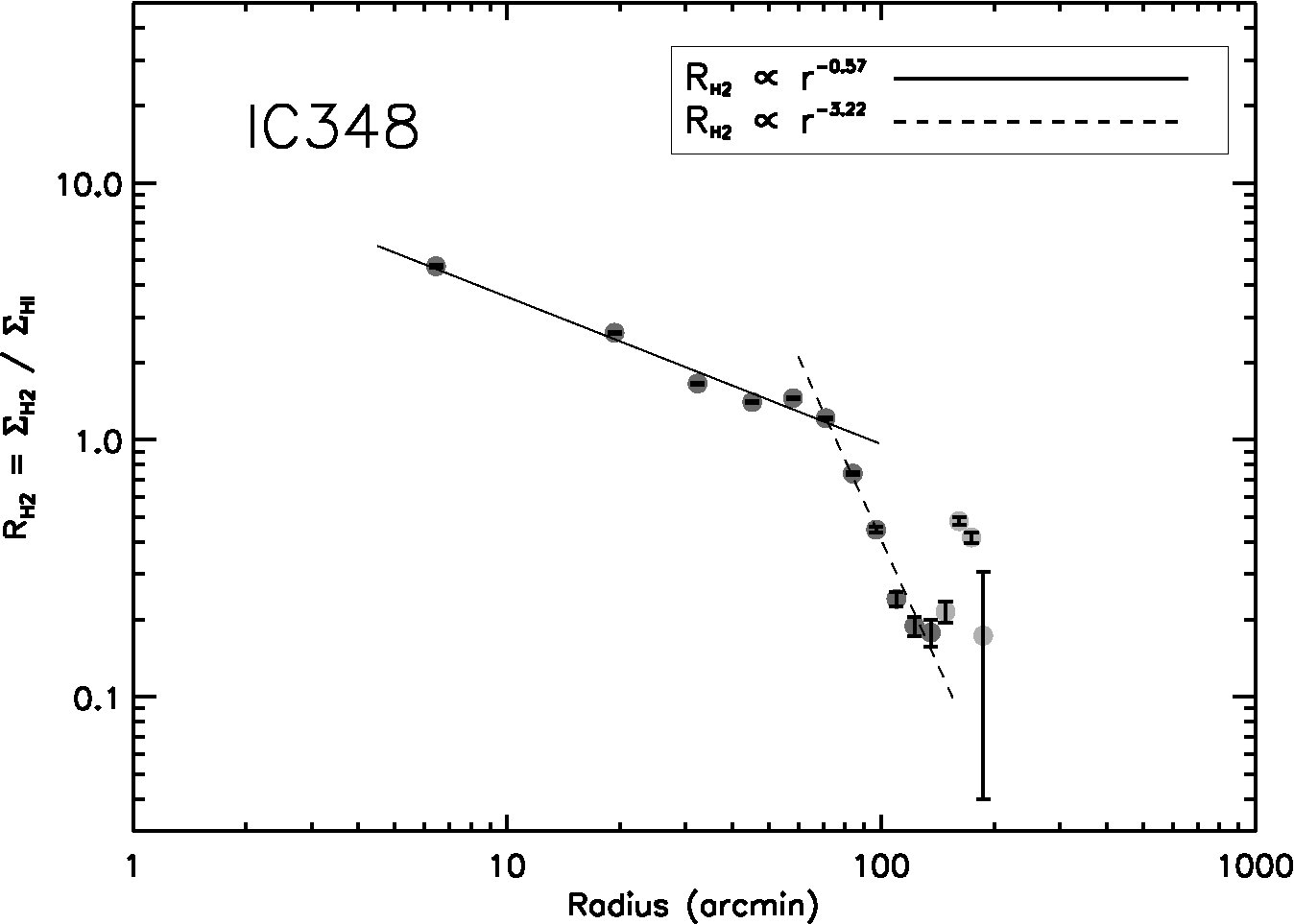}
\includegraphics[scale=0.46]{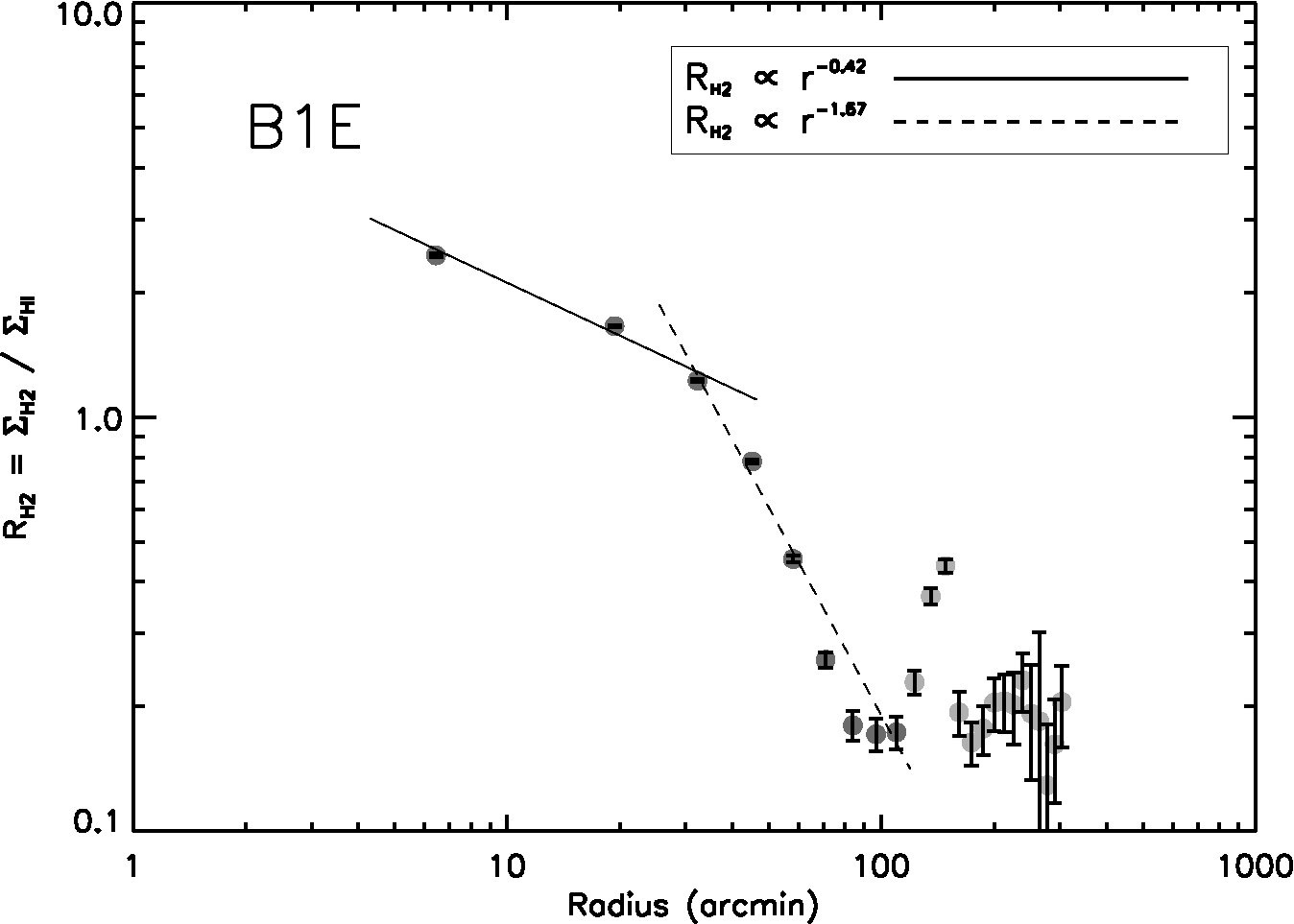}
\caption{\label{f:RH2_radial} $R_{\rm H2}$ radial profiles. 
The data points with SNR of $R_{\rm H2}$ > 1 are selected from the rectangular boxes (see Figure \ref{f:RH2}) 
and the mean $R_{\rm H2}$ values in three-pixel size bins are calculated.
The results from the power-law fitting are summarized in the top right corner of each plot.
The light gray data points are excluded from the power-law fitting. 
1$'$ corresponds to 0.09 pc at the distance of 300 pc.
(Left top) B5. (Right top) IC348. (Left bottom) B1E.}
\end{figure}
\end{landscape}

\begin{figure}
\begin{center}
\includegraphics[scale=0.35]{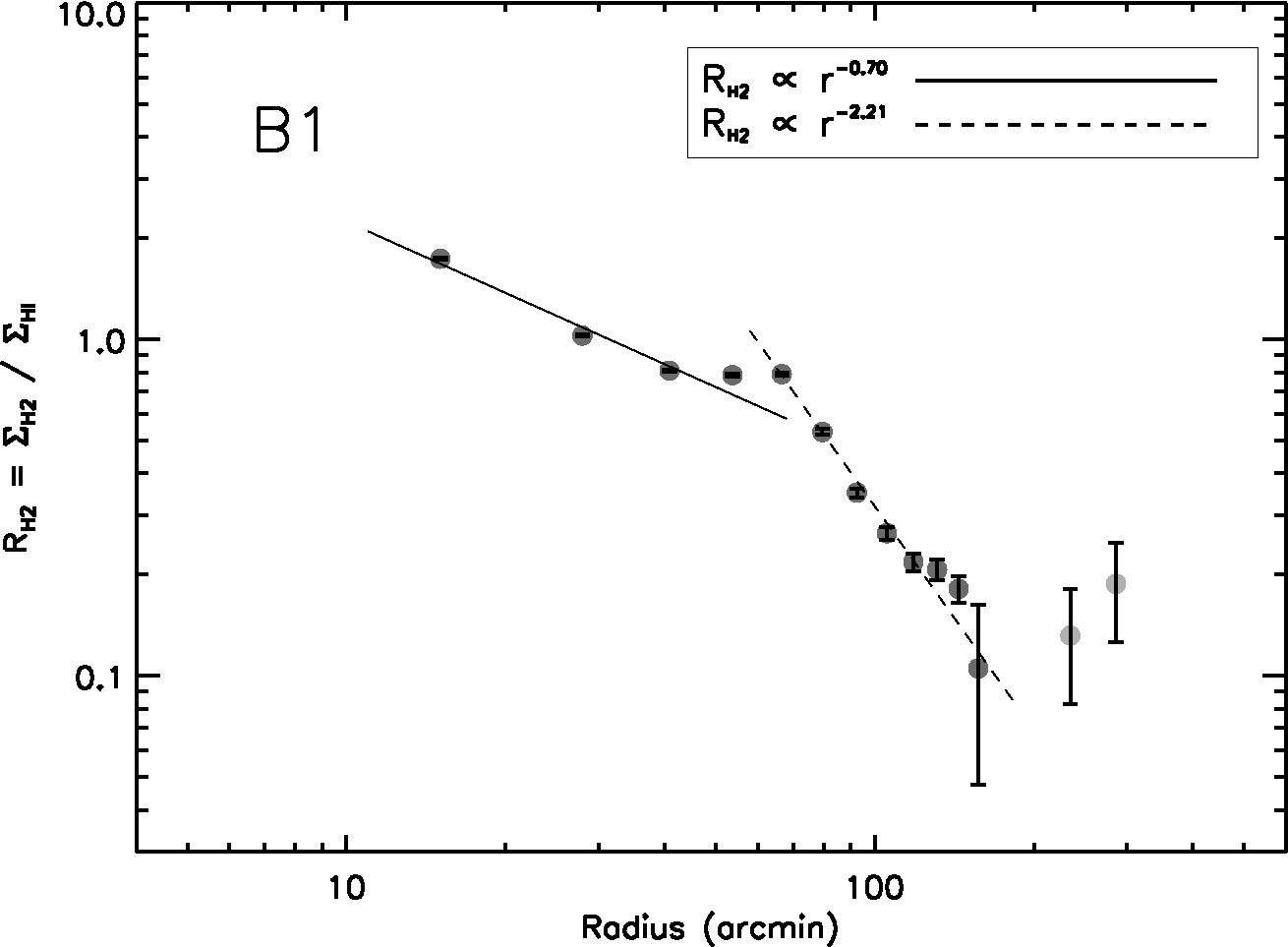}
\includegraphics[scale=0.35]{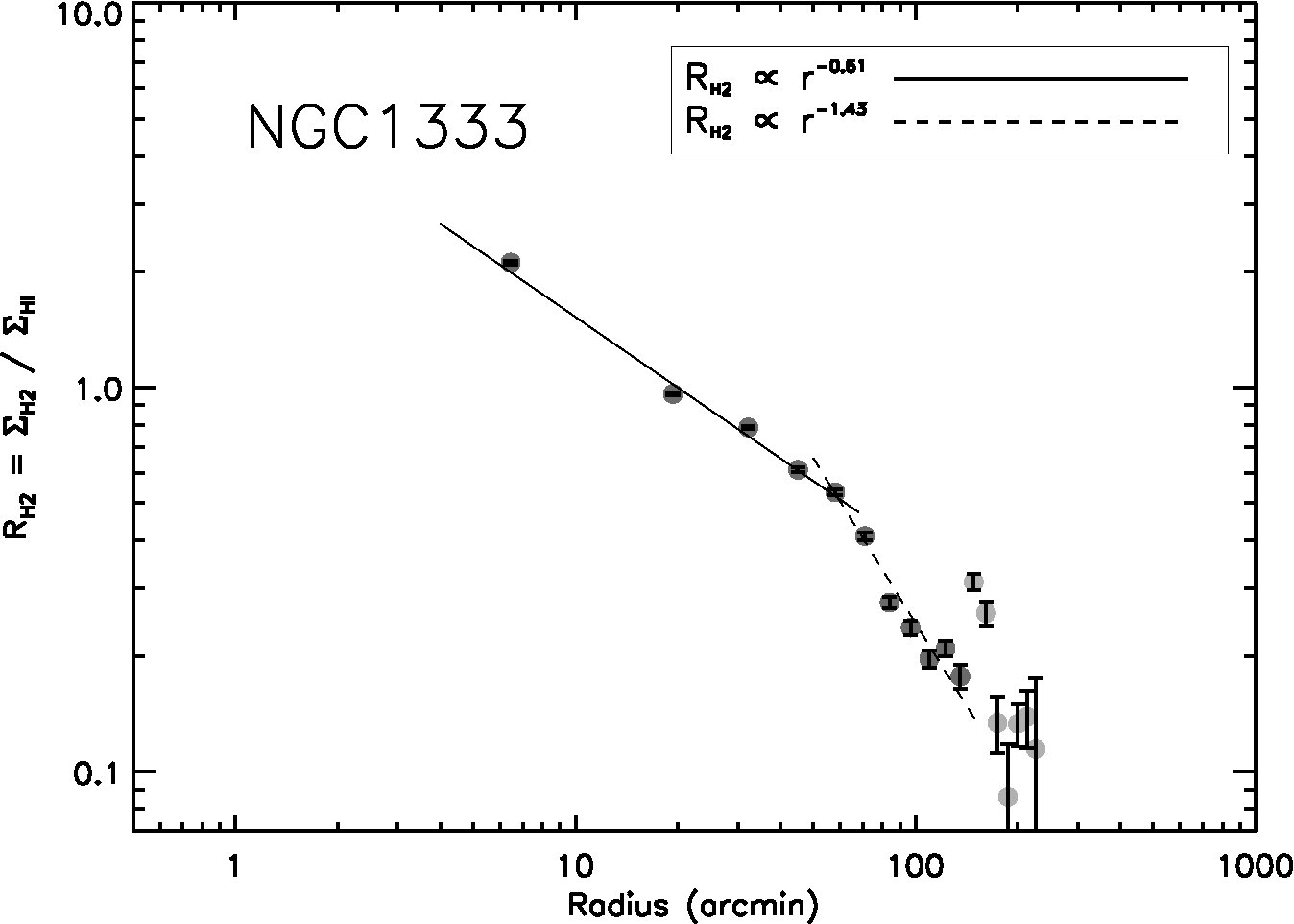}
\vskip 0.3cm
\begin{minipage}{16cm}
Fig. \ref{f:RH2_radial}.--- (Continued) (Left) B1. (Right) NGC1333.
\end{minipage}
\end{center}
\end{figure}

\begin{figure}
\begin{center}
\includegraphics[scale=0.35]{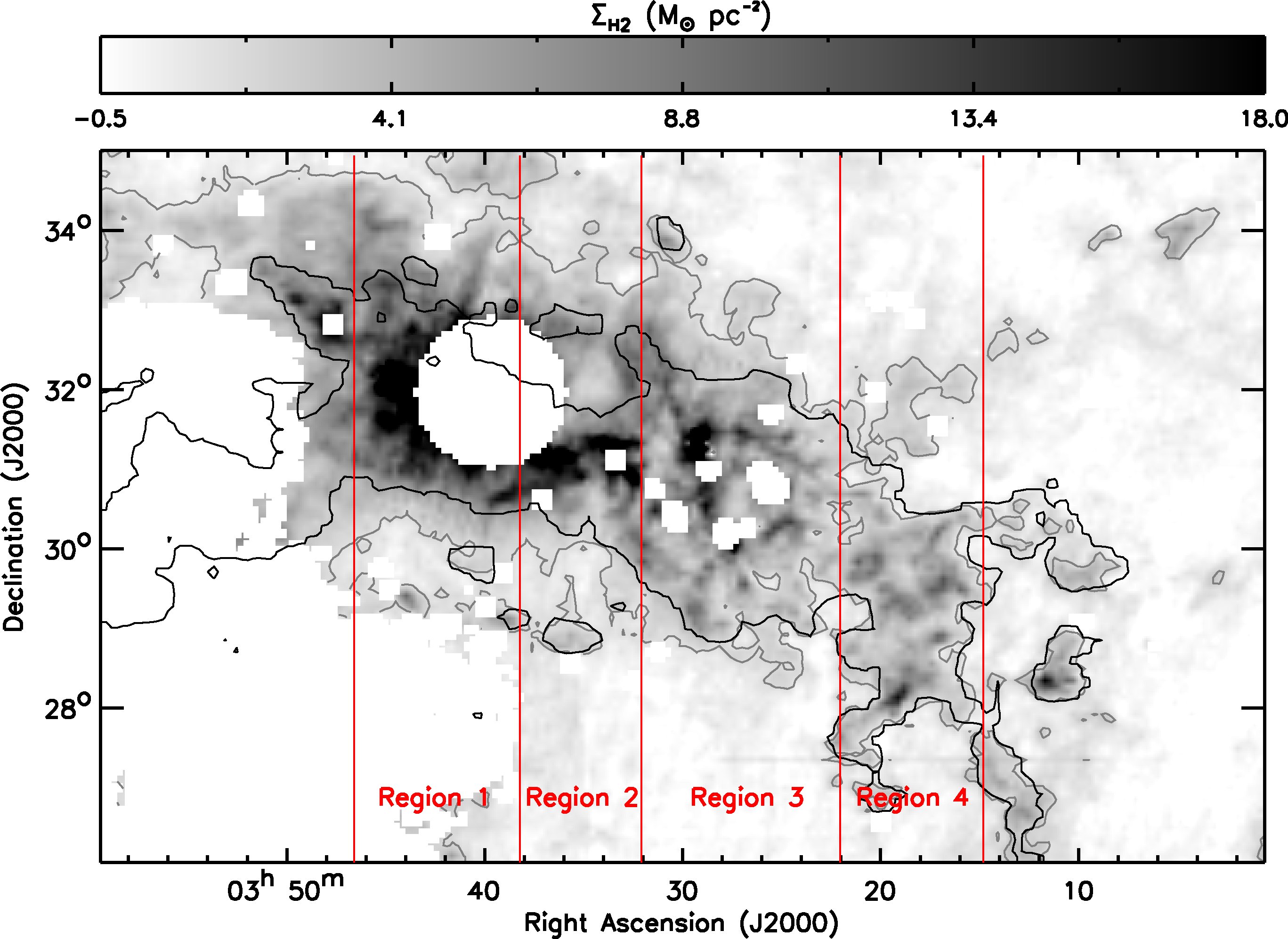}
\caption{\label{f:H2_CO} $\Sigma_{\rm H2}$ image.
The contour for SNR of $\Sigma_{\rm H2}$ = 3 is overlaid in gray.
The contour for SNR of CfA $I_{\rm CO}$ = 3 is also overlaid in black. 
The angular resolution of the $\Sigma_{\rm H2}$ and CfA $I_{\rm CO}$ images is 4.3$'$ and 8.4$'$, respectively.
The $\Sigma_{\rm H2}$ image is divided into four regions and each region has either a dark or a star-forming region.}
\end{center}
\end{figure}

\begin{landscape}
\begin{figure}
\includegraphics[scale=0.46]{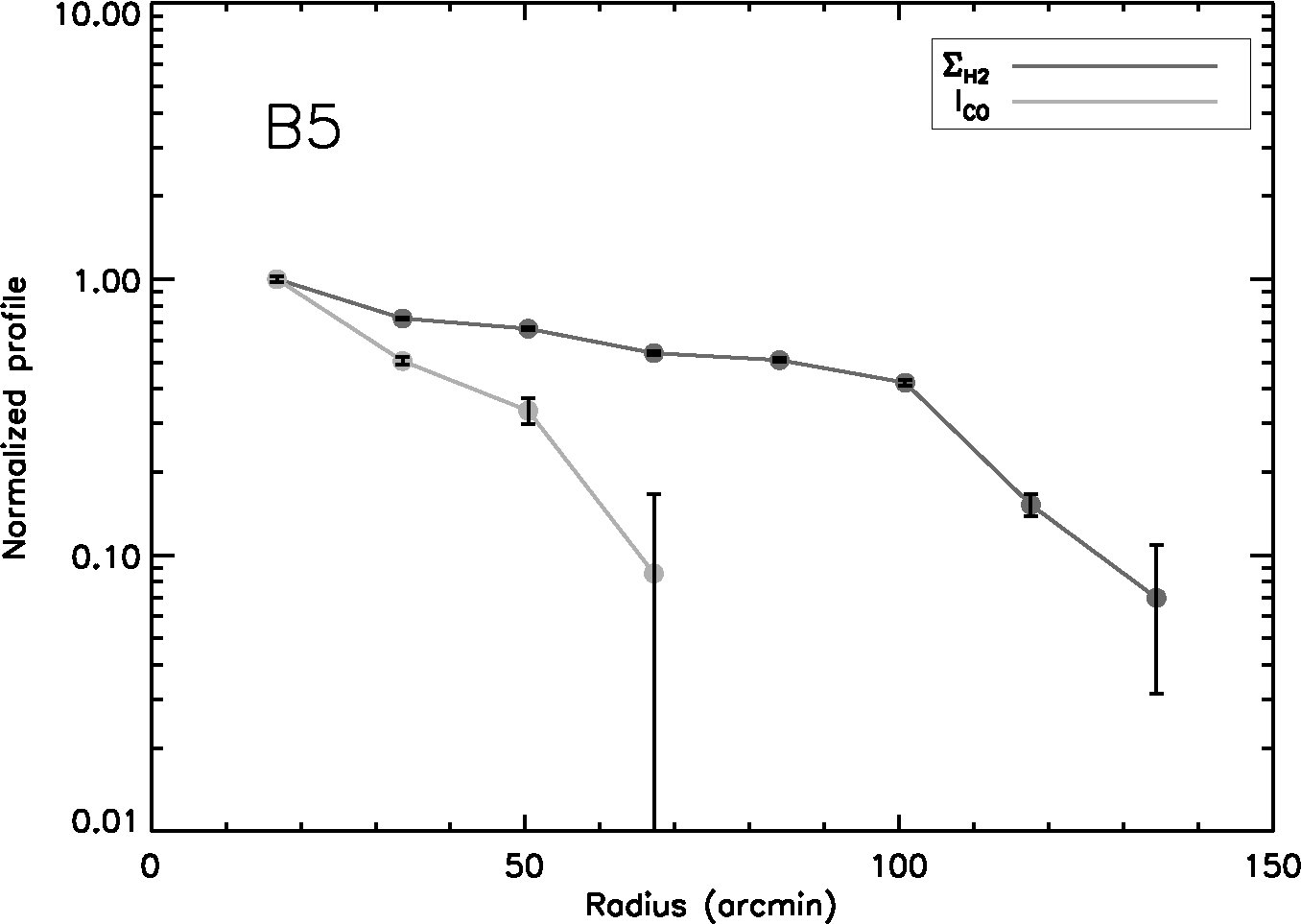}
\includegraphics[scale=0.46]{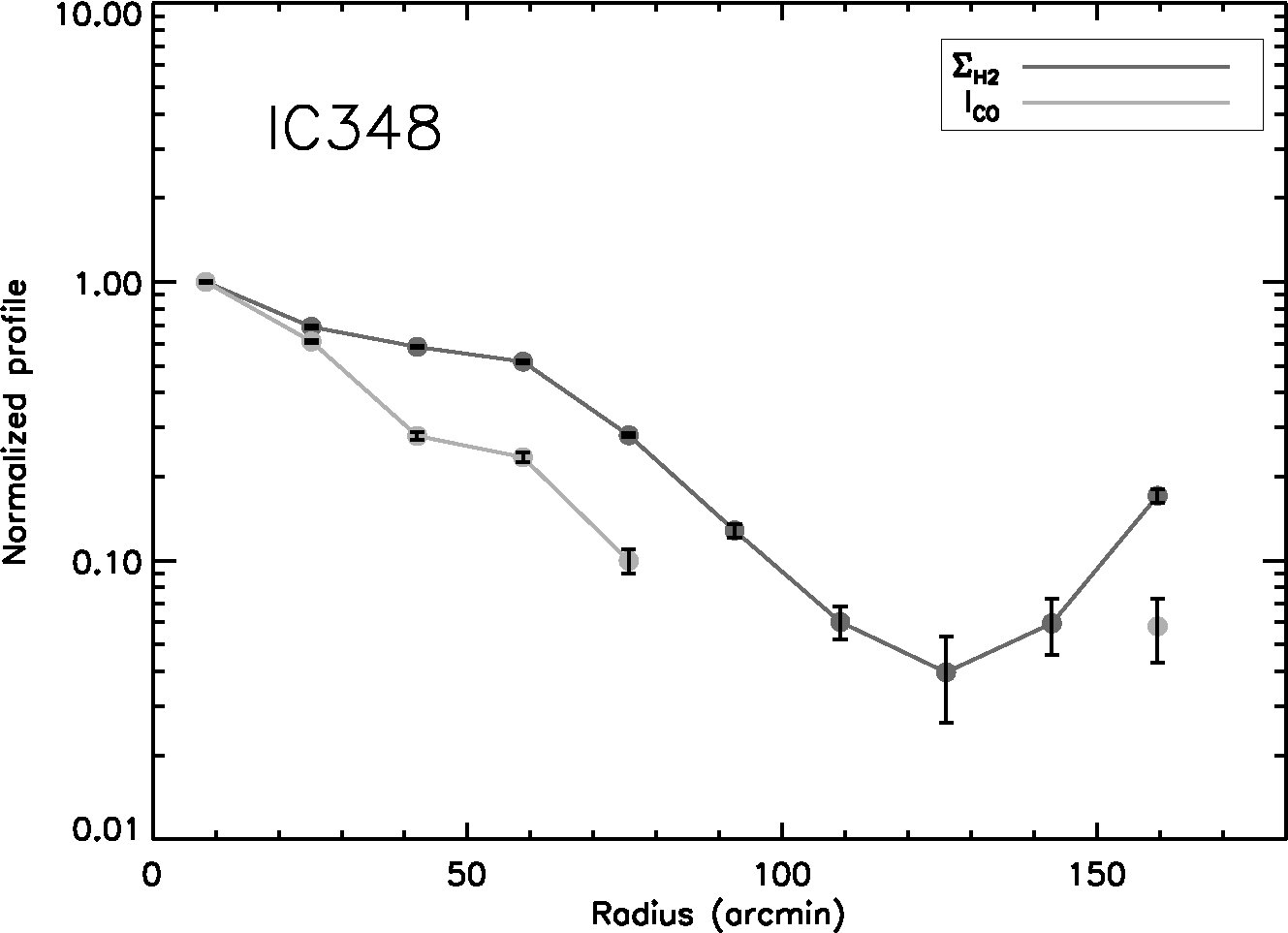}
\includegraphics[scale=0.46]{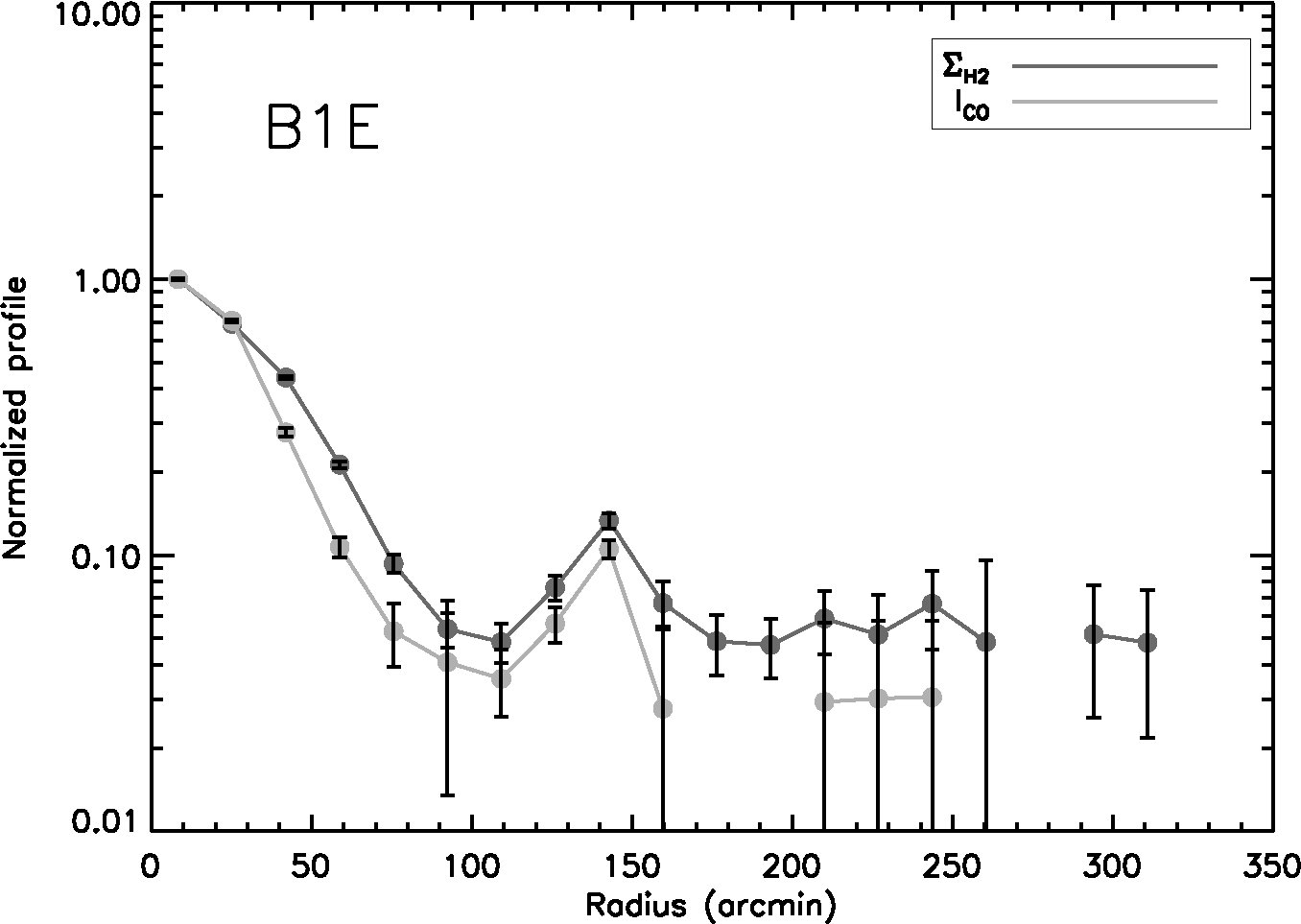}
\caption{\label{f:H2_CO_radial} $\Sigma_{\rm H2}$ (gray) and CfA $I_{\rm CO}$ (light gray) normalized radial profiles. 
For each radial profile, the data points with SNR > 1 are selected from the rectangular boxes (see Figure \ref{f:RH2})
and the mean $\Sigma_{\rm H2}$ and $I_{\rm CO}$ values in two-pixel size bins are calculated. 
(Left top) B5. (Right top) IC348. (Left bottom) B1E.}
\end{figure}
\end{landscape}

\begin{figure}
\begin{center}
\includegraphics[scale=0.35]{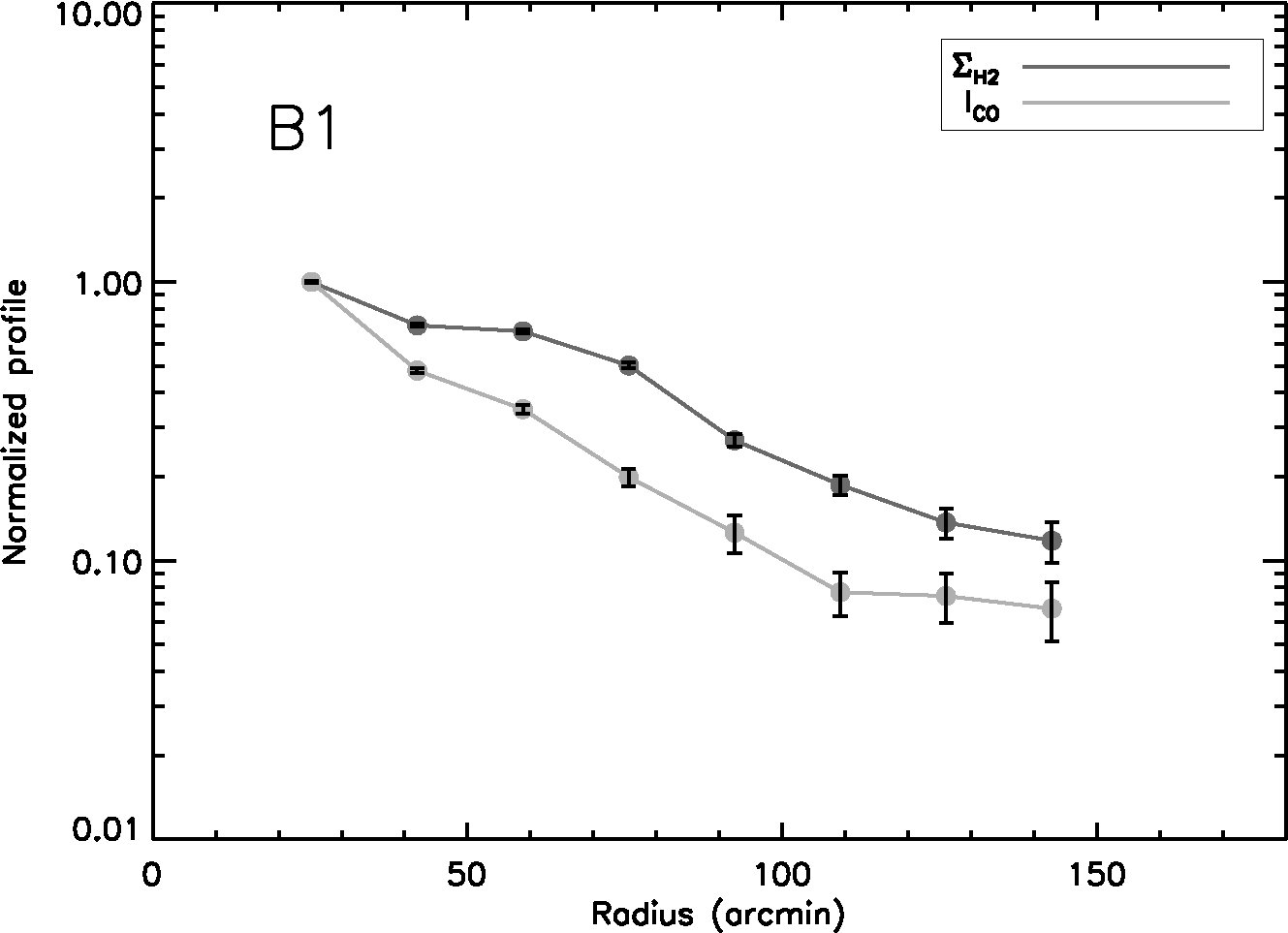}
\includegraphics[scale=0.35]{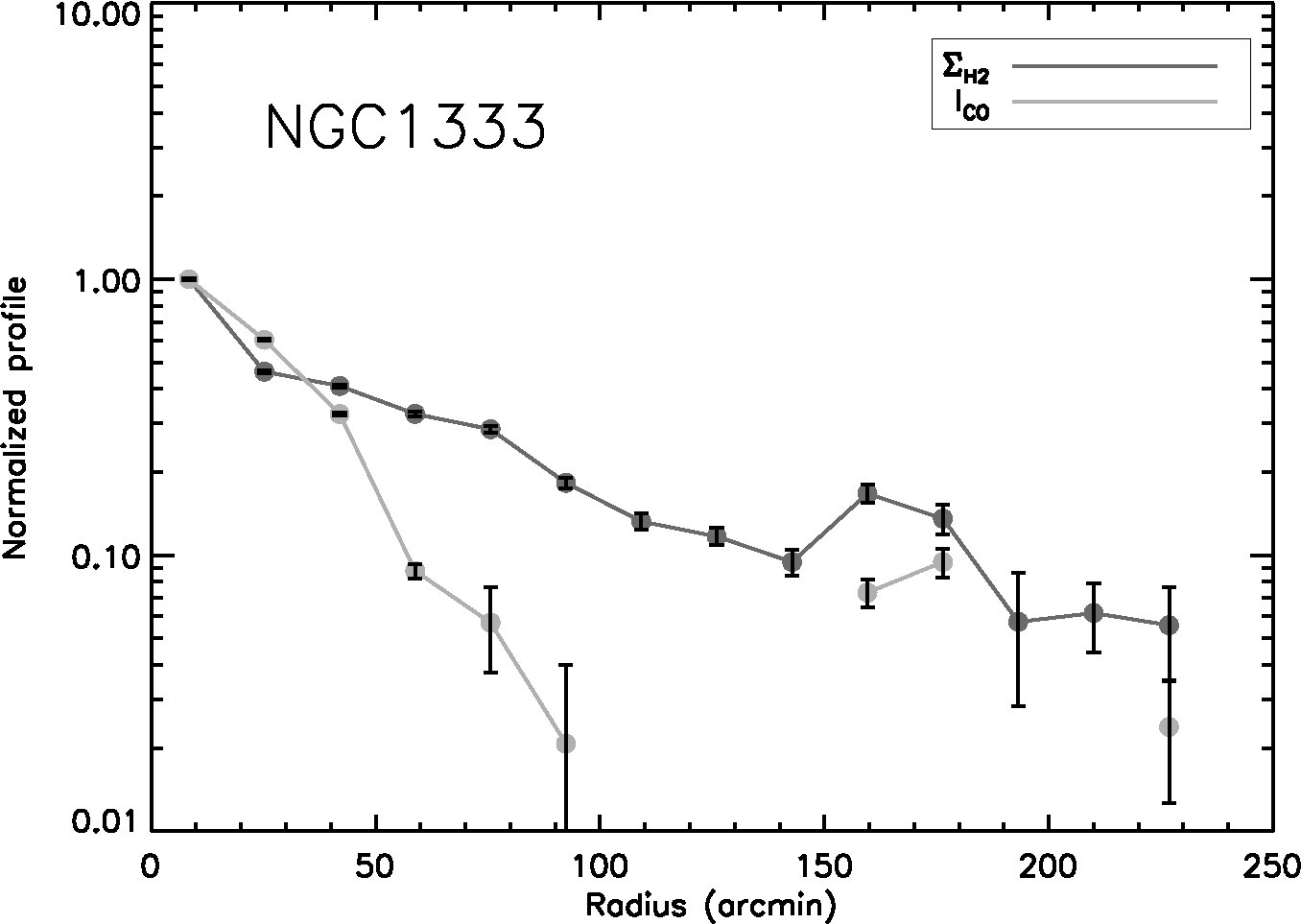}
\vskip 0.3cm
\begin{minipage}{16cm}
Fig. \ref{f:H2_CO_radial}.--- (Continued) (Left) B1. (Right) NGC1333.
\end{minipage}
\end{center}
\end{figure}
\clearpage

\end{document}